%% file: main.tex
\documentclass[12pt,letterpaper,preprint,prd,superscriptaddress,aps,floatfix,nofootinbib]{revtex4-1}

\pdfoutput=1
\usepackage {amsmath, amssymb, graphicx, multirow, rotating, slashed, color, xcolor}
\usepackage{hyperref}
\usepackage{comment}
\usepackage{enumitem}

\renewcommand{\thesection}{\arabic{section}}
\renewcommand{\thesubsection}{\thesection.\arabic{subsection}}

\makeatletter
\renewcommand{\p@subsection}{}
\renewcommand{\p@subsubsection}{}
\makeatother

\newcommand{\eotwash}{E\"ot-Wash}
\newcommand{\meff}{m_\mathrm{eff}}
\def\gsim{ \lower .75ex \hbox{$\sim$} \llap{\raise .27ex \hbox{$>$}} }
\def\lsim{ \lower .75ex \hbox{$\sim$} \llap{\raise .27ex \hbox{$<$}} }

\def\journal#1, #2, #3, #4#5#6#7{
   {\frenchspacing                     
    {#1}~{\bf{#2}},                    
    #3 (#4#5#6#7).}}%

\def\Collab#1{\etal\ (#1)}

\def\multiref#1{\begingroup            
    \frenchspacing                     
    \def\journal##1, ##2, ##3, ##4##5##6##7{%
    {##1}~{\bf{##2}}, 
    ##3 (##4##5##6##7)}
    #1 \endgroup}

\def\prd{\journal Phys. Rev. D, }%
\def\cqg{\journal Class. Quant. Grav., }%
\def\comment#1{}

\def\babar{\mbox{\sl B\hspace{-0.4em} {\small\sl A}\hspace{-0.37em} \sl B\hspace{-0.4em} {\small\sl A\hspace{-0.02em}R}}}

\makeatletter
\def\@seccntformat#1{\csname the#1\endcsname.\quad} 

\renewcommand\section{\@startsection
  {section}{1}{0mm}
  {-\baselineskip}
  {0.5\baselineskip}
  {\normalfont\normalsize\bf}}

\renewcommand\subsection{\@startsection
  {subsection}{2}{-5mm}
  {-\baselineskip}
  {0.5\baselineskip}
  {\normalfont\normalsize\bf}}
\makeatother

\makeatletter
\renewcommand\subsubsection{\@startsection
{subsubsection}{3}{0mm}{-\baselineskip}%
  {0.3\baselineskip}{\normalfont\normalsize\bfseries}}
\makeatother

\newcommand{\beq}{\begin{equation}} \newcommand{\eeq}{\end{equation}}
\newcommand{\bea}{\begin{eqnarray}} \newcommand{\eea}{\end{eqnarray}}

\let\oldbibliography\thebibliography
\renewcommand{\thebibliography}[1]{%
  \oldbibliography{#1}%
  \setlength{\itemsep}{-1pt}%
  }

\input{workshopsymbols}

\begin{document}


\preprint{YITP-SB-36}

\title{Dark Sectors and New, Light, Weakly-Coupled Particles}

\author{{\bf Conveners:} Rouven Essig}
\thanks{rouven.essig@stonybrook.edu}
\affiliation{C.N.~Yang Inst.~for Theoretical Physics, Stony Brook University, NY}

\author{John A.~Jaros}
\thanks{john@slac.stanford.edu}
\affiliation{SLAC National Accelerator Laboratory, Menlo Park, CA}

\author{William Wester}
\thanks{wester@fnal.gov}
\affiliation{Fermi National Accelerator Laboratory, Batavia, IL}

\author{\\\vskip 0.3cm P.~Hansson Adrian}
\affiliation{SLAC National Accelerator Laboratory, Menlo Park, CA}

\author{S.~Andreas}
\affiliation{Deutsches Elektronen-Synchrotron DESY, Hamburg, Germany}
\affiliation{Institut d'Astrophysique de Paris, Paris, France}

\author{T.~Averett}
\affiliation{College of William and Mary, Williamsburg, VA}

\author{O.~Baker}
\affiliation{Yale University, New Haven, CT}

\author{B.~Batell}
\affiliation{University of Chicago, Chicago, IL}

\author{M.~Battaglieri}
\affiliation{	Istituto Nazionale di Fisica Nucleare, Sezione di Genova, Italy}

\author{J.~Beacham}
\affiliation{New York University, NY}

\author{T.~Beranek}
\affiliation{Johannes Gutenberg University Mainz, Mainz, Germany}

\author{J.~D.~Bjorken}
\affiliation{SLAC National Accelerator Laboratory, Menlo Park, CA}

\author{F.~Bossi}
\affiliation{Laboratori Nazionali di Frascati dell'INFN, Frascati, Italy}

\author{J. R.~Boyce}
\affiliation{Thomas Jefferson National Accelerator Facility, VA}
\affiliation{College of William and Mary, Williamsburg, VA}

\author{G.~D.~Cates }
\affiliation{Department of Physics, University of Virginia, Virginia}

\author{A.~Celentano}
\affiliation{	Istituto Nazionale di Fisica Nucleare, Sezione di Genova, Italy}
\affiliation{Dipartimento di Fisica, Universita' di Genova, Genova, Italy}

\author{A.~S.~Chou}
\affiliation{Fermi National Accelerator Laboratory, Batavia, IL}

\author{R.~Cowan}
\affiliation{Laboratory for Nuclear Science, Massachusetts Inst.~of Technology, MA}

\author{F.~Curciarello}
\affiliation{Dipartimento di Fisica e di Scienze della Terra, Università di Messina, Italy}
\affiliation{Istituto Nazionale di Fisica Nucleare, Sezione Catania, Italy}

\author{H.~Davoudiasl}
\affiliation{Brookhaven National Laboratory, Upton, NY}

\author{P.~deNiverville}
\affiliation{Dept.~of Physics and Astronomy, University of Victoria, Canada}

\author{R.~De Vita}
\affiliation{Istituto Nazionale di Fisica Nucleare, Sezione Catania, Italy}

\author{A.~Denig}
\affiliation{Johannes Gutenberg University Mainz, Mainz, Germany}

\author{R.~Dharmapalan}
\affiliation{The University of Alabama, Tuscaloosa, AL}

\author{B.~Dongwi}
\affiliation{Department of Physics, Hampton University, MA}

\author{B.~D\"obrich}
\affiliation{Deutsches Elektronen-Synchrotron DESY, Hamburg, Germany}

\author{B.~Echenard}
\affiliation{California Institute of Technology, CA}

\author{D.~Espriu}
\affiliation{Universitat de Barcelona, Spain}

\author{S.~Fegan}
\affiliation{	Istituto Nazionale di Fisica Nucleare, Sezione di Genova, Italy}

\author{P.~Fisher}
\affiliation{Laboratory for Nuclear Science, Massachusetts Inst.~of Technology, MA}

\author{G.~B.~Franklin}
\affiliation{Department of Physics, Carnegie Mellon University, PA}

\author{A.~Gasparian}
\affiliation{North Carolina A\&T State University, NC}

\author{Y.~Gershtein}
\affiliation{Rutgers University, NJ}

\author{M.~Graham}
\affiliation{SLAC National Accelerator Laboratory, Menlo Park, CA}

\author{P.~W.~Graham}
\affiliation{Department of Physics, Stanford University, Stanford, CA}

\author{A.~Haas}
\affiliation{New York University, NY}

\author{A.~Hatzikoutelis}
\affiliation{University of Tennessee, Knoxville, TN}

\author{M.~Holtrop}
\affiliation{University of New Hampshire, NH}

\author{I.~Irastorza}
\affiliation{Universidad de Zaragoza, Spain}

\author{E.~Izaguirre}
\affiliation{Perimeter Inst.~for Theoretical Physics, Waterloo, Canada}

\author{J.~Jaeckel}
\affiliation{Institut f\"ur theoretische Physik, Universit\"at Heidelberg, Germany}

\author{Y.~Kahn}
\affiliation{Center for Theoretical Physics, Massachusetts Institute of Technology, MA}

\author{N.~Kalantarians}
\affiliation{Department of Physics, Hampton University, MA}

\author{M.~Kohl}
\affiliation{Department of Physics, Hampton University, MA}

\author{G.~Krnjaic}
\affiliation{Perimeter Inst.~for Theoretical Physics, Waterloo, Canada}

\author{V.~Kubarovsky}
\affiliation{Thomas Jefferson National Accelerator Facility, VA}

\author{H-S.~Lee}
\affiliation{College of William and Mary, Williamsburg, VA}
\affiliation{Thomas Jefferson National Accelerator Facility, VA}

\author{A.~Lindner}
\affiliation{Deutsches Elektronen-Synchrotron DESY, Hamburg, Germany}

\author{A.~Lobanov}
\affiliation{Max-Planck-Institut f\"ur Radioastronomie, Bonn, Germany}
\affiliation{Institut f\"ur Experimentalphysik, Universit\"at Hamburg, Germany}

\author{W.~J.~Marciano}
\affiliation{Brookhaven National Laboratory, Upton, NY}

\author{D.~J.~E.~Marsh}
\affiliation{Perimeter Inst.~for Theoretical Physics, Waterloo, Canada}

\author{T.~Maruyama}
\affiliation{SLAC National Accelerator Laboratory, Menlo Park, CA}

\author{D.~McKeen}
\affiliation{Dept.~of Physics and Astronomy, University of Victoria, Canada}

\author{H.~Merkel}
\affiliation{Johannes Gutenberg University Mainz, Mainz, Germany}

\author{K.~Moffeit}
\affiliation{SLAC National Accelerator Laboratory, Menlo Park, CA}

\author{P.~Monaghan}
\affiliation{Department of Physics, Hampton University, MA}

\author{G.~Mueller}
\affiliation{University of Florida, Gainesville, FL}

\author{T.~K.~Nelson}
\affiliation{SLAC National Accelerator Laboratory, Menlo Park, CA}

\author{G.R.~Neil}
\affiliation{Thomas Jefferson National Accelerator Facility, VA}

\author{M.~Oriunno}
\affiliation{SLAC National Accelerator Laboratory, Menlo Park, CA}

\author{Z.~Pavlovic}
\affiliation{Los Alamos National Laboratory, Los Alamos, NM}

\author{S.~K.~Phillips}
\affiliation{University of New Hampshire, NH}

\author{M.~J.~Pivovaroff}
\affiliation{Lawrence Livermore National Laboratory, CA}

\author{R.~Poltis}
\affiliation{University of Cape Town, Cape Town, South Africa}

\author{M.~Pospelov}
\affiliation{Perimeter Inst.~for Theoretical Physics, Waterloo, Canada}

\author{S.~Rajendran}
\affiliation{Department of Physics, Stanford University, Stanford, CA}

\author{J.~Redondo}
\affiliation{Max Planck Institute f\"ur Physik, M\"unchen, Germany}
\affiliation{Arnold Sommerfeld Center, Ludwig-Maximilians-University, Mu\"unchen, Germany}

\author{A.~Ringwald}
\affiliation{Deutsches Elektronen-Synchrotron DESY, Hamburg, Germany}

\author{A.~Ritz}
\affiliation{Dept.~of Physics and Astronomy, University of Victoria, Canada}

\author{J.~Ruz}
\affiliation{Lawrence Livermore National Laboratory, CA}

\author{K.~Saenboonruang}
\affiliation{Kasetsart University, Chatuchak, Bangkok, Thailand}

\author{P.~Schuster}
\affiliation{Perimeter Inst.~for Theoretical Physics, Waterloo, Canada}

\author{M.~Shinn}
\affiliation{Thomas Jefferson National Accelerator Facility, VA}

\author{T.~R.~Slatyer}
\affiliation{Center for Theoretical Physics, Massachusetts Institute of Technology, MA}

\author{J.~H.~Steffen}
\affiliation{Northwestern University, IL}

\author{S.~Stepanyan}
\affiliation{Thomas Jefferson National Accelerator Facility, VA}

\author{D.~B.~Tanner}
\affiliation{University of Florida, Gainesville, FL}

\author{J.~Thaler}
\affiliation{Center for Theoretical Physics, Massachusetts Institute of Technology, MA}

\author{M.~E.~Tobar}
\affiliation{School of Physics, University of Western Australia, WA, Australia}

\author{N.~Toro}
\affiliation{Perimeter Inst.~for Theoretical Physics, Waterloo, Canada}

\author{A.~Upadye}
\affiliation{Argonne National Laboratory, IL}
\affiliation{Institute for the Early Universe, Ewha University, Seoul, Korea}

\author{R.~Van de Water}
\affiliation{Los Alamos National Laboratory, Los Alamos, NM}

\author{B.~Vlahovic}
\affiliation{North Carolina Central University, Durham, NC}

\author{J.~K.~Vogel}
\affiliation{Lawrence Livermore National Laboratory, CA}

\author{D.~Walker}
\affiliation{SLAC National Accelerator Laboratory, Menlo Park, CA}

\author{A.~Weltman}
\affiliation{University of Cape Town, Cape Town, South Africa}

\author{B.~Wojtsekhowski}
\affiliation{Thomas Jefferson National Accelerator Facility, VA}

\author{S.~Zhang}
\affiliation{Thomas Jefferson National Accelerator Facility, VA}

\author{K.~Zioutas}
\affiliation{University of Patras, Greece} 
\affiliation{CERN, Geneva, Switzerland}

\begin{abstract}
\noindent
Dark sectors, consisting of new, light, weakly-coupled particles that do not interact with the known strong, weak, or electromagnetic forces, are a particularly compelling possibility for new physics.  
Nature may contain numerous dark sectors, each with their own beautiful structure, distinct particles, and forces.  
This review summarizes the physics motivation for dark sectors and the exciting opportunities for experimental exploration. 
It is the summary of the Intensity Frontier subgroup ``New, Light, Weakly-coupled Particles'' of the 
Community Summer Study 2013 (Snowmass).  
We discuss axions, which solve the strong CP problem and are an excellent dark matter candidate, and their generalization to axion-like particles. 
We also review dark photons and other dark-sector particles, including sub-GeV dark matter, which are theoretically natural, provide for dark matter candidates or new dark matter interactions, and could resolve outstanding puzzles in particle and astro-particle physics.  
In many cases, the exploration of dark sectors can proceed with existing facilities and comparatively modest experiments. A rich, diverse, and low-cost experimental program has been identified that has the potential for one or more game-changing discoveries. These physics opportunities should be vigorously pursued in the US and elsewhere. 
\end{abstract}

\maketitle

\tableofcontents

\vskip 1cm

\section{Overview and Executive Summary}
\label{sec:nlwcp:overview}

The Standard Model (SM) of particle physics has achieved remarkable success as a result of several decades of exploration, of constantly pushing the boundaries of our knowledge of theory, experiment, and technology.  However, while the SM provides a theoretically consistent description of all known particles and their interactions (ignoring gravity) up to the Planck scale, it is clearly incomplete as it does not address several pieces of evidence for new physics beyond the SM. 
 
One particularly powerful piece of evidence for new physics comes from the existence of dark matter (DM).  DM dominates the matter density in our Universe, but very little is known about it.  Its existence provides a strong hint that there may be a \emph{dark sector}, consisting of particles that do not interact with the known strong, weak, or electromagnetic forces.  Given the intricate structure of the SM, which describes only a subdominant component of the Universe, it would not be too surprising if the dark sector contains a rich structure itself, with DM making up only a part of it.  Indeed, many dark sectors could exist, each with its own beautiful structure, distinct particles, and forces. 
These dark sectors (or ``hidden sectors'') may contain \emph{new light weakly-coupled particles}, particles well below the Weak-scale that interact only feebly with ordinary matter.  Such particles could easily have escaped past experimental searches, but a rich experimental program has now been devised to look for several well-motivated possibilities.  

Dark sectors are motivated also by bottom-up and top-down theoretical considerations.  They arise in many theoretical extensions to the SM, such as moduli that are present in string theory or new (pseudo-)scalars that appear naturally when symmetries are broken at high energy scales.  Other powerful motivations include the strong CP problem, and various experimental findings, including the discrepancy between the calculated and measured anomalous magnetic moment of the muon and puzzling results from astrophysics. 
Besides gravity, there are only a few well-motivated interactions allowed by SM symmetries that provide a ``portal" from the SM sector into the dark sector.  These portals include:

\begin{center}
\begin{tabular}{c|c|c}
Portal & Particles & Operator(s) \\ 
\hline
``Vector''  & Dark photons & $-\frac{\epsilon}{2\cos\theta_W}B_{\mu\nu}F'^{\mu\nu}$ \\
``Axion''  & Pseudoscalars & $
\frac{a}{f_a}F_{\mu\nu}\widetilde F^{\mu\nu},
\frac{a}{f_a}G_{i\mu\nu}\widetilde G_i^{\mu\nu},
\frac{\partial_{\mu}a}{f_{a}} \overline{\psi}\gamma^{\mu}\gamma^{5}\psi$ \\
``Higgs''  & Dark scalars & $ (\mu S + \lambda S^{2})H^{\dagger}H $ \\
``Neutrino'' & Sterile neutrinos & $ y_N LHN $ \\
\end{tabular}
\end{center}
The Higgs and neutrino portal are best explored at high-energy colliders and neutrino facilities, respectively.  Our focus here will be on the 
vector and axion portals.  While these can also be explored at the cosmic and energy frontiers, they present particularly well-motivated 
targets for several low-cost, high-impact experiments at the intensity frontier. 

This paper is the summary of the work of the Intensity Frontier subgroup
 ``New, Light, Weakly-coupled Particles'' of the Community Summer Study 2013 (``Snowmass on the Mississippi''). This paper updates 
 and elaborates the summary included in the Fundamental Physics and the Intensity Frontier workshop report~\cite{Hewett:2012ns}. 
This topic has also been studied in the context of the European strategy~\cite{Baker:2013zta}.  

The outline of this paper is as follows. 
The next subsection contains a brief executive summary.  
\S\ref{sec:nlwcp:axions-&-alps} discusses the QCD axion and more general ``axion-like'' particles (ALPs). \S\ref{sec:nlwcp:dark-photons} reviews dark photons, focusing on sub-MeV and MeV-GeV masses. \S\ref{sec:nlwcp:LDM} describes sub-GeV DM, milli-charged particles, and other dark-sector particles.  \S\ref{sec:nlwcp:chameleons} focuses on chameleons. In all cases, we describe the theoretical and phenomenological motivation, current constraints, and current and future experimental opportunities.  
\S \ref{sec:nlwcp:conclusions} contains our conclusions.  

\subsection{Executive Summary}

A brief executive summary for our subgroup is the following:
\begin{itemize}[leftmargin=*]\setlength{\itemsep}{0pt}
\item The existence of dark sectors, consisting of new, light, weakly-coupled particles that do not interact with the known 
Standard Model forces, is theoretically and phenomenologically motivated.  
A particularly strong motivation is the existence of dark matter.  
Extensions to the Standard Model in various top-down and bottom-up theoretical considerations naturally contain dark sectors, 
which may contain new particles like dark photons, scalars, pseudo-scalars, pseudo-vectors, and fermions.  
The solution to the strong CP problem motivates axions, and understanding the nature of dark energy motivates the search for chameleons.  
Various data ``anomalies'', such as the discrepancy between the measured and calculated 
muon anomalous magnetic moment and some puzzling results from astrophysics provide exciting phenomenological motivations. 
\item New physics need not reside at high energies, at the TeV-scale and above.  
It could well be found at the ``low-energy frontier'', and be accessible with intensity-frontier tools.  
Indeed, experiments that use intense beams of photons, charged particles, and/or sensitive detectors 
may be used to directly produce and study new, feebly-interacting particles that lie well below the Weak scale.  
\item Existing facilities and technologies and modest experiments enable the exploration of dark sectors.
A rich, diverse, and low-cost experimental program is already underway that has the potential for one or more game-changing 
discoveries.  Current ideas for extending  the searches to smaller couplings and higher masses increase this potential markedly.  
\item The US high-energy physics program needs to include these experimental searches, especially when the investment 
is so modest, the motives so clear, and the payoff so spectacular. 
Since we do not know which guiding principle for finding new physics will ultimately bear fruit, 
the support for a diverse experimental program is crucial.  
At present, nearly all the experimental efforts world-wide  have strong US contributions or significant US leadership, a position 
that should be maintained. 
\end{itemize}

\section{Axions and Axion-Like Particles}
\label{sec:nlwcp:axions-&-alps}

\subsection{Theory \& Theory Motivation}
\label{subsec:nlwcp:axions-&-alps:theory}

One of the unresolved puzzles in the SM is the lack of any observed $CP$ violation in the strong interactions described by Quantum Chromodynamics (QCD).  While the weak interactions are known to violate $CP$, 
the strong interactions also contain a $CP$-violating term in the Lagrangian, $\frac{\Theta}{32 \pi^2} G_{\mu\nu} \tilde G^{\mu\nu}$, where $G^{\mu\nu}$ is the gluon field strength.  
For non-zero quark masses, this term leads to (unobserved) $CP$-violating effects of the strong interactions.  
This so-called ``strong $CP$ problem'' is often exemplified by the lack of observation of a neutron electric dipole moment down 
to a present experimental upper limit 10 orders of magnitude smaller than what is expected from a $CP$-violating QCD.

Solutions to this problem are scarce.  Perhaps the most popular suggestion is the so-called Peccei-Quinn (PQ) $U(1)$ approximate global symmetry, which is spontaneously broken at a scale $f_a$. The axion is a hypothetical particle that arises as the pseudo-Nambu-Goldstone boson (PNGB) of this symmetry breaking~\cite{Peccei:1977hh,Weinberg:1977ma,Wilczek:1977pj}.

The axion mass would be $m_a \sim 6~{\rm meV}~(10^9~{\rm GeV}/f_a)$. Its coupling to ordinary matter is proportional to $1/f_a$ and can be calculated in specific models. It couples generically to quarks (or hadrons at low energies) with coupling constants that are uncertain by $\mathcal{O}(1)$ model-dependent factors. The coupling to photons in also generic and has the form 
$\mathcal{L}\supset -\frac{1}{4} \,g_{a\gamma\gamma} \,a \,F_{\mu\nu}\, \tilde{F}^{\mu\nu}$, where the coupling constant 
$g_{a\gamma\gamma} \sim 10^{-13}~{\rm GeV}^{-1}~(10^{10}~{\rm GeV}/f_a)$~\cite{hpsaw:Nakamura:2010zzi} has only a $\mathcal{O}(1)$ model-dependency. The couplings to leptons are not guaranteed but appear in many theoretical realizations of the axion, particularly in models of grand unification.   
All of these interactions can play a role in searches for the axion, and allow the axion to be produced or detected in the laboratory and emitted by the sun or other stars.  

The basic physical mechanism that leads to the axion --- the spontaneous breaking at a high energy scale of a $U(1)$ approximate global symmetry, generating a light PNGB --- also allows for other axion-like particles (ALPs).  Unlike axions, which are linked to the strong interactions and whose masses and couplings are determined by a single new parameter $f_a$, ALPs are much less constrained, and their masses and couplings to photons are independent parameters.  
Searches for ALPs should not therefore be limited to the parameter space of the axion itself.
Both ALPs and axions are generic in string theory~\cite{arXiv:1002.0329,Print-84-0838(PRINCETON),hep-th/0602233,hep-th/0605206,arXiv:0905.4720,Acharya:2010zx,Cicoli:2012sz}, with the natural size of 
their decay constant $f_a$ being the string scale, varying typically between $10^9$ and
$10^{17}$~GeV.

\subsection{Phenomenological Motivation and Current Constraints}
\label{subsec:nlwcp:axions-&-alps:pheno}

\begin{figure}[!t]
\centering
\includegraphics[width=0.7\textwidth]{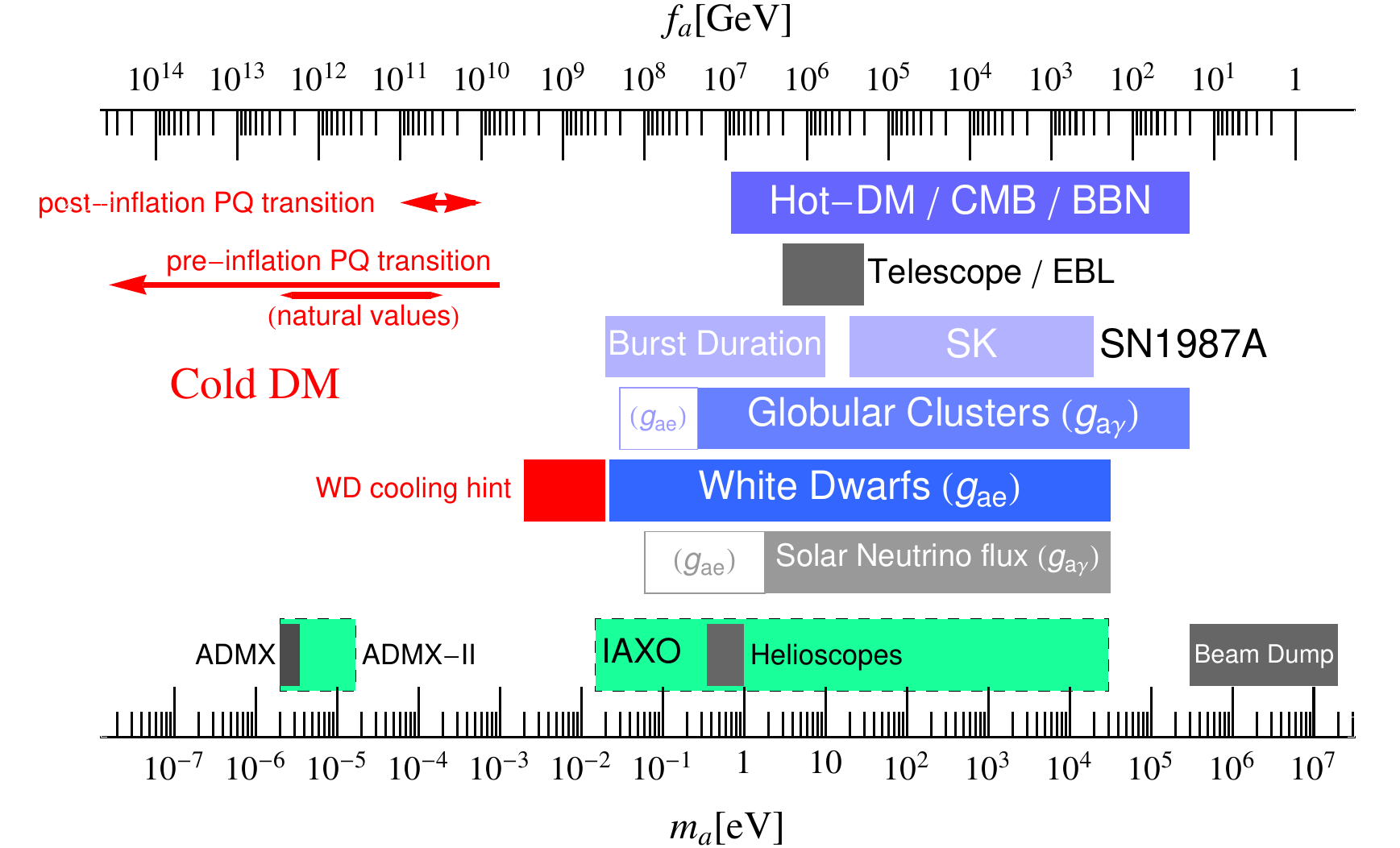} \\
\vskip 2mm
\includegraphics[width=0.7\textwidth]{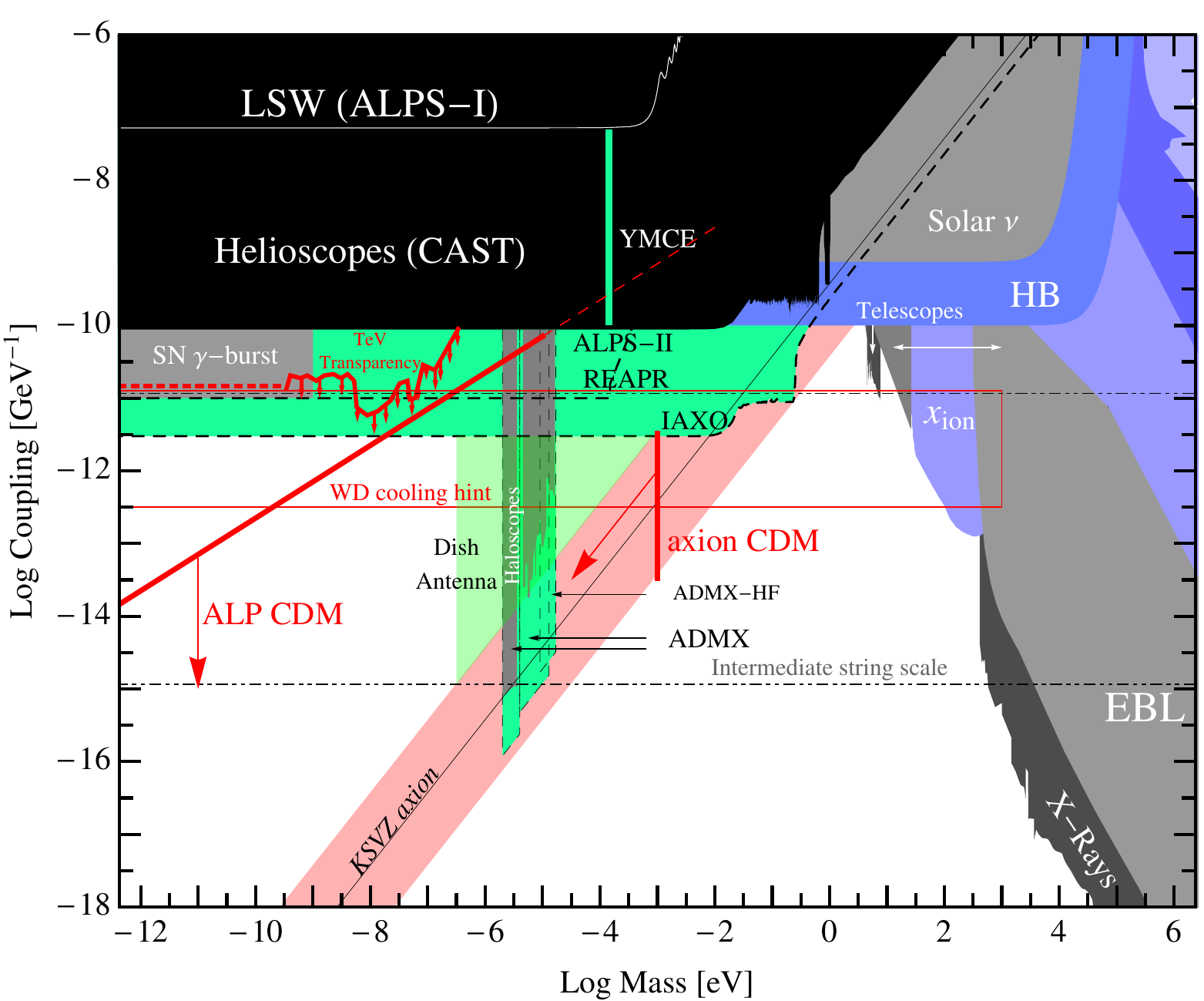} 
\caption{
Parameter space for axions (top) and axion-like particles (ALPs) (bottom).  In the bottom plot, the 
QCD axion models lie within an order of magnitude from the explicitly shown ``KSVZ'' axion line (red band). 
Colored regions are: experimentally excluded regions (dark green), constraints from 
astronomical observations (gray) or from astrophysical or cosmological arguments (blue), and 
sensitivity of planned and suggested experiments (light green)  (ADMX~\cite{Asztalos:2011ei}, ALPS-II~\cite{Bahre:2013ywa}, IAXO~\cite{Irastorza:2011gs,Vogel:2013bta,Irastorza:1567109}, Dish antenna~\cite{Horns:2012jf}). Shown in red are boundaries where ALPs can account for all the dark 
matter produced either thermally in the big bang or non-thermally by the misalignment mechanism.
}\label{fig:hspaw-ALPs}
\end{figure}

Fig.~\ref{fig:hspaw-ALPs} (top) shows the allowed axion parameter space as a function of $f_a$ or, equivalently, $m_a$.  
Direct searches for such particles and calculations of their effect on the cooling of stars and on the supernova SN1987A 
exclude most values of $f_a < 10^9$~GeV.  Some of these constrain only the axion coupling to 
photons ($g_{a\gamma\gamma}$), while others constrain the axion coupling to electrons ($g_{ae}$).  
Recent and future laboratory tests (the latter shown in light green) can probe $f_a\lesssim 10^9$ GeV or $f_a \gtrsim 10^{12}$ GeV (possibly higher) but intermediate values are more challenging.  

The parameter space for ALPs is shown in Fig.~\ref{fig:hspaw-ALPs} (bottom).  The axion parameter space lies within 
an order of magnitude from the line labeled ``KSVZ axion,'' which represents a particular QCD axion model.  
Experimentally excluded regions (dark green), constraints 
from astronomical observations (gray) or from astrophysical, or cosmological arguments (blue) are shown.  Sensitivities of 
a few planned experiments are shown in light green.

\subsubsection{Dark Matter}
\label{subsubsec:nlwcp:axions-&-alps:pheno:DM}

ALPs (including the QCD axion) can naturally serve as the Universe's DM, meaning that the galactic halo may be formed partly or entirely from these particles.  They can be produced thermally or non-thermally in the early Universe.  Thermally produced axions are disfavored by observations of the Universe's large scale structure \cite{Hannestad:2010yi}, but thermally produced ALP DM is still allowed in sizable parts of parameter space ($m_a \gtrsim154$ eV, $g_{a\gamma\gamma}\sim \mathcal{O}(10^{-17}-10^{-13})$ GeV$^{-1}$)~\cite{Cadamuro:2010cz}.  
Non-thermal production can occur through the ``vacuum misalignment mechanism'' or the decay of axionic strings and domain walls. 
Axions with large $f_a$ do not thermalize in the early Universe and their abundance today is set by the initial state set during the Peccei-Quinn phase transition. 
There are two scenarios depending on whether the PQ transition took place after or before inflation. 
In the first case, the dominant contribution arises from the decay of cosmic strings and domain walls into axions. 
This scenario suggests values of $m_a\sim 80-400~\mu$eV with large uncertainties arising from extrapolating the numerical result for the string and domain wall decays~\cite{Sikivie:2006ni,Hiramatsu:2012gg}.   
In the second scenario, inflation homogenizes the initial axion field value in our observable Universe and the 
DM density depends on this value. For values $a_{\rm initial}\sim f_a$ the observed DM density arises for $m_a\sim 12~\mu$eV. 
Smaller values of the mass are possible when $a_{\rm initial}\ll f_a$ and somewhat larger masses (perhaps up to meV~\cite{Wantz:2009it}) can be achieved by tuning towards $a_{\rm initial}= \pi f_a$.  

All in all, the natural values $m_a\sim  10^{-5}-10^{-4}$ eV present a clear experimental target.  The  Axion DM eXperiment (ADMX) will soon probe part of this preferred parameter space.  
Extending these arguments to ALPs, a much larger parameter space needs to be 
explored as indicated in Fig.~\ref{fig:hspaw-ALPs}; see also {\it e.g.},~\cite{Arias:2012az}.  

One important constraint on axion (or ALP) DM is the generation of isocurvature temperature fluctuations in the cosmic microwave background if the axion/ALP exists during inflation. Cosmic microwave background (CMB) probes like the Planck satellite constrain these fluctuations, setting very strong constraints on the Hubble scale during inflation, $H_I\lesssim { O}(10^6)$ GeV. 
Observing tensor modes in the CMB allows one to determine $H_I$, providing a crucial test of axion/ALP DM. 

It is noteworthy that axion or ALP DM may also form a Bose-Einstein condensate~\cite{Erken:2011dz}, which may lead to caustic rings in spiral galaxies, which may already have been observed. This also has detectable consequences in terrestrial direct detection experiments like ADMX. 

Ultra-light ALPs with masses in the $10^{-33}-10^{-18}$ eV range can also contribute to the DM in the Universe, affecting structure formation in a manner distinct from cold DM (CDM). The distinction arises due to a scale dependent sound speed in the ultra-light ALPs fluid \cite{Hu:2000ke,BoylanKolchin:2011de,Amendola:2005ad}. Large scale structure and the CMB thus allow one to constrain the fraction of DM that can be made up of such ultra-light ALPs. 
Future surveys such as Euclid stand to improve constraints with specific improvements at the lowest masses and with discerning differences between ultra-light ALPs and thermal neutrinos of eV mass \cite{Marsh:2011bf,Amendola:2012ys}. The effect of these ALPs on the CMB and weak lensing tomography has been explored in detail in \cite{Marsh:2011bf,Marsh:2013taa}. 
Furthermore, if these ultra-light ALPs are fundamental fields present during inflation they carry isocurvature perturbations. This allows to do consistency checks but also test models of inflation \cite{Marsh:2013taa,Burgess:2013sla}. 

\subsubsection{Hints from astrophysics}
\label{subsubsec:nlwcp:axions-&-alps:pheno:astrophysics}

In the last few years some astrophysical anomalies have found plausible explanations in terms of axion/ALPs 
suggesting target areas in parameter space reachable by near-future experiments. 
We refer here to the apparent non-standard energy loss of white dwarf stars, {\it e.g.},~\cite{arXiv:0806.2807,arXiv:0812.3043,Isern:2012ef,Corsico:2012ki,Corsico:2012sh} (see however~\cite{Melendez:2012iq}) and the anomalous transparency of the Universe for TeV gamma rays, {\it e.g.},~\cite{Horns:2012fx,arXiv:0707.4312,arXiv:0712.2825,arXiv:0905.3270,arXiv:1106.1132,Meyer:2013pny}. 
The required coupling strengths seem within reach in controlled laboratory experiments at the intensity frontier, 
and can serve as useful benchmarks, c.f. Fig.~\ref{fig:hspaw-ALPs}.

In the mass range $10^{-24}$ eV$\leq m_a \lesssim 10^{-20}$  eV, large scale structure formation of ultra-light ALPs is analogous to warm DM (WDM), and is thus relevant to problems with CDM structure formation, such as the cusp-core, missing-satellites, and too-big-to-fail problems~\cite{BoylanKolchin:2011de}. The virtue of ultra-light ALPs is that they avoid the so-called `Catch 22' of WDM~\cite{Maccio:2012qf}. 
The relevance of ultra-light ALPs to these problems in large scale structure is explored in~\cite{Marsh:2013ywa}.

\subsection{Experimental Searches for Axions and ALPs: Status and Plans}
\label{subsec:nlwcp:axions-&-alps:experiments}

\subsubsection{Laser Experiments}
\label{subsubsec:nlwcp:axions-&-alps:experiments:Laser}

The simplest and most unambiguous purely laboratory experiment to
look for axions (or light scalars or pseudoscalars more generally)
is photon regeneration~\cite{vanB87} (``shining light through the
wall''~\cite{Redondo:2010dp}). A laser beam traverses a magnetic field, and the field
stimulates a small fraction of photons to convert to axions of the
same energy. A material barrier easily blocks the primary laser beam;
in contrast, the axion component of the beam travels through the wall
unimpeded and enters a second magnet. There, with the same
probability, the axions are converted back to photons. Because the
photon-regeneration rate goes as $g_{a\gamma\gamma}^{4}$, the
sensitivity of the experiment is poor in its basic form, improved
only by increasing the laser intensity, the magnetic field strength,
or the length of the interaction regions. As initially suggested by
Hoogeveen and Ziegenhagen~\cite{Hoogeveen91} and recently discussed
in detail~\cite{Siki07,mueller09,Sikiviefest,futurelsw} very large gains may be realized
in both the photon-regeneration rate and in the resulting limits on
$g_{a\gamma\gamma}$ by introducing matched optical resonators in
both the axion production and the photon regeneration regions.

Detailed designs for such an experiment exist, including the scheme
for locking two matched high-finesse optical resonators, the signal
detection method, and the ultimate noise
limits~\cite{mueller09,Sikiviefest,Bahre:2013ywa}. Such experiments would improve on
present limits on $g_{a\gamma\gamma}$ by at least a factor of 10. We
note also that these experiments, although challenging, are feasible
using well-established technologies developed for example for laser
interferometer gravitational-wave
detectors~\cite{LIGO-NIM-RPP,LISAUF}. No new technology is needed. 
Two developed designs exist: the Resonantly Enhanced Axion-Photon 
Regeneration (REAPR) experiment, a Florida-Fermilab 
collaboration, and the Any Light Particle Search II (ALPS II) being 
mounted at DESY.

\begin{figure}[!t]
\centering
\includegraphics[width=0.7\textwidth]{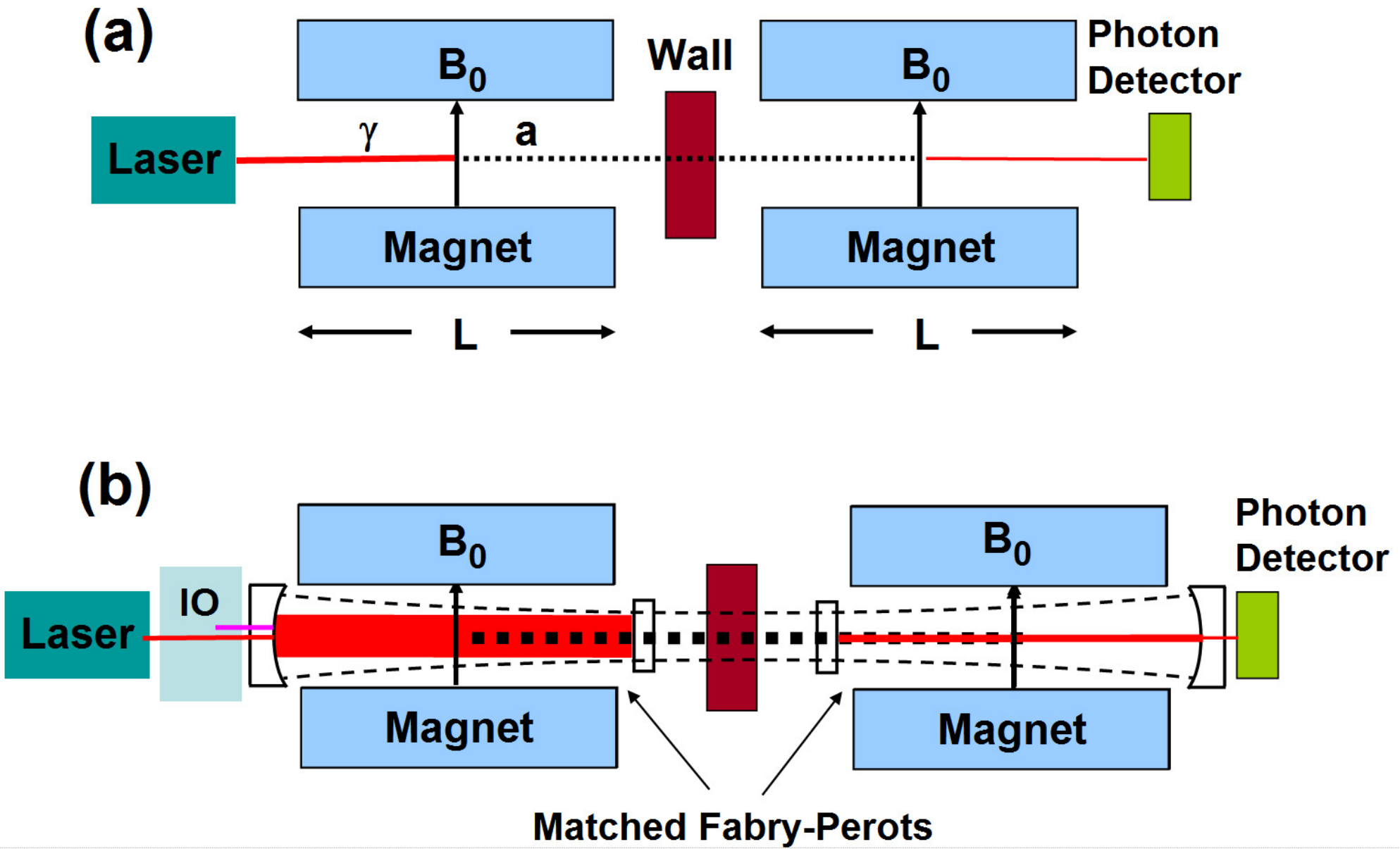} 
\caption{(a) Simple photon regeneration to produce axions or axion-like particles. (b)
Resonant photon regeneration, employing matched Fabry-Perot
cavities. The overall envelope schematically shown by the thin
dashed lines indicates the important condition that the axion wave,
and thus the Fabry-Perot mode, in the photon regeneration cavity
must follow that of the hypothetically unimpeded photon wave from
the Fabry-Perot mode in the axion generation magnet. Between the
laser and the cavity are optics (IO) that manage mode
matching of the laser to the cavity, imposes RF sidebands for
reflection locking of the laser to the cavity, and provides
isolation for the laser. The detection system is also fed by
matching and beam-steering optics. Not shown is the second laser for 
locking the regeneration cavity and for heterodyne readout.}
\label{fig:Reprfig1}
\end{figure}

Figure~\ref{fig:Reprfig1}(a) shows the photon regeneration experiment as
usually conceived. If $E_{0}$ is the amplitude of the laser field
propagating to the right, the amplitude of the axion field traversing
the wall is $E_{0}\sqrt{P}$ where $P$ is the conversion probability
in the magnet on the LHS of 
Fig.~\ref{fig:Reprfig1}a. Let $P^{\prime}$ be the
conversion probability in the magnet on the RHS. 
The field generated on that side is then $E_{S}=E_{0}\sqrt{P'P}$ and 
the number of regenerated photons is $N_{S}=P^{\prime}PN_{0}$ where 
$N_{0}$ is the number of photons in the initial laser beam.

It can be shown~\cite{Siki83,vanB87,Raff88} that the photon to axion
conversion probability $P$ in a region of length $L$ permeated by a
constant magnetic field $B_{0}$ transverse to the direction of
propagation, is given by ($\hbar=c=1$)
\begin{equation}
P=\frac{1}{4} (g_{a\gamma\gamma}B_{0}L)^{2}.
\label{prob}
\end{equation}
This equation is written for the effect in vacuum and for the case
where the difference between the axion and photon momenta
$q= {m_{a}^{2}}/{2\omega}$ is small compared to $1/L$. The axion to
photon conversion probability in this same region is also equal to
$P$.

A number of photon regeneration
experiments have reported results~\cite{Ruos92,Came93,Robi07,Chou08,LIPPS,OSQAR,alps,Ehret:2010mh}, 
with the best limits~\cite{Ehret:2010mh} being
$g_{a\gamma\gamma}<6.5\times10^{-8}~{\rm GeV}^{-1}$. None of these
experiments used cavities on the photon regeneration side of the
optical barrier; recycling on the production side has been used in 
two~\cite{Ruos92,alps}.

Photon regeneration is enhanced by employing matched Fabry-Perot
optical cavities, 
Fig.~\ref{fig:Reprfig1}(b), one within the axion generation magnet
and the second within the photon regeneration
magnet~\cite{Hoogeveen91,Siki07,mueller09}. The first cavity, the
axion generation cavity, serves to build up the electric field on
the input (left) side of the experiment. It is easy to see that when
the cavity is resonant to the laser wavelength, the laser power in
the high-field region is increased by a factor of ${{\cal
F}_{a}}/{\pi}$ where ${\cal F}_{a} = 4\pi T_{1a}/(T_{1a} + V_a)^2$
is the finesse of the cavity, $ T_{1a}$ is the transmittance of the
input mirror, and $V_a$ is the roundtrip loss of the cavity due to
absorption of the coatings, scattering from defects, diffraction
from the finite mirror size, and transmission through the end mirror. 
The increase in the laser power increases the number of created axions 
by a factor of ${{\cal F}_{a}}/{\pi}$.  These
axions propagate through the {}``wall'' and reconvert into photons
in the regeneration cavity one the right side.
The intra-cavity photon field builds up under the
conditions that the second cavity is resonant at the laser
wavelength and that the spatial overlap integral $\eta$ between the
axion mode and the electric field mode is good. This overlap
condition requires that the spatial eigenmodes of the two cavities
are extensions of each other, e.g., when the Gaussian eigenmode in
one cavity propagated to the other cavity is identical to the
Gaussian eigenmode of that cavity.

To detect the regenerated field, a small part is allowed to transmit
through one of the cavity mirrors. The number of detected photons
behind the regeneration cavity
is~\cite{Hoogeveen91,Siki07,mueller09}
\begin{equation}
N_{S}=\eta^{2}\frac{{\cal F}_{\gamma}}{\pi}\frac{{\cal F}_{a}}{\pi}P^{2}N_{in}\,.
\label{finesse}
\end{equation}
Note that resonant regeneration gives an enhancement factor of
$\sim({\cal F}/\pi)^{2}$ over simple photon regeneration. This
factor may feasibly be $10^{10}$, corresponding to an improvement in
sensitivity to $g_{a\gamma\gamma}$ of $\approx300$.

The resonantly-enhanced photon regeneration experiment, involving
the design and active locking of high-finesse Fabry-Perot resonators
and the heterodyne detection of weak signals at the shot-noise limit,
is well supported by the laser and optics technology developed for
LIGO~\cite{LIGO-NIM-RPP}. We mention briefly REAPR and ALPS-II
and then discuss the expected sensitivities of these experiments.

\begin{figure}[!t]
\centering
\includegraphics[width=0.5\textwidth]{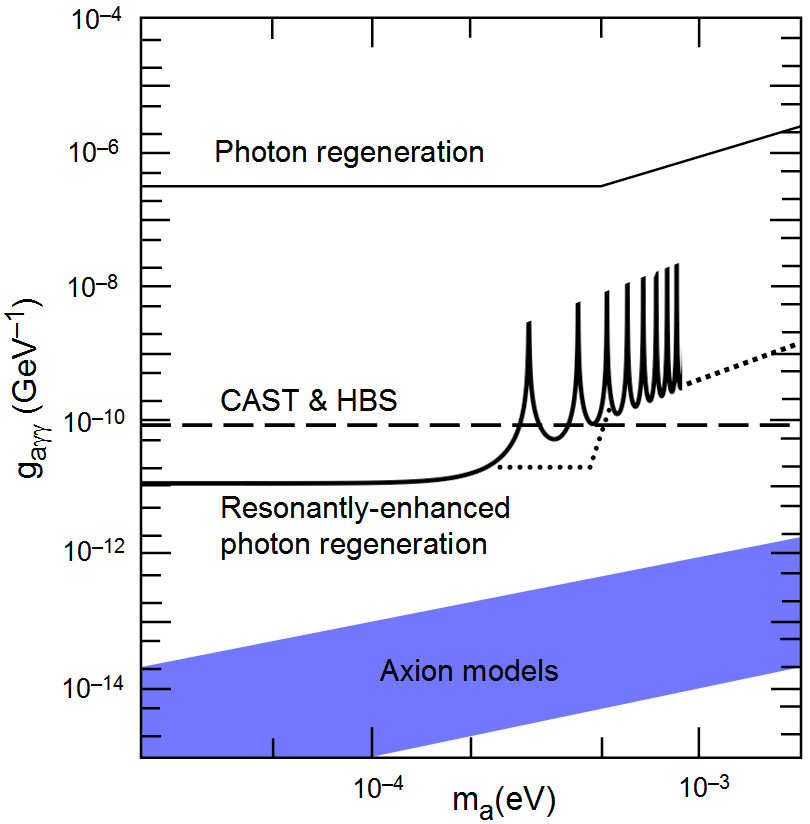}
\caption{ Exclusion plot of mass and photon coupling
$(m_{a},g_{a\gamma\gamma})$ for the axion, and the 95\% CL exclusion
limit for the resonantly enhanced photon regeneration (REPR) 
experiment. The existing exclusion limits indicated on the
plot include the best direct solar axion search (CAST
collaboration)~\cite{cast}, the Horizontal Branch Star
limit~\cite{RaffeltBook}, and previous laser
experiments~\cite{Chou08,alps}. }
\label{fig:REPR}
\end{figure}

For a baseline of 36-m, 5 T, magnets, an input power of $10$~W, a cavity
finesse of ${\cal F}\sim\pi\times10^{5}$ ($T=10\,\mbox{ppm}=V)$ for
both cavities, and 10 days of operation, we find at signal-to-noise 
ratio of unity,
\begin{equation}
g_{a\gamma\gamma}^{min} = \frac{2\times10^{-11}}{\mbox{GeV}}
\left[\frac{0.95}{\eta}\right]
\left[\frac{180\,\mbox{Tm}}{BL}\right]
\left[\frac{3{}\times{}10^5}{{\cal F}}\right]^{1/2}
\left[\frac{10\,\mbox{W}}{P_{in}}\right]^{1/4}
\left[\frac{10\,\mbox{days}}{\tau}\right]^{1/4}.
\label{g-min-1-2}
\end{equation}
The experiment yields a 95\% exclusion limit ($3\sigma$) for axions
or generalized pseudoscalars with
$g_{a\gamma\gamma}^{min}<2.0\times10^{-11}~\mbox{GeV}^{-1}$ after 90
days cumulative running, well into territory unexplored by stellar
evolution bounds or direct solar searches. Note that the exclusion
sensitivity follows the inverse of $\mbox{sinc}(qL/2)$; for REAPR
the first null sensitivity occurs at $2.8\times10^{-4}~\mbox{eV}$ and
for ALPS-II at about half this value. The momentum mismatch between a
massless photon and a massive axion defines the oscillation length
of the process to be $L_{osc}=2\pi/q$. (As pointed out in~\cite{vanB87} however, there is a practical strategy to extend the
mass range upwards if the total magnetic length $L$ is comprised of
a string of $N$ individual identical dipoles of length $l$. In this
case, one may configure the magnet string as a ``wiggler'' to
cover higher regions of mass, up to values corresponding to the
oscillation length determined by a single dipole.) The sensitivity 
of both nonresonant and resonant regeneration experiments, as well 
as other relevent limits, are shown in Fig.~\ref{fig:REPR}.

The optical prototypes being developed for the resonant regeneration
experiment will also have sensitivity to photon-light dark photon 
oscillations~\cite{ITEP-48-1982,ahlers07prd} driven by the  kinetic mixing and the dark photon mass.
Unlike the case of photon-axion oscillations, photon-dark photon 
oscillations do not require the presence of an external magnetic
field, and so can be performed with just the prototype optics and
data acquisition system. On account of the gain from the resonant
cavities, a search with a REAPR or ALPS-II prototype with
meter-length cavities supersede the LIPSS
limit~\cite{Afanasev:2008fv} in less than 1 second of running.  With
a 10-day run, the sensitivity will be improved by a factor of 300,
reaching mixing angles $\chi \approx 10^{-9}$ 
\cite{Jaeckel:2008fi}.  While not the primary goal of the project, a
physics result on dark photons will come for free during the
development phase of a resonantly-enhanced axion-photon regeneration
experiment.

Light shining through walls can also be done with ``light" in the microwave regime~\cite{Hoogeveen:1992nq,Jaeckel:2007ch,Povey:2010hs,Wagner:2010mi,Betz:2013dza}. Allowing for highly sensitive resonant searches for axion-like particles as well as light dark photons.

\subsubsection{Microwave Cavities (Haloscopes)}
\label{subsubsec:nlwcp:axions-&-alps:experiments:Microwave}

Soon after the axion was realized to be a natural DM candidate, a detection concept was proposed that relies on the resonant conversion of DM axions into photons via the Primakoff effect \cite{Sikivie_1}. Though the axion mass is unknown, various production mechanisms in the early Universe point to a mass scale of a few to tens of $\mu$eV if the axion is the dominant form of DM. The detection concept relies on DM axions passing through a microwave cavity in the presence of a strong magnetic field where they can resonantly convert into photons when the cavity frequency matches the axion mass. A 4.13~$\mu$eV axion would convert into a 1~GHz photon, which can be detected with an ultra-sensitive receiver. Axions in the DM halo are predicted to have virial velocities of $10^{-3}~c$, leading to a spread in axion energies 
of $\Delta E_a/E_a \sim 10^{-6}$ (or 1~kHz for our 1~GHz axion example).

Initial experiments run at Brookhaven National Laboratory~\cite{BRF} and the University of Florida~\cite{UofF} came within an order of magnitude of the sensitivity needed to reach plausible axion couplings. 
ADMX~\cite{ADMX_1} was assembled at Lawrence Livermore National Laboratory and consists of a large, 8~T superconducting solenoid magnet with a 0.5 m diameter, 1 m long, open bore. Copper-plated stainless steel microwave cavities are used and have $Q_C \sim 10^5$, low enough to be insensitive to the expected spread
in axion energies. The TM$_{010}$ mode has the largest cavity form factor and is moved to scan axion masses by translating vertical copper or dielectric tuning rods inside the cavity from the edge to the center. TE and TEM modes do not couple to the pseudoscalar axion. 

Using the ADMX setup and an estimated local DM density of $\rho_{DM} = 0.45$ GeV/cm$^3$ \cite{Local_DM}, an axion conversion power $P_a \sim 10^{-23}$~W is expected for plausible DM axions, with the possibility of scanning an appreciable frequency space (hundreds of MHz) in just a few years. Initial data runs were cooled with pumped LHe to achieve physical temperatures of $<$~2~K and used SQUID amplifiers to reach plausible DM axion couplings \cite{SQUID_results}. Recently the ADMX experiment has been moved to the University of Washington where it will be outfitted with a dilution refrigerator that will increase sensitivity and scan rate. A second ADMX site, dubbed ADMX-HF, is being constructed at Yale and will allow access to $>$ 2 GHz while ADMX scans from $0.4 - 2$~GHz. To achieve a greater mass reach, near-quantum limited X-band amplifiers and large volume resonant cavities will have to be developed.

As shown in Fig.~\ref{fig:hspaw-ALPs}, ADMX and ADMX-HF are sensitive to axion and ALP DM in the range of a few to tens of $\mu$eV. The experiments also have exceptional sensitivity to hidden-photons in the same mass region, as shown in Fig.~\ref{fig:hspaw-light-A'}. Additional current microwave cavity efforts are focused at extending the mass range probed by the ADMX 
measurements towards higher particle masses (1.2--1.6 meV) with the
YALE Microwave Cavity Experiment (YMCE)~\cite{ymce} and towards lower masses (0.8--2.0~$\mu$eV)
with the WISP Dark Matter eXperiment (WISPDMX)~\cite{WISPDMX}.

\subsubsection{Oscillating Moments}
\label{subsubsec:nlwcp:axions-&-alps:experiments:OscillatingEDM}

Ultra-light particles such as axions and ALPs can be DM only if they have a large number density, making it possible to describe the DM axion (and ALP) as oscillating classical fields whose energy density is given by the DM density. In addition to single particle scattering that is often used to detect DM (such as WIMP DM), a classical field can give rise to energy and phase shifts. Measurements of such phase shifts can be used to search for the classical DM axion field. This is similar to searches for gravitational waves where the detection is not based on the unobservably small rate at of graviton scattering but rather on the phase shifts caused by the classical gravitational wave field. 

Axion DM causes a time-varying nucleon electric dipole moment which produces oscillating CP-odd nuclear moments \cite{Graham:2011qk,Graham:2013gfa}. In analogy with nuclear magnetic resonance, these moments cause precession of nucleon spins in the presence of a background electric field. The nucleon spin precession can be measured through precision magnetometry in a material sample \cite{Budker:2013hfa}. With current techniques, this experiment has sensitivity to axion masses $m_a \lesssim 10^{-9}$ eV, corresponding to theoretically well-motivated axion decay constants $f_a \gtrsim 10^{16}$ GeV. With improved magnetometry, this experiment could ultimately cover the entire range of masses $m_a \lesssim 10^{-6}$~eV, complementing the region accessible to current axion searches. 

Similarly, ALP DM can give rise to precession of nucleon spins that are not aligned with the direction of the local momentum of the DM \cite{Graham:2013gfa}. Such a precession can also be detected with the nuclear magnetic resonance and precision magnetometry techniques described in~\cite{Budker:2013hfa}.

\subsubsection{Helioscopes }
\label{subsubsec:nlwcp:axions-&-alps:experiments:Helioscopes}

Axions could be produced from blackbody photons in the solar core via the Primakoff effect~\cite{Primakoff51} in the presence of strong electromagnetic fields in the plasma. Since the interaction of these axions with ordinary matter is extraordinarily weak, they can escape the solar interior, stream undisturbed to Earth and reconvert in a strong laboratory transverse magnetic field via the inverse Primakoff effect~\cite{Sikivie83,Sikivie85,vanBibber89}. The minimum requirements for such a helioscope experiment of high sensitivity are a powerful magnet of large volume and an appropriate X-ray sensor covering the exit of the magnet bore. Ideally, the magnet is equipped with a mechanical system enabling it to follow the Sun and thus increasing exposure time. Sensitivity can be further enhanced by the use of X-ray optics to focus the putative signal and therefore reducing detector size and background levels.

\begin{figure}[!t]
\centering
\includegraphics[width=0.6\textwidth]{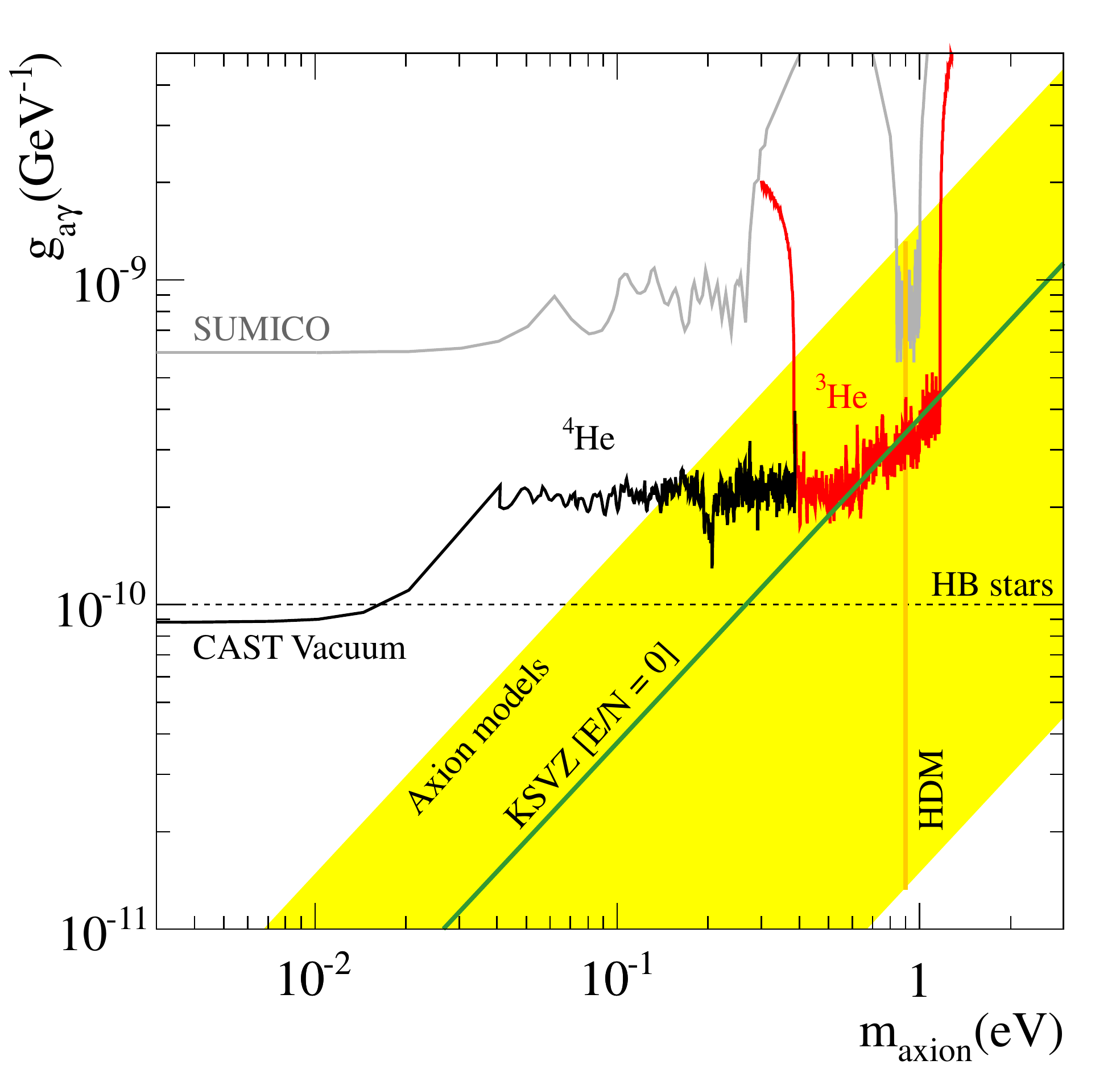} 
\caption{Exclusion regions for axions and axion-like particles in the $m_{a}-g_{a\gamma\gamma}$ plane
achieved by CAST in the vacuum \cite{CAST_PRL05,cast_jcap07}, $^{4}$He \cite{cast_jcap09}, and $^{3}$He
phase~\cite{CAST_PRL11,CAST_PRL13}. We also show constraints from the Tokyo helioscope, horizontal branch (HB) stars~\cite{Raffelt08}, and the hot dark
matter (HDM) bound~\cite{Hannestad10}. The yellow band labeled ``Axion models'' represents typical theoretical models with $\left|E/N-1.95\right| = 0.07-7$. The green solid line inside the band is for $E/N=0$ (KSVZ model).}
\label{fig:CASTexclusion}
\end{figure}

The first axion helioscope search was carried out at Brookhaven National Laboratory in 1992 with a static dipole magnet \cite{Lazarus92}.
A second-generation experiment, the Tokyo Axion Helioscope, uses a more powerful magnet and dynamic tracking of the Sun \cite{Moriyama1998147,Inoue200218,Inoue200893}. The CERN Axion Solar Telescope (CAST), a helioscope of the third generation and the most sensitive solar axion search to date, began data collection in 2003. It employs an LHC dipole test magnet of 10~m length and 10~T field strength \cite{cast99} with an elaborate elevation and azimuth drive to track the Sun. CAST is the first solar axion search exploiting X-ray optics to improve the signal to background ratio (a factor of 150 in the case of CAST) \cite{cast_xrt}. For $m_{a} <  0.02$~eV, CAST has set an upper limit of $g_{a \gamma} < 8.8\times 10^{-11}$~GeV$^{-1}$ and a slightly larger value of $g_{a \gamma}$ for higher axion masses \cite{CAST_PRL05,cast_jcap07,cast_jcap09,CAST_PRL11,CAST_PRL13}. The exclusion plots are shown in Fig.~\ref{fig:CASTexclusion}. 
CAST has also established the first helioscope limits for non-hadronic axion models~\cite{Barth13}. 

\begin{figure}[!t]
\centering
\includegraphics[width=0.55\textwidth]{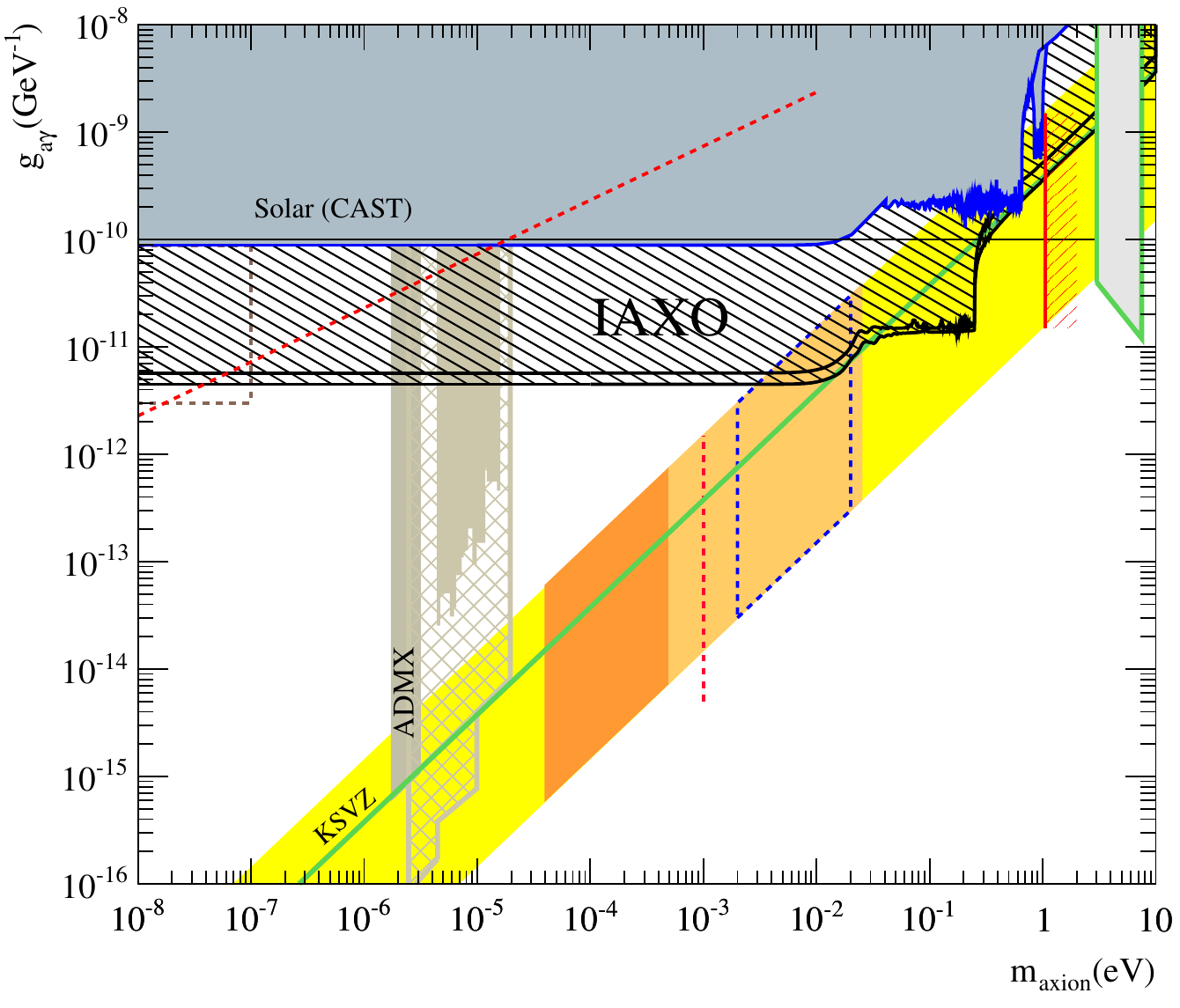} 
\caption{Expected sensitivity of IAXO compared with current bounds from
CAST and ADMX. Also future prospects of ADMX are shown (dashed brown region).}
\label{fig:IAXOprospects}
\end{figure}

So far each subsequent generation of axion helioscopes has resulted in an improvement in sensitivity to the axion-photon coupling constant $g_{a \gamma}$ of about a factor 6 over its predecessors. To date, all axion helioscopes have used ``recycled'' magnets built for other purposes. The IAXO collaboration has recently shown~\cite{Irastorza11} that a further substantial step beyond the current state-of-the-art represented by CAST is possible with a new fourth-generation axion helioscope, dubbed the International AXion Observatory (IAXO). The concept relies on a purpose-built ATLAS-like magnet capable of tracking the sun for about 10 hours each day, focusing X-ray optics to minimize detector area, and low background X-ray detectors optimized for operation in the $0.5-10$~keV energy band. Pushing the current helioscope boundaries to explore the range in $g_{a \gamma}$ down to a few $10^{-12}$ GeV$^{-1}$ (see Fig.~\ref{fig:IAXOprospects}), with sensitivity to QCD axion models down to the meV scale and to ALPs at lower masses, is highly motivated as was shown in previous sections. Lowering X-ray detector thresholds to 0.1~keV would allow IAXO to test whether solar processes can create chameleons \cite{BraxKon10} and further constrain standard axion-electron models.  More speculative, but of tremendous potential scientific gain, would be the operation of microwave cavities inside IAXO's magnet, to allow a simultaneous search for solar and DM axions [{\it e.g.}, \cite{hrelics}]. Searches for solar axions and chameleons that exploit naturally occurring magnetic fields are described in \cite{hrelics,hcham,hnjp} and reviewed in \cite{hpatras}. IAXO can carry out this task as one of the main experimental pathways in the next decade for the axion community. A detection with IAXO would have profound implications for particle physics, with clear evidence of physics beyond the SM.

\subsubsection{Beam Dumps and Colliders}
\label{subsubsec:nlwcp:axions-&-alps:experiments:Colliders}

Axions and ALPs can also be searched for in beam-dump and collider experiments.  We describe these 
type of experiments in greater detail in the next section, although we do not discuss their sensitivity to axions and ALPs.  
See e.g.~\cite{Essig:2010gu} and references therein.

\section{Dark Photons}
\label{sec:nlwcp:dark-photons}

\subsection{Theory \& Theory Motivation}
\label{subsec:nlwcp:dark-photons:theory}

This section describes the theory and motivation for new forces mediated 
by new abelian $U(1)$ gauge bosons $A'$ that couple very weakly to electrically charged particles through 
``kinetic mixing'' with the photon~\cite{Holdom:1985ag,Galison:1983pa}.  
We will usually refer to the $A'$ as a ``dark photon'', but it is also often called a ``U-boson'', ``hidden-sector,'' ``heavy,'' ``dark,'' ``para-,'' or ``secluded'' photon.  Generalizations to other types of couplings beyond those generated by kinetic mixing exist, but our main focus 
will here be on this particularly simple type.  

\begin{figure}[!t]
\centering
\vspace*{-5mm}
\includegraphics[width=0.48\textwidth]{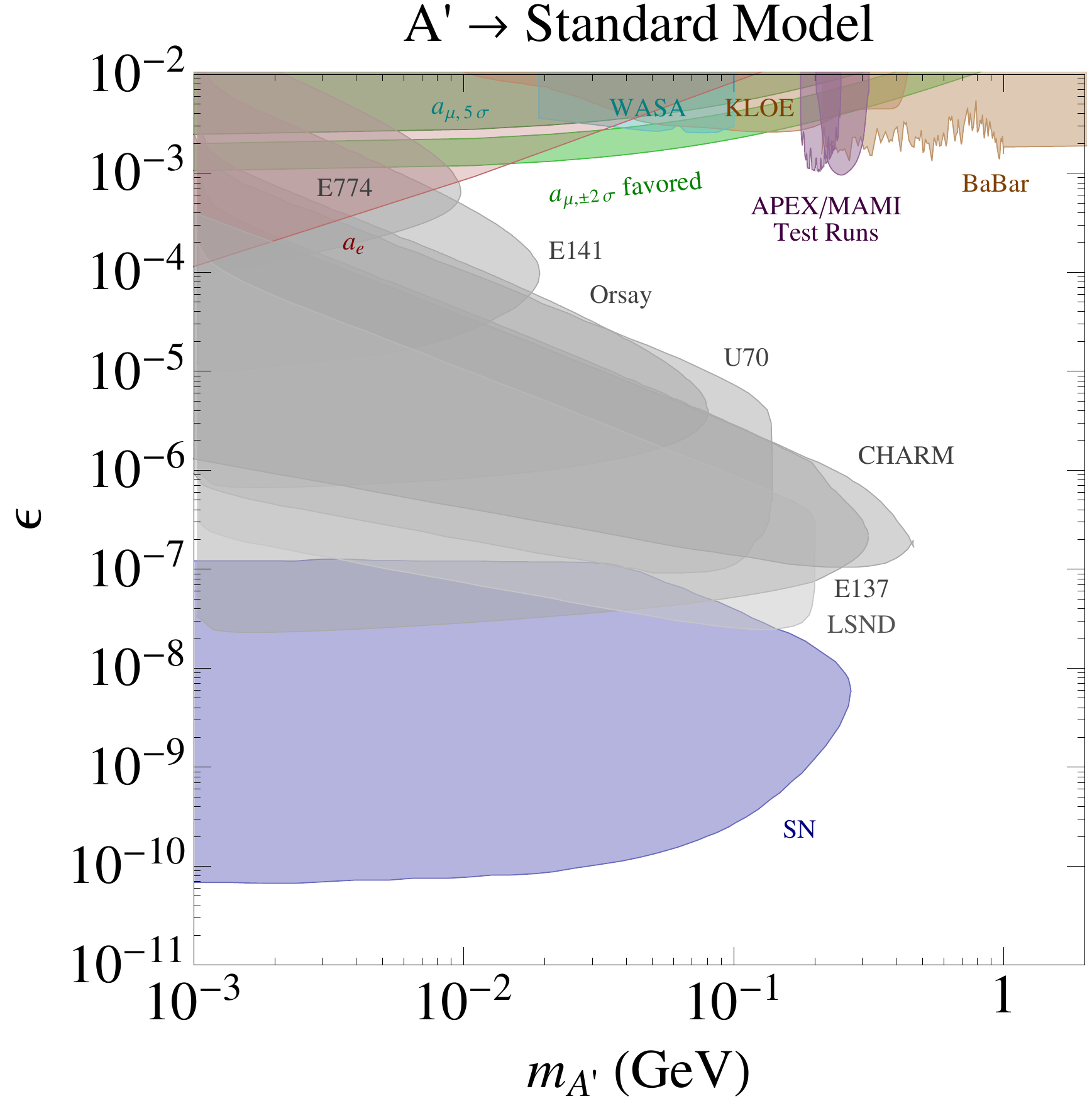}\;\;\; 
\includegraphics[width=0.48\textwidth]{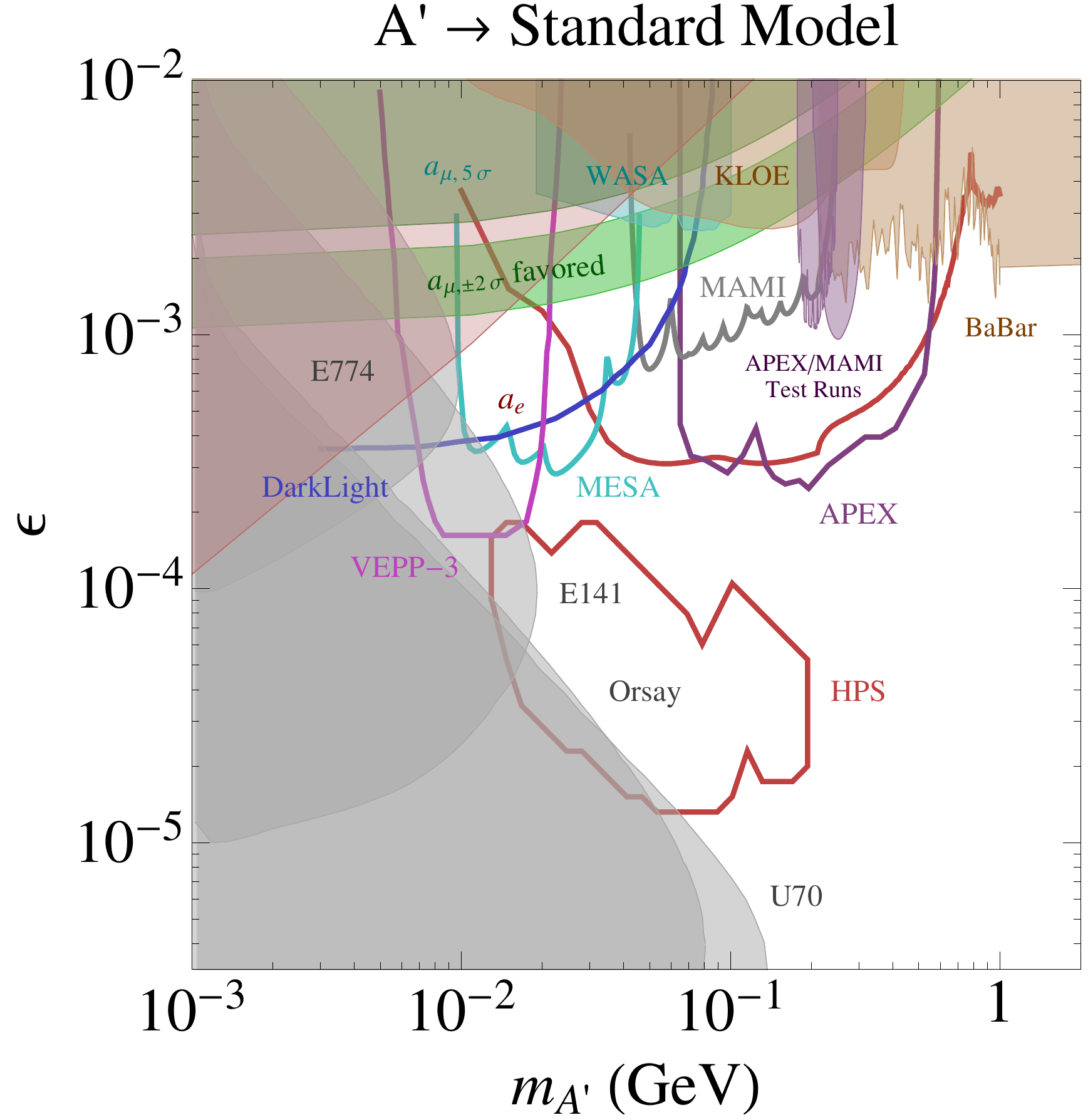} 
\caption{
 Parameter space for dark photons ($A'$) with mass $m_{A'}> 1$~MeV 
(see Fig.~\ref{fig:hspaw-light-A'} for $m_{A'} < 1$~MeV).
Shown are existing 90\% confidence level limits from the SLAC and
Fermilab beam dump experiments 
E137, E141, and
E774~\cite{Bjorken:2009mm,Bjorken:1988as,Riordan:1987aw,Bross:1989mp}   
the electron and muon anomalous magnetic moment $a_\mu$~\cite{Pospelov:2008zw,Davoudiasl:2012ig,Endo:2012hp},  
KLOE~\cite{Babusci:2012cr} (see also~\cite{Archilli:2011zc}), WASA-at-COSY~\cite{Adlarson:2013eza}, 
the test run results reported by APEX~\cite{Abrahamyan:2011gv} and MAMI~\cite{Merkel:2011ze}, 
an estimate using a BaBar result~\cite{Bjorken:2009mm,Reece:2009un,Aubert:2009cp}, and 
a constraint from supernova cooling~\cite{Bjorken:2009mm,Dent:2012mx,Dreiner:2013mua}.
In the green band, the $A'$ can explain the observed discrepancy between the
calculated and measured muon anomalous magnetic moment~\cite{Pospelov:2008zw} 
at 90\% confidence level.
On the right, we show in more detail the parameter space for larger values of 
$\epsilon$.  This parameter space can be probed by several proposed experiments, 
including APEX~\cite{Essig:2010xa}, HPS~\cite{HPS},
DarkLight~\cite{Freytsis:2009bh}, VEPP-3~\cite{Wojtsekhowski:2009vz,Wojtsekhowski:2012zq}, MAMI, 
and MESA~\cite{Beranek:2013yqa}.   
Existing and future $e^+e^-$ colliders such as \babar, BELLE, KLOE, Super$B$, BELLE-2, and KLOE-2  can also probe large 
parts of the parameter space for $\epsilon > 10^{-4}-10^{-3}$; their reach is not explicitly shown.}
\label{fig:hspaw-heavy-A'}
\end{figure}

Kinetic mixing produces an effective parity-conserving interaction
$\epsilon e A'_\mu J^\mu_{\rm EM}$ of the $A'$ to the 
electromagnetic current $J^\mu_{EM}$,  suppressed relative to the electron charge 
$e$ by the parameter $\epsilon$, which theoretically is not required to be small.  
In fact, $\epsilon$ can theoretically be $\mathcal{O}(1)$, as the vector portal is a dimension-four operator and unsuppressed by any 
high mass scale.  
In particular models, however, $\epsilon$ can be calculated and can be naturally small 
(we often write the coupling strength as $\alpha' \equiv  \epsilon^2 \alpha$  where $\alpha= e^2/4\pi\simeq 1/137$). 
In particular, if the value of $\epsilon$ at very high energies is zero, 
then $\epsilon$ can be generated by perturbative or non-perturbative effects.  Perturbative contributions can include heavy messengers that carry both hypercharge and the new $U(1)$ charge, and quantum loops of various order can 
generate $\epsilon \sim 10^{-8} -10^{-2}$~\cite{ArkaniHamed:2008qp,Essig:2009nc}.   
Non-perturbative and large-volume effects common in string theory constructions can generate much smaller 
$\epsilon$.  
While there is no clear minimum for $\epsilon$, values in the $10^{-12}-10^{-3}$ range have been predicted in the 
literature~\cite{Abel:2008ai,Goodsell:2009xc,Cicoli:2011yh,Goodsell:2011wn}. 

\begin{figure}[!t]
\centering
\includegraphics[width=\textwidth]{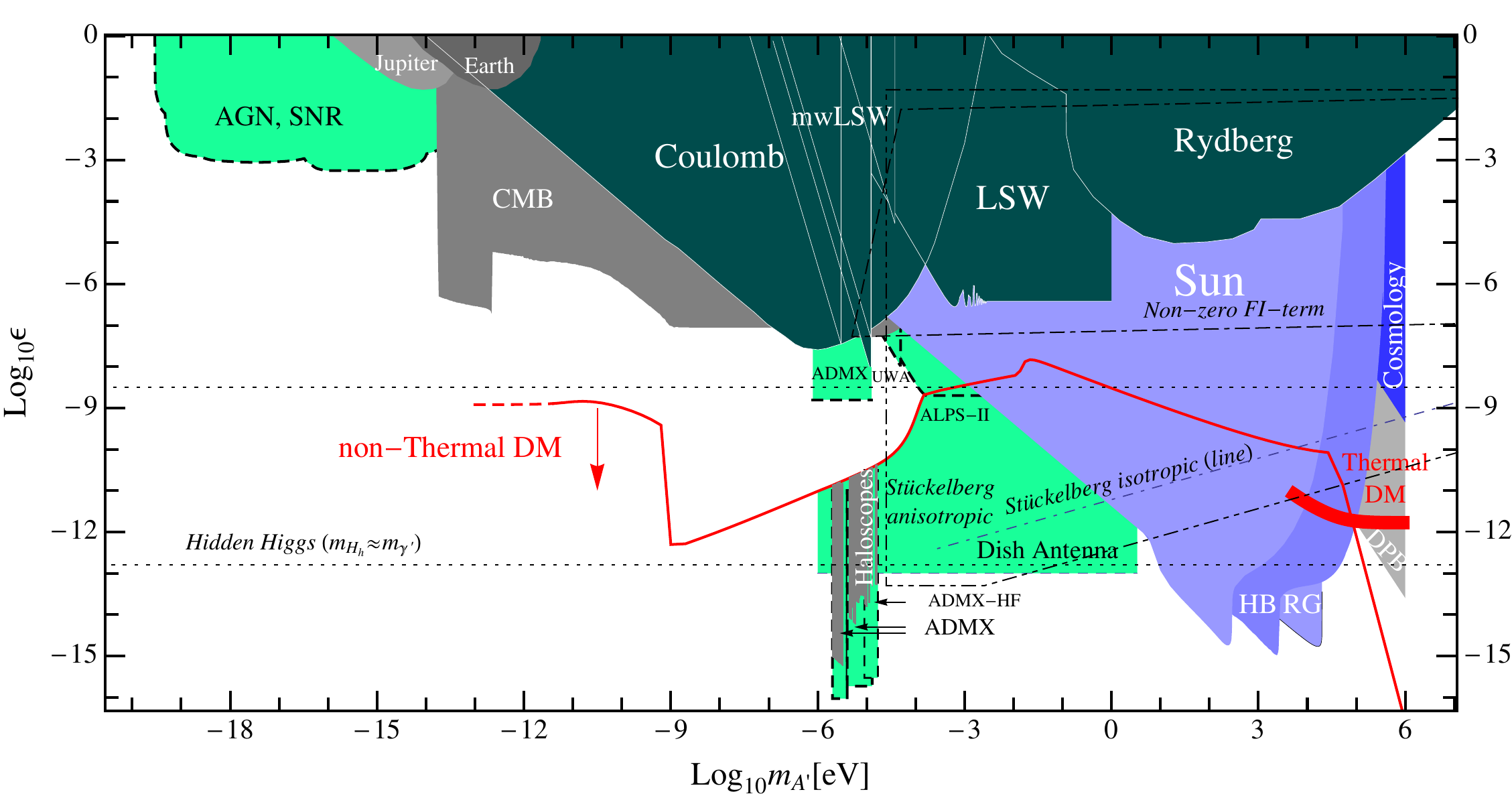} 
\caption{
Parameter space for hidden-photons ($A'$) with mass $m_{A'} < 1$~MeV (see Fig.~\ref{fig:hspaw-heavy-A'} for 
$m_{A'}>1$~MeV). 
Colored regions are: experimentally excluded regions (dark green), constraints from astronomical 
observations (gray) or from astrophysical, or cosmological arguments (blue), and 
sensitivity of planned and suggested experiments (light green) (ADMX~\cite{Asztalos:2011ei}, ALPS-II~\cite{Bahre:2013ywa}, Dish antenna~\cite{Horns:2012jf}, AGN/SNR~\cite{Lobanov:2012pt}). Shown in red are boundaries where the $A'$ would account for all the DM produced either thermally in the Big Bang or non-thermally by the misalignment mechanism (the corresponding line is an upper bound).  Regions bounded by dotted lines show predictions from string theory corresponding to different possibilities for the nature of the $A'$ mass: Hidden-Higgs, a Fayet-Iliopoulos term, or the St\"uckelberg mechanism. Predictions are uncertain by $\mathcal{O}(1)$-factors. }
\label{fig:hspaw-light-A'}
\end{figure}

A dark sector consisting of particles that do not couple to any of the known forces and containing an $A'$ is commonplace 
in many new physics scenarios.  Such hidden sectors can have a rich structure, consisting of, for example, fermions 
and many other gauge bosons.  The photon coupling to the $A'$ could provide the only non-gravitational window into 
their existence.   
Hidden sectors are generic, for example, in string theory 
constructions~\cite{Goodsell:2010ie,NSF-ITP-84-170,PRINT-86-0084(PRINCETON),Andreas:2011in}. 
and recent studies have drawn a very clear picture of the different possibilities obtainable in type-II compactifications 
(see dotted contours in Fig.~\ref{fig:hspaw-light-A'}).  
Several portals beyond the kinetic mixing portal are 
possible, many of which can be investigated at the intensity frontier. 

Masses for the $A'$ can arise via the Higgs mechanism and can take on a large range of values. 
$A'$ masses in the MeV--GeV range arise in the models 
of~\cite{Fayet:2007ua,ArkaniHamed:2008qp,Cheung:2009qd,Morrissey:2009ur} (these models often involve supersymmetry).  
However, much smaller (sub-eV) masses are also possible.  
Masses can also be generated via the St\"uckelberg mechanism, which is 
especially relevant in the case of large volume string compactifications with branes~\cite{Goodsell:2009xc}. 
In this case, the mass and size of the kinetic mixing are typically linked through one scale, the string
scale $M_s$, and therefore related to each other.  In Fig.~\ref{fig:hspaw-light-A'}, various theoretically motivated 
regions are shown~\cite{Goodsell:2009xc,Cicoli:2011yh}.  
The $A'$ mass can be as small as $M_s^2/M_{\rm Pl}$, i.e.~$m_{A'} \sim$~meV (GeV) for $M_s\sim$~TeV ($10^{10}$~GeV).  
Note that particles charged under a \emph{massive} $A'$ do not have an electromagnetic millicharge, but a \emph{massless} 
$A'$ can lead to millicharged particles (see \S\ref{subsubsec:nlwcp:other:theory:MCP}).  

The previous discussion focused on kinetic mixing 
between the hypercharge $U(1)_Y$ and the dark $U(1)$ gauge bosons, 
parametrized by $\epsilon$.   
As we mentioned above, many generalizations exist.  One generalization is obtained by 
allowing for the possibility of mass matrix mixing, 
parametrized by $\epsilon_Z$, between the dark photon and the 
heavy $Z$ boson of the SM~\cite{Davoudiasl:2012ag}.  
Because of its expanded properties, the dark $U(1)$ vector boson has
been dubbed the ``dark $Z$" and labeled $Z_d$ in such a picture, in order 
to emphasize its $Z$-like properties~\cite{Davoudiasl:2012ag}.  
Overall, the $Z_d$ couples to both the electromagnetic ($J^\text{EM}_\mu$) and
the weak neutral ($J^\text{NC}_\mu$) currents of the SM, via~\cite{Davoudiasl:2012ag}
\begin{equation}
{\cal L}_\text{int} = - \left( \epsilon \,e \, J^\text{EM}_\mu + 
\epsilon_Z \, \frac{g}{2 \cos\theta_W} J^\text{NC}_\mu \right) Z_d^\mu\,.
\end{equation}                           
The additional interactions involving $\epsilon_Z$ 
violate parity and current conservation.  Consequently, new phenomena such as ``Dark Parity
Violation" in atoms and polarized electron scattering can occur~\cite{Davoudiasl:2012ag,Davoudiasl:2012qa}.  
Enhancements in rare ``dark" decays of the Higgs as well as $K$ and $B$ mesons into $Z_d$
particles can also occur, suggesting new experimental areas of 
discovery~\cite{Davoudiasl:2012ag,Davoudiasl:2012ig,Davoudiasl:2013aya}. Note however that for small $m_{A'}$, $\epsilon_Z$ is suppressed by $(m_{A'}/m_Z)^2$ (if it originates exclusively from kinetic mixing) and these effects are extremely small.

\subsection{Phenomenological Motivation and Current Constraints}
\label{subsec:nlwcp:dark-photons:pheno}

A natural dividing line is $m_{A'} \sim 2 m_e \sim 1$~MeV.  For $m_{A'}>1$~MeV, 
an $A'$ can decay to electrically charged particles ({\it e.g.}, $e^+e^-$, $\mu^+\mu^-$, or $\pi^+ \pi^-$) or to light hidden-sector particles (if available), which can in turn decay to ordinary matter.   
Such an $A'$ can be efficiently produced in 
electron or proton fixed-target experiments \cite{Bjorken:2009mm,Freytsis:2009bh,Essig:2010xa,Abrahamyan:2011gv,HPS,Merkel:2011ze,Batell:2009di,Essig:2010gu,Wojtsekhowski:2009vz,Wojtsekhowski:2012zq,MeijerDrees:1992kd,Beranek:2013yqa} 
and at $e^+e^-$ and hadron colliders  
\cite{ArkaniHamed:2008qp,Essig:2009nc,Cheung:2009qd,Strassler:2006im,Reece:2009un,AmelinoCamelia:2010me,Aubert:2009cp,Batell:2009yf,Aubert:2009pw,Abazov:2009hn,Abazov:2010uc,Babusci:2012cr,Archilli:2011zc,Lees:2012ra,Baumgart:2009tn,Adlarson:2013eza}, see \S\ref{subsec:nlwcp:dark-photons:experiments}.  
Hidden-sector particles could be directly produced through an off-shell $A'$ and decay to ordinary matter.  
An $A'$ in this mass range is motivated by the theoretical considerations discussed above, by anomalies related 
to DM~\cite{ArkaniHamed:2008qn,Pospelov:2008jd}, 
and by the discrepancy between the measured and calculated value of the anomalous magnetic moment of the 
muon~\cite{Pospelov:2008zw,Davoudiasl:2012ig,Endo:2012hp}.  

Fig.~\ref{fig:hspaw-heavy-A'} shows existing constraints for $m_{A'}>1$~MeV~\cite{Bjorken:2009mm} and the sensitivity of 
several planned experiments that will explore part of the remaining allowed parameter space. 
These include the future fixed-target experiments APEX~\cite{Essig:2010xa,Abrahamyan:2011gv}, HPS~\cite{HPS}, 
DarkLight~\cite{Freytsis:2009bh} at Jefferson Laboratory, 
an experiment using VEPP-3~\cite{Wojtsekhowski:2009vz,Wojtsekhowski:2012zq}, and 
experiments using the MAMI and MESA~\cite{Beranek:2013yqa} at the University of Mainz.  
Existing and future $e^+e^-$ colliders can also probe large parts of
the parameter space for $\epsilon > 10^{-4}-10^{-3}$, and include \babar, Belle, KLOE, Super$B$, Belle II, and KLOE-2 
(Fig.~\ref{fig:hspaw-heavy-A'} only shows existing constraints, and no future sensitivity, for these experiments).  
Proton colliders such as the LHC and Tevatron can also see remarkable signatures for light hidden-sectors~\cite{Strassler:2006im}.  
This rich experimental program is discussed in more detail in \S\ref{subsec:nlwcp:dark-photons:experiments}.  

For $m_{A'}<1$ MeV, the $A'$ decay to $e^+e^-$ is kinematically forbidden, and only a much slower decay to 
three photons is allowed. For most of the parameter space shown in  Fig.~\ref{fig:hspaw-light-A'}, the $A'$ lifetime is longer than the age of the Universe.  This figure shows the constraints, theoretically and phenomenologically motivated regions, and some soon-to-be-probed parameter space.
At these low masses, the mixing of $A'$ with the photon can reveal itself in 
 the phenomenon of photon $\leftrightarrow$ $A'$ oscillations~\cite{ITEP-48-1982}. This happens in general when the propagation and the interaction eigenstates are misaligned, the most famous case being neutrino flavor oscillations. Oscillatory patterns de-cohere rapidly 
 with increasing mass and are typically not relevant for $m_{A'}>1$ MeV. 
Photons emitted from a source can transform to an $A'$, which, being weakly interacting, might leave no trace.  The effective photon disappearance is frequency dependent and distorts continuous spectra with a characteristic sinusoidal pattern in photon wavelength.

Like axions or ALPs, $A'$ bosons can also be DM through the vacuum-misalignment mechanism~\cite{Nelson:2011sf,Arias:2012az}. This intriguing possibility can be realized in a wide range of values for $m_{A'}$ and $\epsilon$, see Fig.~\ref{fig:hspaw-light-A'}. Experiments such as ADMX, looking for axion DM, is sensitive to $A'$s as well.  In this case, the use of magnetic fields to trigger the $A' \to$ photon conversion is not required.  This is another example, where the same experimental apparatus can often look for several kinds of particles.  
A few experimental searches are planned and discussed in \S\ref{subsubsec:nlwcp:dark-photons:experiments:Laser}, but a large parameter space still remains to be experimentally explored.

\subsubsection{Hints for MeV-GeV mass Dark Photons from Dark Matter}
\label{subsubsec:nlwcp:dark-photons:pheno:DM}

Couplings between DM and dark photons at the MeV-GeV scale can drastically modify the phenomenology of DM. In direct detection experiments, the scattering cross section can be increased due to the light mediator, or alternatively the kinematics of the scattering can be altered if the mediator couples to nearly-degenerate states. In indirect searches, the self-annihilation and self-scattering rates for the DM can both be enhanced at low velocities; the former can lead to striking signals in cosmic rays, photons, and neutrinos, while the latter can significantly modify the internal structure of DM halos. While the search for dark photons has strong motivations entirely independent from their possible link to DM, their existence could potentially provide an entirely new window on the dark sector. 

{\bf Cosmic rays:} 

In 2008, the PAMELA experiment reported an unexpected rise in the ratio of cosmic-ray (CR) positrons to CR electrons, beginning at $\sim 10$ GeV and extending to above 100 GeV~\cite{Adriani:2008zr}. This result was later confirmed by the Fermi Gamma-Ray Space Telescope~\cite{FermiLAT:2011ab} and most recently by AMS-02~\cite{Aguilar:2013qda}. The (largely model-independent) expectation from standard CR propagation models is that the positron fraction should fall with increasing energy.  While there are proposals for generating the positron excess by modifications to CR propagation, they require non-trivial changes to the usual propagation paradigm, e.g. that the positrons do not suffer significant radiative losses over kpc distances~\cite{Blum:2013zsa}, or that the positron production by proton scattering occurs primarily within the original CR acceleration site~\cite{Blasi:2009hv, Cowsik:2009ga}.  Complementary measurements of the total $e^+ e^-$ spectrum by the Fermi Gamma-Ray Space Telescope~\cite{Abdo:2009zk} are consistent with a new source of $e^+ e^-$ pairs in the 10-1000 GeV energy range.

The annihilation of weak-scale DM provides an attractive hypothesis for the origin of this signal, but there are several difficulties with the conventional WIMP interpretation, e.g.~\cite{Cholis:2008hb}.  (Non-DM explanations involving a new $e^+e^-$ source have also been advanced, with the most popular being a population of pulsars; see e.g.~\cite{Yin:2013vaa,Linden:2013mqa}.) DM annihilating to a dark photon which subsequently decays, however, naturally yields (i) an enhanced signal (by up to 2-3 orders of magnitude) and (ii) a sufficiently hard positron spectrum to match the observations, as well as forbidding the production of antiprotons, if the dark photon is lighter than twice the proton mass (an antiproton excess was searched for, and not observed)~\cite{ArkaniHamed:2008qn, Pospelov:2008jd}. Benchmark models of this type were computed for $m_{A'}\sim 200-900$~MeV in~\cite{Finkbeiner:2010sm}, and found to provide a good fit to the data.

The AMS-02 data, with their much smaller uncertainties, prefer a somewhat softer spectrum of positrons than PAMELA. In turn, this favors dark photon models where the dark photon is heavy enough to decay to muons and charged pions, or possibly multi-particle final states (e.g.~via decays through the dark sector); the spectrum due to dark photon decay to an $e^+ e^-$ pair is (as the sole channel) somewhat harder than preferred by the data~\cite{Cholis:2013psa}. Direct leptophilic annihilation to SM particles no longer appears to provide a good explanation for the signal: the softer spectrum favors $\tau^+ \tau^-$ final states, which are constrained by searches for gamma-rays from dwarf galaxies~\cite{Yuan:2013eja}.

There are also gamma-ray bounds on $\mu^+\mu^-$, $\pi^+\pi^-$, and $e^+ e^-$ final states, but gamma-ray production in these decays is small, and so the bounds are generally much weaker (unless upscattering of ambient starlight by electrons is included, but this contribution also depends on the electron propagation). Constraints from the inner Galaxy are dependent on the slope of the DM density profile, which is not well-constrained by the data or theory; constraints from the outer halo and extragalactic gamma-ray background depend sensitively on the amount of small-scale substructure present, which is also poorly known. There is tension between gamma-ray observations and the predictions from models fitting the PAMELA signal, e.g.~\cite{Ackermann:2012rg, Zavala:2011tt, Abazajian:2011ak}, but stronger statements are limited by the astrophysical uncertainties.

A more robust constraint arises from measurements of the CMB. DM annihilation during the epoch of recombination can inject electrons and photons, which modify the ionization history of the Universe; this in turn modifies the scattering of CMB photons at late times and perturbs the observed anisotropy spectrum \cite{Padmanabhan:2005es}. The current constraints probe relevant regions of parameter space~\cite{Madhavacheril:2013cna,Lopez-Honorez:2013cua}, and the Planck polarization data should improve the sensitivity by another factor of two, e.g.~\cite{Madhavacheril:2013cna,Galli:2013dna}.  The constraints are weaker if the local DM density is higher by a factor of $\sim \sqrt{2}$, or by permitting an $\mathcal{O}(1)$ contribution to the signal from local clumps of DM. This second option is particularly attractive for lighter dark photons ($m_{A'} \ll 1$ GeV), where the annihilation cross section continues to grow at velocities smaller than that of the main Milky-Way halo, and so the constraints from the CMB (originating from an epoch when the DM was extremely slow-moving) grow even stronger; this conclusion can be evaded if the excess observed by AMS-02 largely originates from DM clumps with small internal velocity dispersions~\cite{Slatyer:2011kg}.

The constraints discussed above do not apply if the signal originates from decaying DM (e.g.~\cite{Ibe:2013jya}). In this case the size of the signal is not a difficulty, but the lack of antiprotons and the hard spectrum still motivate scenarios with decay through dark photons.

{\bf Light DM (with mass $\sim 1-10$~GeV):}

There have been several experimental results that might hint at the presence of $\mathcal{O}(1-10)$ GeV DM. The CDMS experiment has recently reported three events in their signal region \cite{Agnese:2013rvf}, with the best fit WIMP hypothesis being favored over the background-only hypothesis at $99.8\%$ confidence. The best-fit WIMP mass is $8.6$ GeV/cm$^2$, with a $68\%$ confidence contour extending from $6.5-20$ GeV. This region is in good agreement with earlier hints of a signal from CoGeNT \cite{Aalseth:2012if}; it appears in tension with limits from XENON100, but the comparison does depend on the response of xenon to low-energy nuclear recoils and on the DM velocity distribution \cite{Hooper:2013cwa}.

The preferred DM-nucleon scattering cross section for the CDMS events, $\sigma \approx 2 \times 10^{-41}$ cm$^2$, is quite large. The two SM particles which might be expected to mediate such a scattering are the $Z$ boson and the Higgs, both of which are constrained (for light DM) by bounds on the invisible decay width of the $Z$ and the Higgs; the cross section preferred by CDMS seems clearly ruled out for Higgs portal DM~\cite{Djouadi:2012zc}, and barely consistent for scattering through the $Z$ \cite{Choi:2013fva}. This observation motivates the existence of a new mediator particle, in the event that the signal is indeed due to DM, e.g.~\cite{Andreas:2013iba}. A dark photon mediator naturally enhances the cross section; if the mass of the dark photon is inherited from the weak scale, the relation $m_{A'} \sim \sqrt{\epsilon}~ m_Z$ naturally predicts a DM-nucleon cross section comparable to that mediated by the $Z$, but the constraints on invisible decays no longer apply.

There have also been hints of possible annihilation signals from $\sim 10$ GeV DM in the Galactic Center and inner Galaxy \cite{Hooper:2011ti,Linden:2011au,Abazajian:2012pn,Hooper:2013rwa}; these signals can be accommodated by light DM annihilation to dark photons which subsequently decay to SM particles \cite{Hooper:2012cw}.

{\bf Self-interacting DM:}

Any coupling between MeV-GeV dark photons and DM will also give rise to a long-range self-interaction for the DM. This in turn can modify DM structure formation, flattening the cusps at the centers of halos \cite{Spergel:1999mh} and reducing the concentration of subhalos \cite{Vogelsberger:2012ku}. These are two areas in which there are marked disagreements between the predictions of collisionless cold DM simulations and observations of galaxies, and the effect of self-interaction is to bring the two into closer agreement.

Recent work on the cross section required to achieve agreement has pointed to a low-velocity cross section in the range of $\sigma/m_\mathrm{DM} \sim 0.1-1$ cm$^2$/gram~\cite{Zavala:2012us}. In dark-photon scenarios where the potential due to self-interaction can be approximated as a Yukawa potential, the maximum transfer cross section  (in the classical regime, see \cite{Tulin:2012wi}) is given by $\sigma_T \approx 22.7/m_{A'}^2$ (e.g.~\cite{Vogelsberger:2012ku}). Setting $\sigma_T/m_\mathrm{DM} \simeq 1$ cm$^2$/gram, we require $m_{A'} \approx 70$ MeV $\times \sqrt{\mathrm{GeV}/m_\mathrm{DM}}$, in agreement with similar estimates in \cite{Slatyer:2011kg}. It is remarkable that this entirely independent line of enquiry suggests a mass scale in the range accessible by dark photon searches.

\subsubsection{Ultra-light Dark Photons}

In recent years, fits to cosmological data including large-scale structure and the CMB anisotropies (WMAP and Planck) have suggested the existence of dark radiation, i.e.~a relic population of relativistic particles decoupled from ordinary matter. 
$A'$s with meV mass and $\epsilon\sim \mathcal{O}(10^{-6})$ would be produced 
in the early Universe in the right amount to account for this tendency~\cite{Jaeckel:2008fi} but a recent examination of the stellar evolution constraints showed that this region is ruled out~\cite{An:2013yfc}. 
If dark radiation exists and it is due to $A'$s the relic abundance has to be produced by decays or annihilation of dark-sector particles. Constraints on this scenario are very mild so there is motivation to explore a large range of masses and $\epsilon$.

More interesting is the possibility that light $A'$s constitute the DM of the Universe. 
There are two possibilities depending on the origin of the relic density of $A'$s.   
If $m_{A'} \sim 100$ keV and $\epsilon \sim 10^{-12}$ (thick red band labelled ``Thermal DM'' in Fig.~\ref{fig:hspaw-light-A'}) the right amount of DM $A'$s is produced by oscillations from the photon thermal bath before big bang nucleosynthesis~\cite{Redondo:2008ec}. This hypothesis can be tested in direct DM detection experiments or indirectly through the $A'$ decay into three photons, which could be observed above the astrophysical diffuse X-ray backgrounds~\cite{Pospelov:2008jk,Redondo:2008ec}. 
In this case, DM $A'$s have larger velocities than standard ``cold'' DM qualifying as ``warm'' DM. 
This possibility is very appealing in the light of the unresolved issues with structure formation 
present in the cold DM paradigm mentioned in Sec.~\ref{subsubsec:nlwcp:axions-&-alps:pheno:astrophysics}.   

Alternatively, a CDM relic of $A'$ can be produced by the ``misalignment mechanism'' in a way analogous to  axions and ALPs~\cite{Nelson:2011sf,Arias:2012az}. In this case, a small part of the DM $A'$s possibly oscillates into photons in the early Universe leaving a fingerprint in cosmological observables like the CMB spectrum, abundances of light elements created during BBN, and the isotropic diffuse photon background~\cite{Arias:2012az}. The region above the line labelled ``non-Thermal DM'' in Fig.~\ref{fig:hspaw-light-A'}) is free from any of such constraints and thus perfectly viable for CDM $A'$. 

\subsection{Experimental Searches for Dark Photons: Status and Plans}
\label{subsec:nlwcp:dark-photons:experiments}

For $m_{A'} > 1~$MeV, our discussion here focuses on the case where the dark photon can only decay into SM matter, 
with $\epsilon$-suppressed decay width.  Another possibility is that the dark photon has $\epsilon$-unsuppressed couplings to 
some new species ``$\chi$'' of fermions or bosons (dark-sector matter), which are neutral under the SM gauge group, and 
in particular are electrically neutral.  
The latter will be discussed in detail in \S \ref{sec:nlwcp:LDM}.  
We also comment on searches for ultralight $A'$, i.e.~$m_{A'} < 1~$MeV.

\subsubsection{Electron Beam Dump Experiments}
\label{subsubsec:nlwcp:dark-photons:experiments:Beam-dumps-electron}

In electron beam dump experiments, a high-intensity electron beam dumped onto a fixed target provides the large luminosities needed to 
probe the weak couplings of dark photons. When the electrons from the beam scatter in the target, the dark photons can be emitted in a 
process similar to ordinary bremsstrahlung because of the kinetic mixing. The dark photons are highly boosted carrying most of the 
initial beam energy and get emitted at small angles in the forward direction. The detector is placed behind a sufficiently long shield 
to suppress the SM background. Dark photons can traverse this shielding due to their weak interactions with 
the SM and can then be detected through their decay into leptons (mostly $e^+e^-$ for the mass range of interest). Therefore, a decay 
length of $\mathcal{O}(\mathrm{cm}-\mathrm{m})$ is needed in order for the dark photons to be observable by decaying behind the shield 
and before the detector. This is possible for dark photons with masses larger than $2m_e$ up to $\mathcal{O}(100)$ MeV and small values 
of $\epsilon$ (roughly $10^{-7} \lesssim \epsilon \lesssim 10^{-3}$). Electron beam dump experiments are thus well 
suited to probe this region of the parameter space. 

Depending on the specific experimental set-up with respect to the decay length of the dark photon, the possible reach of an experiment 
is determined not only by the collected luminosity but also by the choice of the beam energy, the length of the shield, and the distance 
to the detector. Large values $\epsilon$, for which the lifetime is very short, are not accessible, since 
the dark photon decays within the shield.  At very small values of $\epsilon$, the sensitivity of these experiments is limited by statistics 
as there are very few dark photons that will be produced and that decay before the detector. The total number 
of events expected in such experiments from decays of dark photons has been determined in~\cite{Bjorken:2009mm,Andreas:2012mt}.  

Several electron beam dump experiments were operated in the last decades to search for light metastable pseudoscalar or scalar 
particles (e.g.\ axion-like particles or Higgs-like particles). Examples are the experiments E141~\cite{Riordan:1987aw} and 
E137~\cite{Bjorken:1988as} at SLAC, the E774~\cite{Bross:1989mp} experiment at Fermilab, an experiment at KEK~\cite{Konaka:1986cb} 
and an experiment in Orsay~\cite{Davier:1989wz}. The measurements performed by the experiments at SLAC and Fermilab have been
reanalysed in~\cite{Bjorken:2009mm} to derive constraints on the dark photon mass and coupling. Updated limits for all experiments 
were presented in~\cite{Andreas:2012mt}, where the acceptances obtained 
with Monte Carlo simulations for each experimental set-up have been included. 
These limits are shown in Fig.~\ref{fig:hspaw-heavy-A'} together with all current constraints. Electron beam dump experiments cover 
the lower left corner of the parameter space in which the lifetime of the dark photon is sufficiently large to be observed behind 
the shield. In order to extend these limits with future experiments to smaller values of $\epsilon$ large luminosities and/or a 
long distance to the detector are needed, since the lower limit of an experiment's reach scales only with the fourth root of those 
two parameters.    

\subsubsection{Electron Fixed-Target Experiments}
\label{subsubsec:nlwcp:dark-photons:experiments:Fixed-target}

Fixed-target experiments using high-current electron beams are an excellent place to search for $A^\prime$s with 
masses $2m_e<m_{A'} <$~GeV and couplings down to $\epsilon^2 \equiv \alpha'/\alpha > 10^{-10}$.   In these experiments,
the $A^\prime$ is radiated off electrons that scatter on target nuclei. Radiative and Bethe-Heitler trident production 
give rise to large backgrounds.   
Generally speaking, three experimental approaches have been proposed:   
dual-arm spectrometers, forward vertexing spectrometers, and full final-state reconstruction.  
In most cases, the detectors are optimized to detect the $e^+e^-$ daughters of the $A'$. The complementary approaches
map out different regions in the mass-coupling parameter space. General strategies for $A'$ searches 
with electron fixed-target experiments were laid out in \cite{Bjorken:2009mm}. The reach for recently proposed dark photon searches
is shown in Fig.~\ref{fig:hspaw-heavy-A'}.

Several experiments have been proposed to search for dark photons: APEX, HPS, and DarkLight at Jefferson Lab (JLab), 
and A1 using MAMI and MESA at Mainz.  

Existing dual-arm spectrometers at Hall A at JLab and MAMI at Mainz have already been used to 
search for dark photons.  
Two groups, the $A'$ Experiment (APEX) collaboration at JLab and the A1 collaboration at Mainz, have performed short test runs 
(few days of data taking) and published search results with sensitivity down to  $\alpha'/\alpha>10^{-6}$ over narrow mass ranges~\cite{Merkel:2011ze,Abrahamyan:2011gv}.  
These results clearly demonstrate the high sensitivity that can be reached in fixed-target experiments.
These experiments use high-current beams ($\sim 100~\mu$A) on relatively thick targets 
(radiation length $X_{0} \sim$ 1-10\%)  to overcome the low geometric acceptance of the detectors ($\sim 10^{-3}$).  
Beam energy and spectrometer angles are varied to cover overlapping regions of invariant mass.  Searches for $A'$ involve looking for 
a bump in the $e^+e^-$ invariant mass distribution over the large trident background, which requires excellent mass 
resolution.

APEX has been approved for a month-long run, tentatively in 2016.  
Using high-current beams ($\sim 100 \mu$A) at four different beam energies on relatively 
thick targets (1-10\% of a radiation length), the proposed full APEX experiment will probe $A'$ masses from 65 to 550 MeV 
for couplings  $\alpha'/\alpha>10^{-7}$~\cite{Essig:2010xa}.  

A1 has already taken some more data, with the expectation that they will probe $A'$ masses in the range $50-120$~MeV 
with a sensitivity in $\alpha'/\alpha$ similar to the test run published in 2011.  
Furthermore, A1 is developing a new experiment to search for dark photon decay vertices displaced from the target by approximately 
$10$~mm. They hope to cover the $A'$ mass range $40<m_{A'}<130$~MeV with a sensitivity in $\alpha'/\alpha$ from  $10^{-9}$ 
down to $10^{-11}$.

The Heavy Photon Search (HPS) collaboration~\cite{HPS} has proposed an experiment for 
Hall~B at JLab using a Si-strip based vertex tracker 
inside a magnet to measure the invariant mass and decay point of  $e^+e^-$ 
pairs and a PbWO$_{4}$ crystal calorimeter to trigger.
HPS uses lower beam currents and thinner targets than the dual arm 
spectrometers, but compensates with large ($\sim 20\%$) forward acceptance. HPS has
high-rate data acquisition and triggering to handle copious beam backgrounds and high-rate trident production.  
Because it can discriminate $A'$ decays displaced more 
than a few millimeters from the large, prompt, trident background, HPS has enhanced sensitivity to small couplings, 
roughly $10^{-7} >  \alpha'/\alpha > 10^{-10}$ for masses $30<m_{A'}<500$~MeV. For prompt decays, HPS will simultaneously explore 
couplings $\alpha^\prime/\alpha>10^{-7}$ over the same mass range.  HPS has conducted a successful test run at JLab during the spring of 2012, 
which demonstrated technical feasibility and confirmed simulations of the background rates. The proposal for ``full'' HPS was approved and funded
by DOE in Summer, 2013. JLab has scheduled commissioning and running during 2014 and 2015. HPS is being constructed during 2013-2014, will be
installed at JLab in September, 2014, and will begin running thereafter at the upgraded CEBAF accelerator.

The DarkLight detector is a compact, magnetic spectrometer
designed to search for decays to lepton pairs 
of a dark photon $A'$ in the mass range $10~{\rm MeV} < m_{A'} 
< 90~{\rm MeV}$ at coupling strengths of $10^{-7} < \alpha'/\alpha < 10^{-4}$. 
The experiment will use the 100 MeV beam of the JLaB FEL incident on a hydrogen gas target at the center of a solenoidal detector, 
comprising silicon detectors (for the recoil proton), a low mass tracker (for the leptons), and shower counters (for photon detection).
By measuring all the final state particles, Darklight can provide  full kinematic reconstruction. The available information also permits 
searching for invisible $A'$ decays via a missing mass measurement. A series of beam tests in summer 2012 verified that sustained, 
high-power transmission of the FEL beam through millimeter-size apertures is feasible~\cite{Balewski:2013oza,Alarcon:2013yxa}. 
JLab has  
approved Darklight. A full technical design is underway and funding is being sought. The goal is to begin data taking in 2016.

The MESA accelerator~\cite{Aulenbacher:2012tg}, which recently has been approved
for funding within the PRISMA cluster of excellence at the University of Mainz, hopes to cover a mass range comparable to that
covered by Darklight~\cite{Beranek:2013yqa}. The MESA accelerator ($155$~MeV beam energy) 
will be operated in the energy recovering linac mode with one recirculating arc as well as a windowless gas jet target.
The Mainz group is considering to use two compact high-resolution spectrometers rather than a high-acceptance tracking
detector. The project is several years off.

\subsubsection{Proton Beam Dump Experiments}
\label{subsubsec:nlwcp:dark-photons:experiments:Beam-dumps-proton}

Proton beam dump experiments can also search for dark photons which decay to visible channels. Several reinterpretations of past experimental 
analyses from LSND~\cite{Batell:2009di,Essig:2010gu,Athanassopoulos:1997er}, 
$\nu$-Cal I~\cite{Blumlein:2011mv,Blumlein:1990ay,Blumlein:1991xh}, 
NOMAD~\cite{Gninenko:2011uv,Astier:2001ck}, PS191~\cite{Gninenko:2011uv,Bernardi:1985ny}, 
and CHARM~\cite{Gninenko:2012eq,Bergsma:1985is} 
have resulted in limits on dark photons that are complementary to those coming from electron fixed target experiments, 
precision QED, and B-factories.
One can take advantage of the large sample of pseudoscalar mesons ({\it e.g.}, $\pi^0$, $\eta$) produced 
in the proton-target collisions, which will decay to $\gamma A'$ with a branching ratio proportional 
to $\epsilon^2$ if kinematically allowed~\cite{Batell:2009di}.  These experiments probe a similar 
region in $A'$ mass and coupling parameter space as past electron beam dumps discussed in 
Section~\ref{subsubsec:nlwcp:dark-photons:experiments:Beam-dumps-electron}, but do have unique 
sensitivity in certain cases. It remains to be investigated whether future proton beam dump experiments 
can cover new regions of $A'$ parameter space. 

Proton beam dump experiments also have significant sensitivity to invisible decays of $A'$, 
particularly when the decay products are stable and can re-scatter in the detector, ({\it e.g.}, 
as in the case of $A'$ decaying to dark matter). In fact, there is a proposal to do a 
dedicated beam dump mode run at MiniBooNE to search for light dark matter~\cite{Dharmapalan:2012xp}. 
This subject is discussed in more detail in \S\ref{sec:nlwcp:LDM}.

\subsubsection{Electron-Positron Colliders}
\label{subsubsec:nlwcp:dark-photons:experiments:Colliders-e+e-}

During the past 15 years, high luminosity $e^+e^-$ flavor factories have been producing an enormous amount of data 
at different center-of-mass energies. In Frascati (Italy), 
the KLOE experiment running at the DA$\Phi$NE collider, has acquired about 2.5 fb$^{-1}$ of data at the $\phi$(1020) 
peak. B-factories at PEP-II (USA) and KEK-B (Japan) have 
delivered an integrated luminosity of 0.5-1 ab$^{-1}$ to \babar\ and Belle, respectively. In China, the Beijing BEPC 
collider is currently running at various energies near the charm threshold and has already delivered several inverse 
femtobarn of data to the BESIII experiment.  

These large datasets have been exploited to search for dark photon production in the following processes:

\begin{itemize}

\item{The radiative production of a dark photon ($A'$) followed by its decay into a charged lepton pair, 
$e^+e^- \rightarrow \gamma A', A' \rightarrow l^+l^-$ ($l=e,\mu$)~\cite{Fayet:2007ua}. (If instead of an $A'$ one produces 
a new scalar or pseudo-scalar particle, then decays of this particle to two photons are also possible.)}

\item{The associated production of a dark photon with a new light scalar particle, generally dubbed as $h'$. 
The existence of the latter is postulated in models where the hidden symmetry is broken by some 
Higgs mechanism~\cite{Batell:2009yf}. Similarly to the SM Higgs, the mass of the $h'$ is not 
predictable by first principles and could be at the $\sim\gev$ scale as well. The phenomenology 
is driven by the mass hierarchy. While scalar bosons heavier than two dark photons decay promptly, 
giving rise to events of the type $e^+e^- \rightarrow A' h' \rightarrow 3 A', A' \rightarrow l^+l^-, \pi^+\pi^-$, 
their lifetime becomes large enough to escape undetected for $m_{h'} < m_{A'}$, resulting in 
$e^+e^- \rightarrow A' h' \rightarrow l^+l^- + missing~energy$ events.}

\item{Radiative meson decays, which could also produce a dark photon with a branching ratio suppressed by a 
factor $\epsilon^2$~\cite{Reece:2009un}.}

\end{itemize}

The search for a light CP-odd Higgs ($A^0$) in $\Upsilon(2S,3S) \rightarrow \gamma A^0, A^0 \rightarrow 
\mu^+\mu^-$ conducted by \babar~\cite{Aubert:2009cp} has been reinterpreted in terms of constraints on dark photon 
production~\cite{Bjorken:2009mm,Reece:2009un,Essig:2010xa}, 
as its signature is identical to that of $e^+e^- \rightarrow \gamma A', A' \rightarrow \mu^+\mu^-$. 
Limits on the coupling $\epsilon^2$ at the level of $10^{-5}$ have been set. Future analyses based on the 
full \babar\ and Belle datasets are expected to increase the sensitivity by an order of magnitude.

A search for a dark photon and an associated scalar boson has been performed at \babar\ in the range 
$0.8 < m_{h'} < 10.0 ~\gev$ and $0.25 < m_{A'} < 3.0 ~\gev$, with the constraint $m_{h'} > 2 m_{A'}$~\cite{Lees:2012ra}. 
The signal is either fully reconstructed into three lepton or pion pairs, or partially reconstructed as two 
dileptonic resonances, assigning the remaining dark photon to the recoiling system. No significant signal is 
observed, and upper limits on the product $\alpha_D \epsilon^2$ are set at the level $10^{-10} - 10^{-8}$. 
These bounds are translated into constraints on the mixing strength in the range $10^{-4} - 10^{-3}$, 
assuming $\alpha_D = \alpha \simeq 1/137$. A similar search currently performed by Belle should improve 
these limits by a factor of two. 

KLOE has searched for $\phi(1020) \rightarrow \eta A', A' \rightarrow e^+e^-$ decays, in which the $\eta$ 
was tagged with either the 3$\pi^{0}$ or the $\pi^{+}\pi^{-}\pi^{0}$ final states~\cite{Archilli:2011zc,
Babusci:2012cr}. The $A' \rightarrow \mu^{+}\mu^{-},\pi^{+}\pi^{-}$ channels were not included due to a higher 
background level. After subtraction of the $\phi$ Dalitz decay background, no evident peak is observed, and the 
following limits are set at 90$\%$ CL: $\epsilon^{2} <1.5 \times 10 ^{-5}$ for $30 <m_{A'}< 420 ~\Mev$, 
$\epsilon^{2}< 5\times 10^{-6}$ for $60 <m_{A'}< 190 ~\Mev$. 

The BESIII Collaboration has published a search for invisible decays of the $\eta$ and $\eta'$ mesons, 
motivated by the possible existence of light neutral dark matter particles~\cite{Ablikim:2012gf}. Events 
are selected from $J/\psi \rightarrow \phi\eta(\eta')$ decays, where the $\phi$ is tagged by its charged 
kaon decay mode. No significant signal is observed, and 90\% CL limits on the branching ratio $BR(\eta \rightarrow 
invisible) < 1.0 \times 10^{-4}$ and $BR(\eta' \rightarrow invisible) < 5.3 \times 10^{-4}$ are set. These bounds 
constrain the invisible dark photon decay through $\eta (\eta') \rightarrow A'A', A' \rightarrow invisible$.

\noindent {\bf Future Searches using Current Datasets}

Current datasets have not been fully exploited to search for signatures of a dark sector. Current studies 
of the $e^+e^- \rightarrow \gamma A', A' \rightarrow l^+l^-,\pi^+\pi^-$ based on the full \babar\ and Belle 
datasets are expected to probe values of the coupling $\epsilon^2$ down to $\sim 10^{-6}$, and extend the 
coverage down to $\sim 20 ~\Mev$, covering the full region favored by the $g-2$ discrepancy. KLOE is 
expected to probe values of $\epsilon^2$ between $\sim 10^{-5}$ and $\sim 7 \times 10^{-7}$ in the range 
$ 500 < m_{A'} < 1000 ~\Mev $ using the $e^+e^- \rightarrow \mu^+\mu^- \gamma$ sample selected for the 
study of the hadronic contribution to the muon magnetic anomaly. Similarly, invisible dark photon decays 
could be studied in the $e^+e^- \rightarrow \gamma + invisible$ final state, using data collected at \babar\ 
with a specific single-photon trigger (see also \S\ref{sec:nlwcp:LDM}). 
This search could probe dark photon masses $ 0 < m_{A'} < 5 ~\gev$, 
significantly extending the parameter space covered by proposed searches in neutrino experiments~\cite{Dharmapalan:2012xp}. 
The calorimeter hermeticity and energy resolution play a crucial role for this study, as well as the amount of 
accidental background. Similar considerations apply to searches for purely neutral 
dark photon decays. 

A search for a light $h'$, pair produced with a dark photon is being performed at KLOE using $e^+e^- \rightarrow 
A' h' \rightarrow l^+l^- + missing~energy$ events. This search fully complements the analysis performed by \babar\, 
covering a totally different parameter space. Extensions to non-Abelian model could easily be probed using current 
datasets. The simplest scenario include four gauge bosons, one dark photon and three additional dark bosons, 
generically denoted $W'$. A search for di-boson production has been performed at \babar\ in the four lepton 
final state, $e^+e^- \rightarrow W' W', W' \rightarrow l^+l^-$ ($l=e,\mu$), assuming both bosons have similar 
masses~\cite{arXiv:0908.2821}. More generic setups could easily be investigated.

The existence of a dark scalar or pseudo-scalar particle can also be investigated in $B \rightarrow K^{(*)} l^+l^-$ decays. 
The sensitivity of \babar\ and Belle searches to the SM Higgs--dark scalar mixing angle and pseudo-scalar couplings 
constants are projected to be at the level of $10^{-4} - 10^{-3}$ and $10^3~\tev$, respectively \cite{Batell:2009jf}.

\subsubsection{Rare Kaon decays}
\label{subsubsec:nlwcp:dark-photons:experiments:Kaons}

Many experiments produce an enormous number of kaons and other mesons.  Searches for rare kaon decays, like 
$K^+ \to \mu^+ \nu A'$ and $K^+\to \pi^+ A'$, with $A' \to e^+e^-$, can potentially be sensitive to new regions in parameter space.  
Examples include ORKA~\cite{E.T.WorcesterfortheORKA:2013cya}, NA62~\cite{NA62}, and TREK/E36~\cite{TREK/E36,Strauch}.  
We do not show their reach explicitly.

\subsubsection{Proton Colliders}
\label{subsubsec:nlwcp:dark-photons:experiments:Colliders-proton}

Proton colliders have the ability to reach high center-of-mass energy, making it possible to produce $Z$ bosons, Higgs bosons, 
and perhaps other new, heavy particles (such as supersymmetric particles, $W'$/$Z'$ states, or hidden-sector particles) 
directly. As pointed out in many theoretical 
studies~\cite{Strassler:2006im,ArkaniHamed:2008qp,Baumgart:2009tn,Ruderman:2009tj,Cheung:2009su}, if new states are produced, they could decay to $A'$ bosons and other hidden-sector states with very large branching ratios. 
For GeV-scale $A'$ masses, the $A'$ would be highly boosted when produced in such decays and its 
decay products would form collimated jets, mostly composed of leptons (``lepton-jets'' \cite{ArkaniHamed:2008qp}).

The general-purpose proton collider experiments at the Tevatron and LHC have all presented first searches for 
lepton-jets in heavy-particle decays \cite{Abazov:2009hn,Abazov:2010uc,cdflj,atlaslj,Chatrchyan:2011hr}. 
The searches usually employ a specialized lepton-jet identification algorithm to distinguish them from the large multi-jet background. 
Events with additional large missing transverse energy (from other escaping hidden-sector particles) or a particular di-lepton 
mass (corresponding to the $A'$ mass) have also been searched for~\cite{Chatrchyan:2012cg}. Results have often been interpreted in supersymmetric 
scenarios; the updated ATLAS analysis using 7 TeV pp data from 2011 excludes di-squark production with a squark mass up to about 1000 GeV or a weakly-produced state with mass up to about 400 GeV, decaying through cascades to two lepton-jets~\cite{Aad:2012qua}.
Current searches have mostly focused on $A'$ bosons heavy enough to decay to muon pairs, since this offers a cleaner signal than electron pairs, but good sensitivity has also been seen down to $\sim 50$~MeV (limited by photon conversions to $e^+e^-$ pairs). 

ATLAS has recently searched for decays of the Higgs boson to electron lepton-jets, excluding a branching ratio of about 50\%~\cite{Aad:2013yqp}.
Searches have mostly focused so far on prompt decays of dark photons, 
but ATLAS has now searched for decays of the Higgs boson to long-lived $A'$ bosons decaying to muons in the muon chambers, constraining the branching ratio to be less than 10\% for a proper lifetime between 10 and 100 mm~\cite{Aad:2012kw}.

\subsubsection{Photon Regeneration Experiments (ultra-light dark photons)}
\label{subsubsec:nlwcp:dark-photons:experiments:Laser}

Light dark photons ($\sim$~meV) may be searched for in photon regeneration experiments. 
Experiments using laser light as REAPR and ALPS-II have already been explained in \S\ref{subsubsec:nlwcp:axions-&-alps:experiments:Laser}. The sensitivity of ALPS-II (in its final phase) 
is shown in Fig.~\ref{fig:hspaw-light-A'} (REAPR would be similar). 
It reaches the region where meV mass $A'$ can be cold DM. This region is motivated by 
type-II string models, in particular anisotropic compactifications (with one very large dimension), 
where the $A'$ arises via the Stuckelberg mechanism.  

Photon regeneration experiments have been performed in the microwave range ($\mu$eV $A'$ masses) 
by attempting the transfer of power between isolated microwave cavities in tune following the concept of~\cite{Hoogeveen:1992nq,Jaeckel:2007ch} (dark green sharp triangles labelled ``mwLSW''). 
The current constraints correspond to proof-of-concept experiments by groups in the University of Western Australia (UWA)~\cite{Povey:2010hs} and ADMX~\cite{Wagner:2010mi} and a more mature experiment performed at CERN~\cite{Betz:2013dza}. 
The axion-sensitivity improvements expected by ADMX (see~\ref{subsubsec:nlwcp:axions-&-alps:experiments:Microwave}) will also 
allow a much more sensitive parasitic search of $A'$s, see curve labelled ADMX in Fig.~\ref{fig:hspaw-light-A'}~\cite{Wagner:2010mi} 
(see also~\cite{Parker:2013fba}). 

\subsubsection{Helioscopes (ultra-light dark photons)}

Helioscopes looking for solar axions can detect transversely polarized $A'$s emitted from the solar interior~\cite{Redondo:2008aa}. If the $A'$ mass is due to the Stuckelberg mechanism, the region in which they are sensitive is already excluded by our knowledge of the Sun~\cite{An:2013yfc} (the emission of longitudinally polarized $A'$s would require a too high solar core temperture to be compatible with the measured solar neutrino fluxes~\cite{Redondo:2013lna}). If the $A'$ mass comes from a Higgs mechanism, the situation is different and is currently being explored~\cite{An:2013yua}. The SHIPS helioscope in the Hamburg Sternwarte~\cite{Schwarz:2011gu} is currently looking for this possibility but more theoretical input is required to assess the impact of its measurements and future directions. 

A more promising endeavor is to detect the solar flux of longitudinally polarized $A'$s, which dominates for $m_{A'}\lesssim10$~eV. These $A's$ have typical energies 10-300 eV and produce ionization in DM detectors such as XENON10~\cite{An:2013yua}. The first rough analysis of~\cite{An:2013yua} shows a very 
promising venue to search for Stuckelberg $A'$s near the solar limit. 
This line shall benefit of future experiments with more exposure and more detailed analysis of the low energy 
single-electron ionization events. The parameter-space regions that can be explored are motivated from string theory and from DM. 

\subsubsection{Cold DM searches (ultra-light dark photons)}

If DM comprises cold $A'$s produced by the misalignment mechanism it will produce a signal in 
Haloscopes looking for axion DM if they are tuned to the $A'$ mass~\cite{Arias:2012az}. 
The signal has two characteristics that make it different from an axion signal. 
First, it is not proportional to the magnetic field and thus survives when this is switched off (which could lead one to think it is a systematic). Second, the $A'$ polarization vector changes its orientation 
with respect to the cavity eigenmodes producing an oscillation of the effectiveness of the DM-cavity coupling and thus of the power output. 
If the $A'$ polarization is homogeneously aligned in the DM halo only a daily modulation is expected, but if this is not the 
case, more complicated patterns are to be expected. 
The sensitivity of the next phases of ADMX to axion DM can be directly translated into sensitivity to 
$A'$ DM (green regions labelled ADMX and ADMX-HF in the low-$\epsilon$ regime of Fig.~\ref{fig:hspaw-light-A'}).  The sensitivity of these experiments is impressive, as a moderate coverage of the axion line in axion searches implies orders of magnitude 
in $\epsilon$ of pristine unexplored parameter space. 
The Yale search of 0.1 meV ALP DM~\cite{Slocum:2013qi} is a good example. With its sensitivity goal of $g_{a\gamma\gamma}\sim \mathcal{O}(10^{-10})$ GeV$^{-1}$ it can barely surpass the strong astrophysical constraints on ALPs but its sensitivity to $A'$s is $\chi\simeq 10^{-12}$, two orders of magnitude deep into unexplored territory. 

The vast amount of parameter space to be explored has inspired new DM \emph{broadband} experiments. 
 The idea is avoid the time-consuming (and technology dependent) mode-tuning of resonant cavities 
 and employ a dish antenna (which triggers $A'$ DM conversion into light of frequency given by the $A'$ mass)  and a broadband receiver~\cite{Jaeckel:2013sqa,Jaeckel:2013eha}. With this setup one can study the 3-D momentum distribution of DM in the halo~\cite{Jaeckel:2013sqa}. 
 One can also search for axion/ALP DM by embedding it in a strong magnetic field parallel to the dish surface~\cite{Jaeckel:2013sqa}. Currently there are no experiments of this type being planned but a few groups in Hamburg (DESY), CERN, and University of Zaragoza have expressed interest.

\subsection{Opportunities for Future Experiments: New Ideas, Technologies, \& Accelerators}
\label{subsec:nlwcp:dark-photons:future}

The physics motivations for dark photons, as outlined in \S\ref{subsec:nlwcp:dark-photons:theory}, 
easily motivate extending searches far beyond the present 
generation of experiments. Large parts of the mass-coupling parameter space, shown in Fig.~\ref{fig:hspaw-heavy-A'}, will remain uncovered after the 
experiments at JLab and Mainz have run, and after data from the B- and $\Phi$-factories will have been fully analyzed.  If something 
is found in the present generation of experiments, it will be incumbent on future experiments to confirm the findings, 
explore the detailed properties of the 
new particle, and to seek any cousins. That exercise will demand experiments with improved performance and reach.   If nothing 
is found in the present searches, there remains a vast and viable region of parameter space to explore. Specific models for 
dark photons have been advanced, which populate the virgin territory, and general considerations from theory and phenomenology 
do so as well.  So extending searches  for dark photons throughout the whole of the parameter space is a high priority in either case.

Future experiments can improve upon coverage of the dark photon parameter space significantly.
Future fixed target electro-production experiments 
and future $e^+e^-$ colliding beam facilities can extend the search for visible dark photon decays, and future beam dump experiments, 
discussed below in \S\ref{sec:nlwcp:LDM}, will extend the search for invisible decays, ultralight dark photons, and millicharged particles.

\subsubsection{Future Fixed Target Experiments}

\begin{figure}[!t]
\centering
\vspace*{-5mm}
\includegraphics[width=0.49\textwidth]{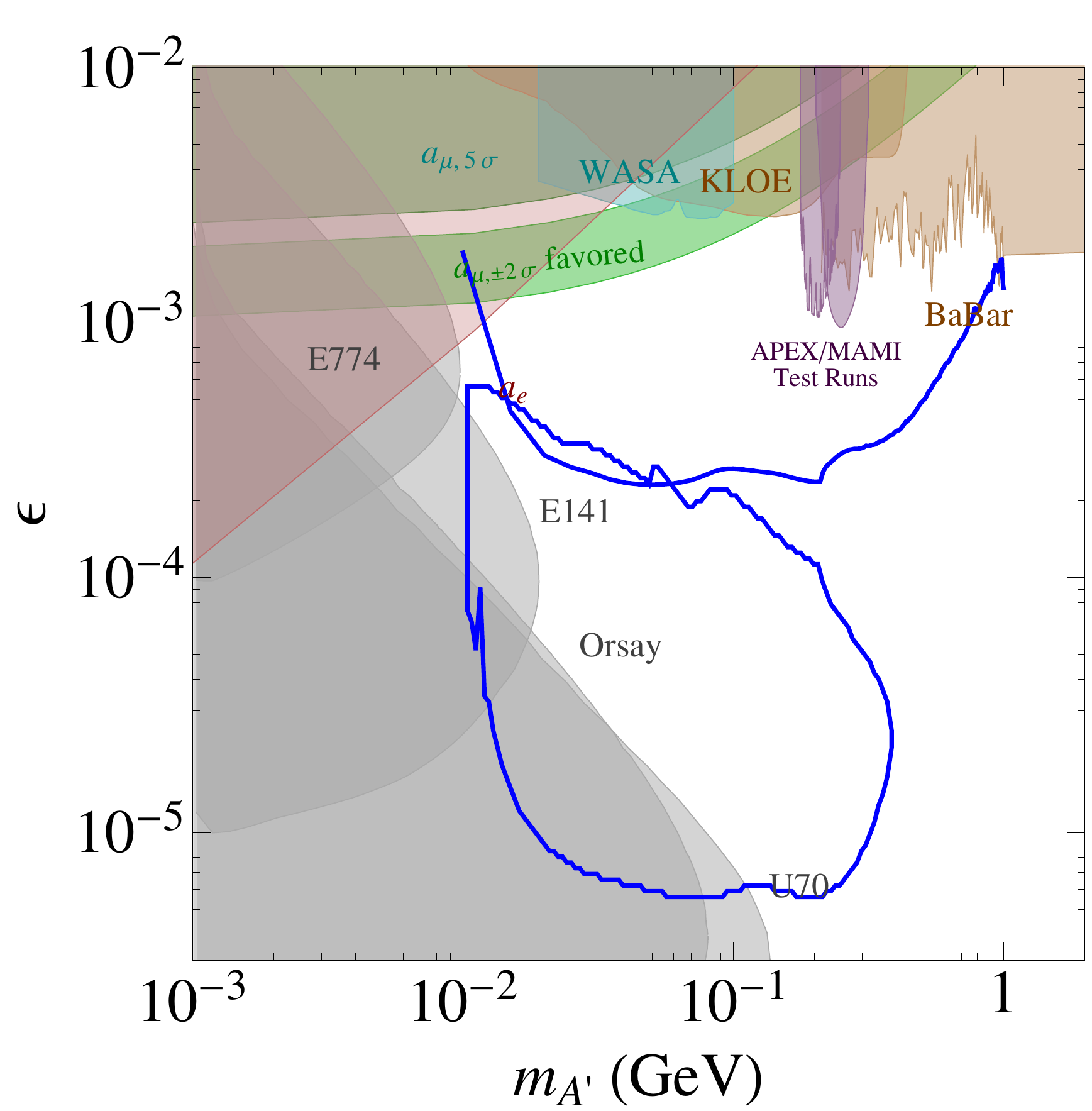}
\caption{
Parameter space for hidden-photons ($A'$) with mass $m_{A'}> 1$~MeV 
(see Fig.~\ref{fig:hspaw-heavy-A'} for present constraints and presently planned experiments), showing in blue a reach projection possible in a future HPS-style 
fixed target experiment.  The projected reach here assumes a factor of two improvement in the vertex resolution, a factor of four improvement in mass resolution, a factor of 30 times more luminosity, and higher-energy running with improved particle ID.}
\label{fig:future-heavy-A'}
\end{figure}

At fixed target machines, several generic improvements look possible which can expand the reach significantly.  First, in 
HPS-like experiments, it should be possible to boost the integrated luminosity by at least one order of magnitude by boosting
the product of the current on target and the run time. Boosting the current on target 
will require tracking detectors that avoid the highest occupancy/radiation 
damage environments yet preserve most of the acceptance, or new pixellated and rad hard detectors that can tolerate the 
higher rates. Second, studies have shown that catching the recoil electron in addition to the $A'$ decay products, 
will boost the mass resolution by a factor $\sim 2$, and will reduce Bethe-Heitler backgrounds
by as much as a factor of 2 to 3, improving the significance. Third, triggering on pions and muons can boost the sensitivity 
for $A's$ for masses beyond the 
dimuon mass by large factors where the $\rho$ dominates the $A'$ decays,  and will help significantly for higher 
masses too. Fourth, using low-$Z$ nuclear targets and maximal beam energies improves the reach in the $300-1000$~MeV range 
too, since the radiative $A'$ cross section increases with higher beam energies and form-factor suppression will 
be mitigated by going to smaller, lower-$Z$ nuclei. CEBAF12 at JLab will provide 12 GeV beams where these effects can boost
production by a factor of 5 or more. Fifth, since  the sensitivity of bump hunt searches 
depend inversely with the square root of the invariant mass resolution and directly with the square root of the acceptance, and since
vertex searches depend critically on the vertex resolution, improving these parameters can lead to significant gains. 
It is not unreasonable to assume a factor of 2 improvement in the acceptance and the mass resolution. 
The vertex resolution can be improved by using thinner targets (with 
compensating higher currents), shorter extrapolation distances from the first detector layer to the target, and thinner 
detectors, yielding better impact parameter resolution. 
Taken together, future experiments may be able to discriminate $A'$ decays just a few mm from the target (vs 15 mm 
in the current version of  HPS). Finally, optimized  analysis procedures and multivariate analyses 
may buy factors of two improvement in sensitivity.  An estimate from Snowmass studies of the reach of a future 
experiment~\cite{GrahamSnowmass}, which assumes a factor 4 improvement in mass resolution, a factor 2 in vertex resolution, 
30 times the luminosity, and higher energy running with particle ID, gives Fig.~\ref{fig:future-heavy-A'}.  
Note that this exercise exploits just one of 
the current approaches for fixed target electroproduction.
 
New approaches may be even more powerful~\cite{NelsonSnowmass}. While detailed performance estimates for new experimental 
layouts are not yet available, several new ideas are being discussed and the possibilities for improving the reach of dark photon searches
significantly look very real.

\subsubsection{Searches at Future $e^+e^-$ Colliding Beam Facilities}

The $e^+e^-$ colliding beam machines have conducted sensitive searches for dark photons over a wide range of masses. These searches, 
using existing data sets, are continuing. Since future facilities are already approved, it is comparatively straight-forward to 
extrapolate their performance for future searches. The coupling accessible by current datasets from $e^+e^-$ colliders is at 
the level $\epsilon^{2} \sim 10^{-6} - 10^{-5}$ for dark photon masses below a few hundred $\Mev$. This limitation comes essentially
from the available statistics and thus the luminosity that can be delivered by the accelerators. The luminosity typically scales 
quadratically with their center of mass energy, basically compensating the inverse scaling of the relevant production cross-sections.

Current factories reach instantaneous luminosities of a few times $10^{32}$ ($10^{34}$) cm$^{-2}$s$^{-1}$ at 1 (10) $\gev$. 
Several next generation flavor factories have been proposed or are currently under construction. The upgraded KEK B-factory, 
SuperKEKB, is expected to start taking data in 2016 and should collect 50 ab$^{-1}$ by 2022, about two orders of magnitude 
larger than the dataset collected by Belle. Several tau-charm factories operating between $2-5 ~\gev$ with instantaneous 
luminosities at the level of $10^{35} - 10^{36}$ cm$^{-2}$s$^{-1}$ have been proposed, but remain to be funded at the time 
of this writing. Their expected sensitivity would roughly be at the level SuperKEKB should reach. At Frascati, KLOE-2 will 
install a new inner tracker, a cylindrical GEM detector, to improve the momentum resolution of charged particles while 
keeping the amount of material at a minimum. This approach will hopefully reduce the background from photon conversions 
produced in $e^{+}e^{-} \rightarrow \gamma\gamma, \gamma \rightarrow e^+e^-$ events, allowing KLOE-2 to explore the very 
low mass region.

An alternative approach has been proposed by the authors of~\cite{Wojtsekhowski:2012zq}, colliding a {\it single} intense 
positron beam on an internal target. Specifically, the VEPP-3 collaboration has proposed to use a $500 ~\Mev$ positron beam 
of VEPP-3 on a hydrogen gas internal target. The search method is based on the study of the missing mass spectrum in the 
reaction $e^+e^- \rightarrow A' \gamma$, which allows the observation of a dark photon independently of its decay modes 
and lifetime in the range $m_{A'}=5-20 ~\Mev$, see Fig.~\ref{fig:hspaw-heavy-A'}. 

In summary, next generation flavor factories could probe values of the coupling $\epsilon^2$ down to a level comparable 
to presently planned fixed target experiments for prompt decays, while  extending their mass coverage to significantly higher
masses. Should a signal be 
observed, $e^+e^-$ colliders will be ideally suited to investigate in detail the structure of a hidden sector, 
complementing dedicated experiments.  

\subsubsection{Future Searches at the LHC}
The large datasets expected at the LHC in the near-term future (300 fb$^{-1}$ at 14 TeV) will contain 
billions of $Z$ and millions of Higgs bosons, allowing branching ratios to lepton-jets as low as $10^{-7}$ (or $\epsilon\simeq 10^{-3}$) 
to be probed for $Z$ decays and $10^{-3}$ for Higgs decays. Strongly-produced SUSY particles with masses up to 1 (2.5) TeV 
are another potential source of lepton jets, if decays proceed through the hidden sector. In fact, the lepton jet signature 
could even be critical for their discovery. In the longer term with its very high luminosity option (3000 fb$^{-1}$), the LHC
will allow ever more sensitive $Z$ and Higgs decay searches, and extend the mass reach for SUSY production even higher.

\section{Light Dark-Sector States (incl.~Sub-GeV Dark Matter)}
\label{sec:nlwcp:LDM}

\subsection{Theory \& Theory Motivation}
\label{subsec:nlwcp:other:theory}

DM and neutrino mass provide strong empirical evidence for physics beyond the SM. Arguably, rather than suggesting any specific mass scale for new physics, they point to a hidden (or dark) sector, weakly-coupled to the SM. Dark sectors containing light stable degrees of freedom, with mass in the MeV-GeV range, are of particular interest as DM candidates as this regime is poorly explored in comparison to the weak scale. Experiments at the intensity frontier are ideally suited to explore this light dark-sector landscape, as discussed in this section.

\subsubsection{Light Dark Matter}
\label{subsubsec:nlwcp:other:theory:LDM}

DM provides one of the strongest empirical motivations for new particle physics, with evidence coming from various disparate sources in astrophysics and cosmology. While most activity has focused on the possibility of weakly-interacting massive particles (WIMPs) with a weak-scale mass, this is certainly not the only possibility. The lack of evidence for non-SM physics at the weak scale from the LHC motivates a broader perspective on the physics of DM, and new experimental strategies to detect its non-gravitational interactions are called for. A wider theoretical view has also been motivated in recent years by anomalies in direct and indirect detection~\cite{Adriani:2008zr,DAMA,CoGent}, possible inconsistencies of the $\Lambda$CDM picture of structure formation on galactic scales~\cite{Weinberg:2013aya}, 
and the advent of precision CMB tests of light degrees of freedom during recombination. 

The mass range from the electron threshold $\sim$0.5 MeV up to multi-TeV  characterizes the favored range for DM candidates with non-negligible SM couplings and is on the scale accessible to terrestrial particle physics experiments. The simple thermal relic framework, with abundance fixed by freeze-out in the early Universe, allows DM in the MeV-GeV mass range if there are light (dark-force) mediators that control the annihilation rate \cite{Boehm:2003hm}. Related scenarios, such as asymmetric DM, also require significant annihilation rates in the early Universe, and thus light mediators are a rather robust prediction of models of MeV-GeV-scale DM that achieve thermal equilibrium. Current direct detection experiments searching for elastic nuclear recoils lose sensitivity rapidly once the mass drops below a few GeV, 
although several ideas have been proposed to look for 
DM scattering off electrons or molecules~\cite{Essig:2011nj,Essig:2012yx,Graham:2012su}.  
Experiments at the intensity frontier provide a natural alternative route to explore this light MeV-GeV scale DM regime. Crucially, the light mediators required for DM annihilation to the SM provide, by inversion, an accessible production channel for light DM that can be exploited in high luminosity experiments. 

Models of sub-GeV DM are subject to a number of terrestrial and cosmological constraints, as discussed below. However, simple models interacting through one or more of the portal couplings are viable over a large parameter range; e.g.~an MeV-GeV mass complex scalar charged under a massive dark photon can be thermal relic DM, with SM interactions mediated by the kinetic mixing portal.

\subsubsection{Light Dark-Sector States}
\label{subsubsec:nlwcp:other:theory:LDSS}

There is no compelling argument, beyond simplicity, for cold DM to be composed of a single species, or even a small number of species. Light stable thermal relics require the presence of additional light mediators as discussed above, and the dark sector may be quite complex. Indeed, the annihilation channels required for (thermalized) DM in the early Universe could occur within the dark sector itself if there are additional light states, subject to constraints from cosmology on the number of relativistic degrees of freedom.
Since SM neutrinos do contribute a (highly sub-dominant) fraction of hot DM, we already know that in the broadest sense DM must be comprised of multiple components. Thus care is required to assess the experimental sensitivity according to the underlying assumptions about the stability of the dark sector state in cosmological scales, and whether or not stable dark sector states under study comprise some or all of DM.

\subsubsection{Millicharged Particles}
\label{subsubsec:nlwcp:other:theory:MCP}

Particles with small un-quantized electric charge, often called mini- or milli-charged particles (MCPs), also arise 
naturally in many extensions of the SM.
MCPs are a natural consequence of extra $U(1)$s and the kinetic mixing discussed in 
\S\ref{subsec:nlwcp:dark-photons:theory} for massless $A'$ fields. 
In this case, any matter charged (solely) under the hidden $U(1)$ obtains a small electric charge.
MCPs can also arise in extra-dimensional scenarios or as hidden magnetic monopoles receiving their mass from a magnetic mixing effect~\cite{Batell:2005wa,Brummer:2009cs,Bruemmer:2009ky}.
Milli-charged fermions are particularly attractive because chiral symmetry protects their masses against quantum corrections, making it more natural to have small or even vanishing masses.
MCPs have also been suggested as DM candidates~\cite{Goldberg:1986nk,Cheung:2007ut,Feldman:2007wj}. 

Terrestrial experiments as well as astrophysical and cosmological observations provide interesting bounds on MCPs. These limits, in addition to comments on future prospects, are summarized in \S\ref{subsubsec:nlwcp:other:pheno:MCP}.

\subsection{Current Constraints}
\label{subsec:nlwcp:other:pheno}

\subsubsection{Constraints on Light Dark Matter and Dark Sectors}
\label{subsubsec:nlwcp:other:pheno:LDM}

\begin{figure*}[tbp]
\centering
\includegraphics[width=0.44\textwidth]{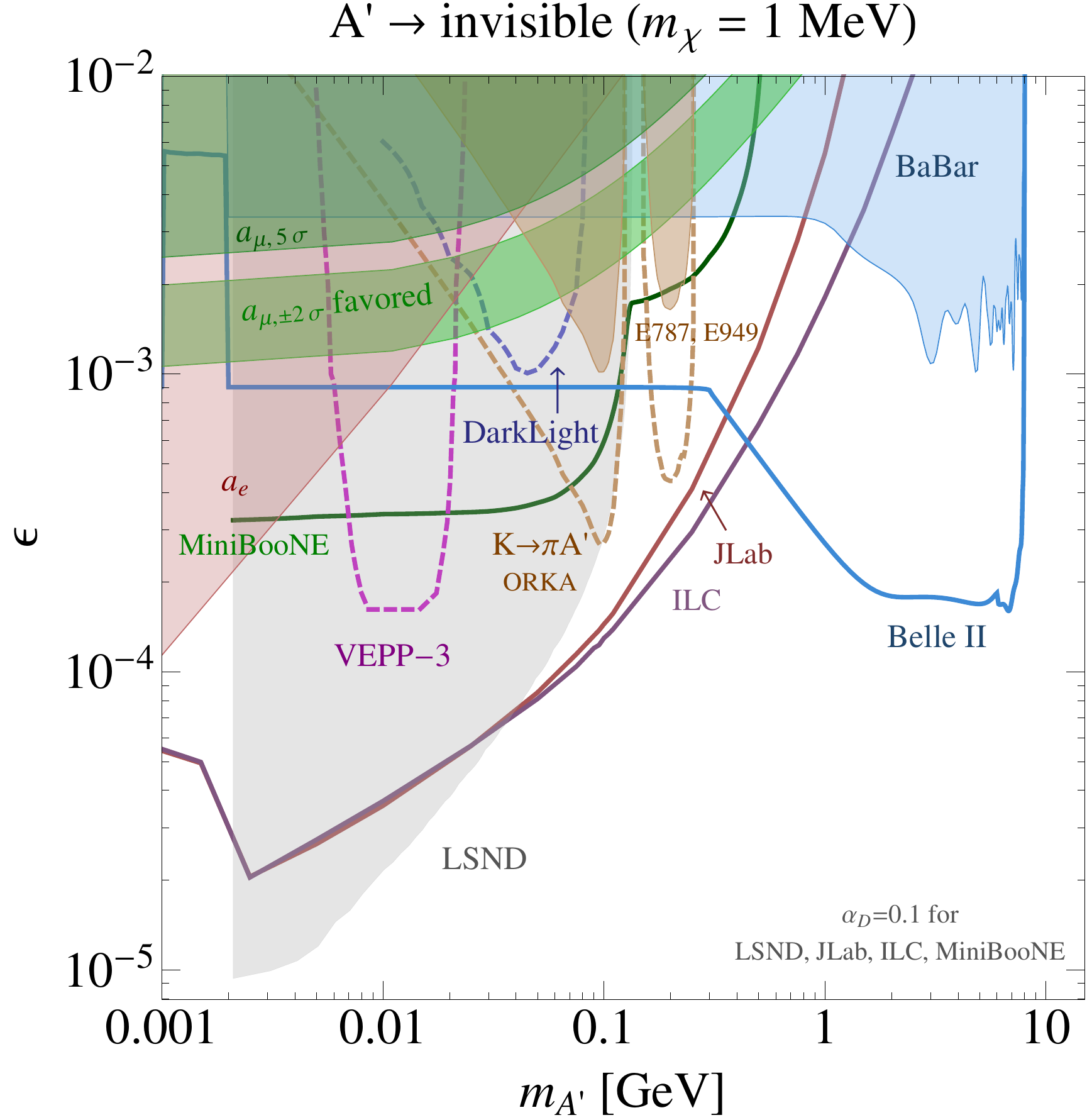}~~
\includegraphics[width=0.44\textwidth]{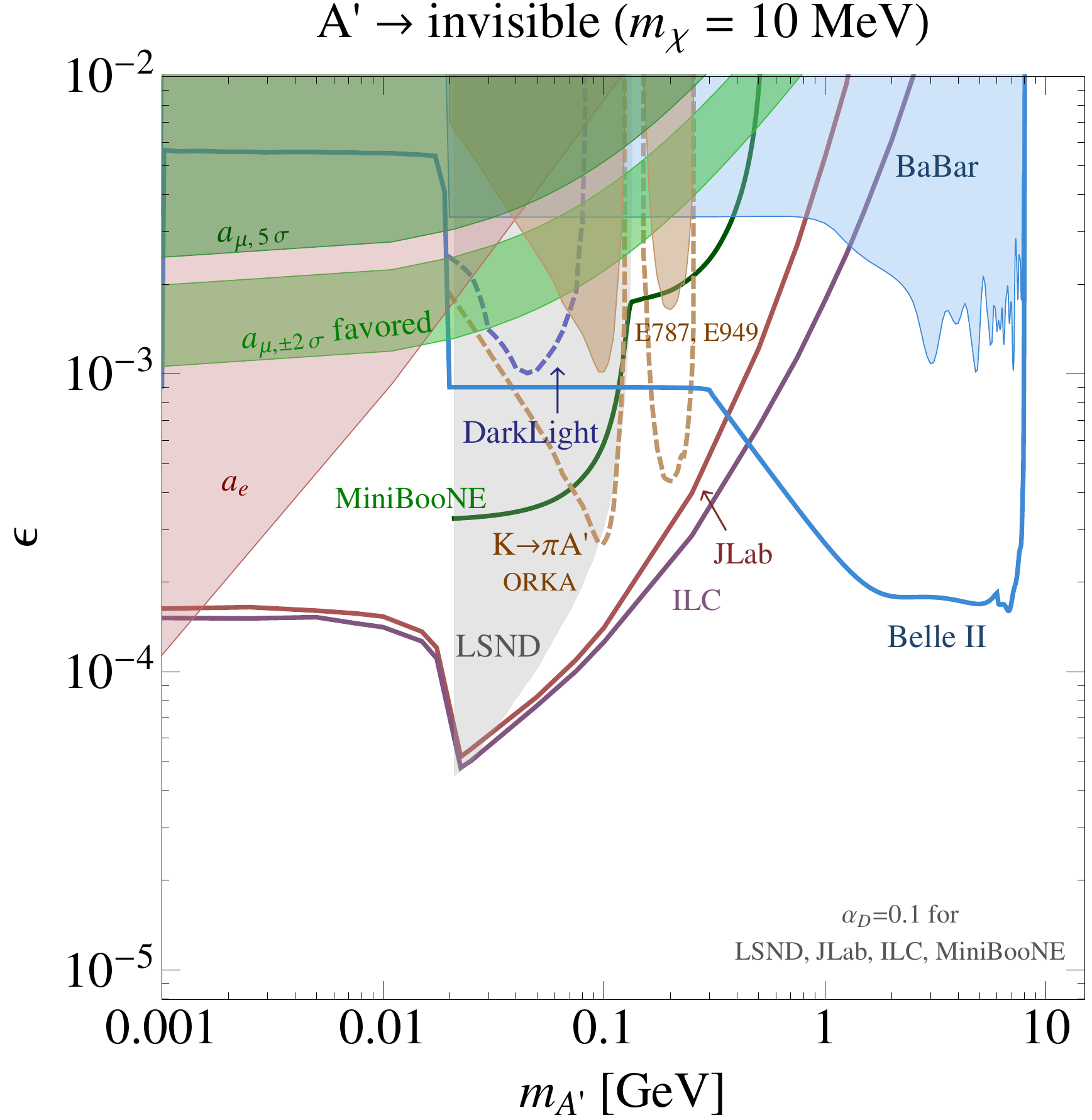} \\
\vskip 1mm
\includegraphics[width=0.44\textwidth]{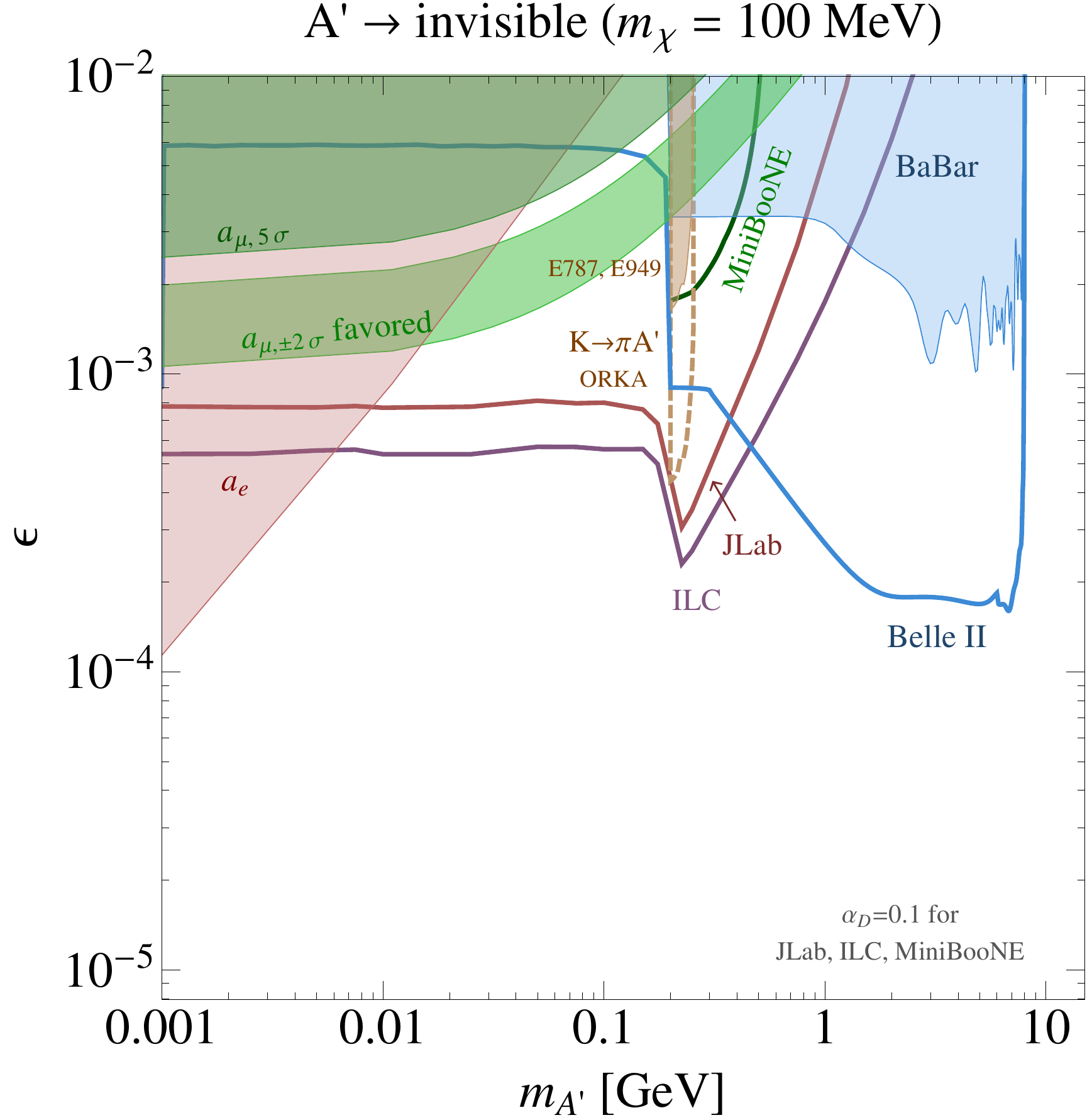}
\vskip -4mm
\caption{Parameter space for dark photons ($A'$) 
decaying invisibly to dark-sector states $\chi$ for various $m_\chi$. 
Constraints from the electron (red) and muon (green) anomalous magnetic moment~\cite{Pospelov:2008zw}
are independent of the $A'$ decay mode (see also Fig.~\ref{fig:hspaw-heavy-A'}). 
Constraints from (on-shell) $A'$ decays to \emph{any} invisible final state arise from 
the measured $K^+ \rightarrow \pi^+ \nu \bar \nu$  branching ratio~\cite{Artamonov:2009sz,Pospelov:2008zw,Dharmapalan:2012xp} (brown) and from a \babar\ mono-photon search~\cite{Aubert:2008as,Essig:2013vha,Izaguirre:2013uxa} (blue); significant improvements are possible with DarkLight~\cite{Kahn:2012br} (dark blue dashed), VEPP-3~\cite{Wojtsekhowski:2009vz,Wojtsekhowski:2012zq} (magenta dashed), ORKA~\cite{Essig:2013vha} (brown dashed), and BELLE II~\cite{Essig:2013vha} (light blue solid).  
If the $\chi$ are long-lived/stable and re-scatter in a downstream detector, constraints 
arise also from LSND (gray) for $m_A' < m_{\pi^0}$, $m_\chi <  m_A'/2$~\cite{deNiverville:2011it}.  
Additional parameter space can then also be probed at existing/future proton beam-dump facilities like Project X, LSND etc., 
(the solid dark green line shows a proposed MiniBooNE beam-off-target-run~\cite{Dharmapalan:2012xp}), and at 
electron-beam dumps at JLab (dark red), the ILC (purple), and other facilities like SLAC, SuperKEKB 
etc.~(not shown)~\cite{Izaguirre:2013uxa}.  Supernova constraints are applicable for lower $\epsilon$~\cite{Dreiner:2013mua} (not shown). 
 }  
\label{nlwcp:fig:invisible-A'}
\end{figure*}

A variety of terrestrial, astrophysical, and cosmological constraints exist on light DM and dark sector states, which we now summarize. 
We focus on the scenario with dark sector states $\chi$ (including DM) interacting with the SM through a dark photon, emphasizing the assumptions going into each limit.
For example, the limits depend on the dark photon and light DM mass.  We also note that generalizations to beyond the kinetic mixing 
portal, including for example leptophilic DM, can drastically alter the limits.  
The limits on dark photons that couple to light DM, along with prospects for various future experiments to be discussed below, are displayed in Fig.~\ref{nlwcp:fig:invisible-A'}.  

Several constraints are common to both visibly (see \S\ref{sec:nlwcp:dark-photons}) and invisibly decaying dark photons.  
In particular, precision QED measurements~\cite{Pospelov:2008zw} and precision tests of the fine-structure constant $\alpha$ 
(including the electron anomalous magnetic moment) constrain the kinetic mixing parameter $\epsilon \lesssim 10^{-4}~(10^{-2})$ for a dark photon mass  $m_{A'} \sim 1\, {\rm MeV}~(100\, {\rm MeV})$. The muon anomalous magnetic moment provides a stronger constraint for heavier dark photons, with 
$\epsilon \lesssim {\rm few} \times 10^{-3}~(10^{-2})$ for $m_{A'} \sim 50\, {\rm MeV} ~(300\, {\rm MeV})$. Furthermore, model-independent constraints from the measurements of the $Z$-boson mass, precision electroweak observables, and $e^+e^-$ reactions at a variety of c.o.m.~energies constrain $\epsilon \lesssim 3 \times 10^{-2}$ independent of $m_{A'}$~\cite{Hook:2010tw} (not shown).

The next class of constraints relevant to this scenario relies on the assumption that the dark photon decays invisibly (but not necessarily to stable states like DM). 
Measurements of the $K^+ \rightarrow \pi^+ \nu \bar \nu$  branching ratio~\cite{Artamonov:2009sz} place limits on $\epsilon$ in the range $10^{-2}-10^{-3}$ if the decay $K^+ \rightarrow \pi^+ A'$ is kinematically allowed.  Strong constraints on $\epsilon$ exist in a narrow region $m_{A'} \sim m_{J/\psi}$, in which case the decay $J/\psi \rightarrow $ invisible is resonantly enhanced.  Furthermore, a limit on the branching fraction $\Upsilon(3S) \rightarrow \gamma + A^0$, $A^0 \rightarrow$ invisible (with $A^0$ a scalar~\cite{Aubert:2008as}) can be recast as a limit on the continuum process $e^+ e^-  \rightarrow \gamma A'$, $A'\rightarrow$ invisible, leading to $\epsilon \lesssim {\rm few} \times 10^{-3}$~\cite{Essig:2013vha,Izaguirre:2013uxa}.

If the dark sector states $\chi$ are stable ({\rm e.g.}, if $\chi$ is the DM), or at least metastable with lifetimes of $O(100~{\rm m})$, then proton- and electron-beam fixed target experiments are sensitive to the scattering of $\chi$ with electrons or nuclei, with a rate that depends on $\alpha_D$. The LSND measurement of the electron-neutrino elastic scattering cross section~\cite{Auerbach:2001wg} places a limit in the range $\epsilon \lesssim 10^{-5} - 10^{-3}$ for $\alpha_D = 0.1$, $m_A' < m_{\pi^0}$, $m_\chi <  m_A'/2$~\cite{deNiverville:2011it}. Furthermore, the SLAC MilliQ search for milli-charged particles~\cite{Prinz:1998ua} is senstive to $A'$s heavier than $\pi^0$, and constrains values of  $\epsilon$ as low as $10^{-3}$~\cite{Diamond:2013oda} (not shown).
Constraints from the SLAC beam-dump experiment E137 are also applicable~\cite{E137-invisible} (not shown).  

Constraints on dark photons that couple to dark-sector particles with masses up to a few 10's of keV 
can also be obtained from white dwarf star cooling bounds~\cite{Dreiner:2013tja}.  In addition, 
supernova cooling can constrain dark-sector particles with masses as high as 100~MeV~\cite{Dreiner:2013mua}.  

Direct detection experiments can probe light DM $\chi$ in the halo through its scattering with 
electrons~\cite{Essig:2011nj,Essig:2012yx,Graham:2012su}. An analysis of the XENON10 dataset has placed limits on the $\chi$-electron scattering cross section $\sigma_e < 10^{-37}\, {\rm cm}^2$ for $\chi$ masses in the range 20 MeV - 1 GeV~\cite{Essig:2012yx}, and more recent direct detection experiments as well as 
dedicated future experiments could probe significant new parameter space.  

Late-time DM annihilation to charged particles can distort the CMB. Assuming $\chi$ saturates the observed relic density and annihilates through an $s$-wave reaction, then the CMB essentially rules out this scenario~\cite{Padmanabhan:2005es,Madhavacheril:2013cna,Lopez-Honorez:2013cua,Galli:2013dna}. 
These bounds are, however, model-dependent and may be avoided in several ways, for example: 1) $\chi$ may annihilate through a $p$-wave process, e.g.~scalar DM annihilating through an $s$-channel dark photon to SM fermion pairs~\cite{deNiverville:2011it}, 2) the dark sector may contain new light states, opening up new annihilation modes for $\chi$, which do not end with electromagnetic final states, 3) the DM may be matter-asymmetric~\cite{Lin:2011gj}, and 4) $\chi$ may comprise a sub-dominant component of the DM.

\subsubsection{Constraints on Millicharged Particles }
\label{subsubsec:nlwcp:other:pheno:MCP}

\begin{figure}[t!]
 \begin{center}
\includegraphics[width=0.7\textwidth]{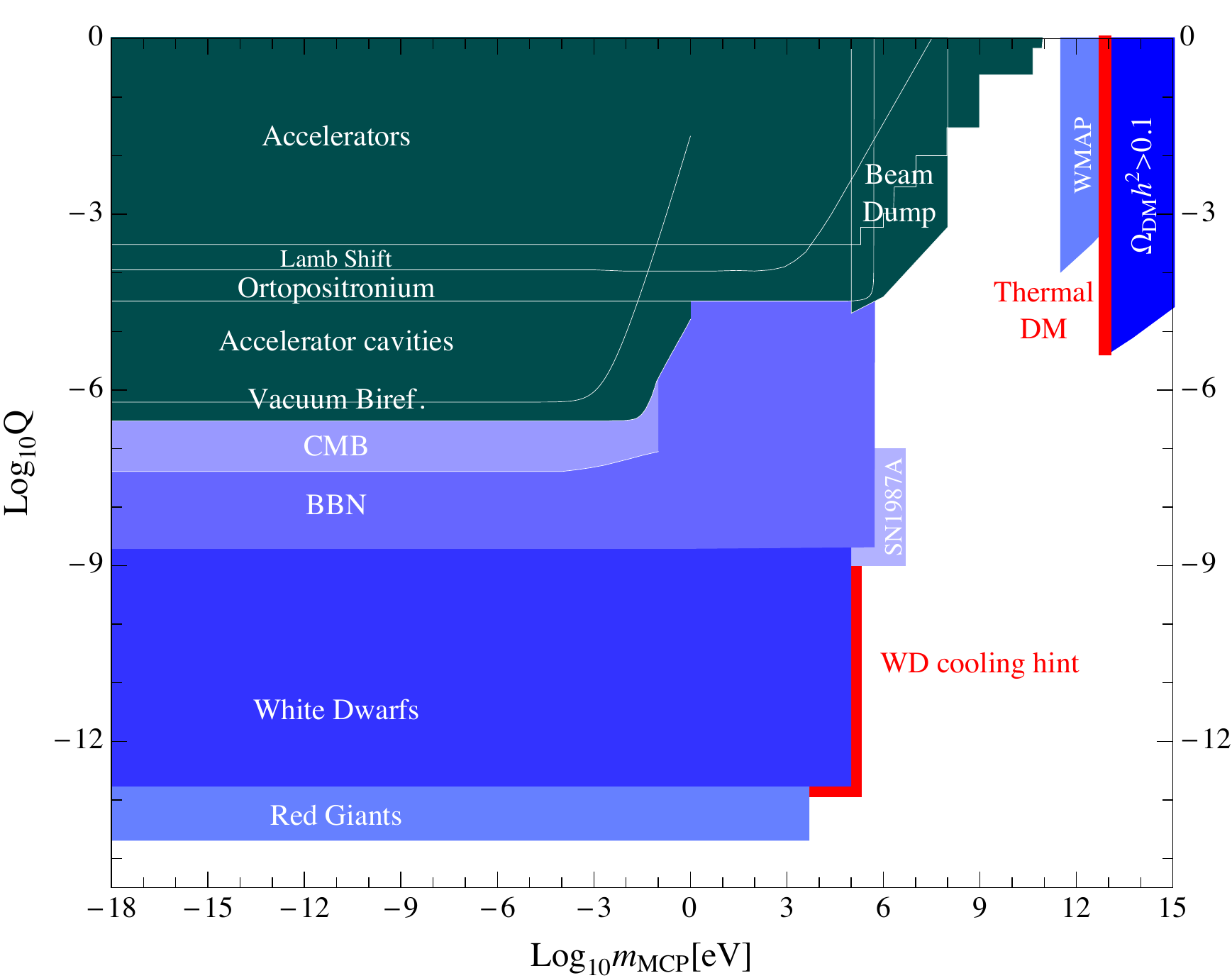}
\caption{Bounds on the millicharge $Q$ vs mass $m_{\rm MCP}$ from astrophysics and various experiments.}\label{fig:hspaw-MCP}
\end{center}
\end{figure}

Several portions of the charge-vs-mass ($Q$-vs-$m_{\rm MCP}$) parameter space for MCPs can be excluded based upon available experimental results. Some of these bounds, e.g.~direct measurements, rely on relatively few assumptions, while others are dependent on the accuracy of astrophysical and cosmological models. Fig.~\ref{fig:hspaw-MCP} illustrates the parameter space for MCPs and a brief summary of the most stringent bounds follows. 

Direct measurements cover large parts of the parameter space of MCPs for $Q\sim e$. The SLAC ASP (Anomalous Single Photon) search  for $e^+ e^-\to\gamma X$ final states, where $X$ is any weakly interacting particle, excludes $Q>0.08e$ for $m_{\rm MCP}~\lsim$ 10~GeV \cite{Davidson:2000hf,Davidson:1993sj}.  Data from a proton beam dump experiment, E613, at Fermilab excludes charges between $10^{-1} e$ and $10^{-2} e$ for $m_{\rm MCP}<200$ MeV~\cite{Golowich:1986tj}. A SLAC electron beam dump experiment looking for trident production $e^- N\to e^- N Q^+ Q^-$ excludes $Q>0.03e$ for $m_{\rm MCP}< 1$ GeV~\cite{Davidson:2000hf}. Moreover, the SLAC MilliQ experiment excludes $Q>5.8\times10^{-4}e$ for $m_{\rm MCP}<100$ MeV~\cite{Prinz:1998ua}. In addition to these accelerator-based experiments, the results of a search for orthopositronium decays into invisible particles can be recast into bounds on MCPs. This measurement excludes $Q  < 8.6\times10^{-5}e$ for $m_{\rm MCP}<500$ keV~\cite{Mitsui:1993ha}. Finally, the precise agreement between the measured and calculated values for the Lamb shift can be used to exclude 
$Q < (1/9)e$ for $m_{\rm MCP} \gtrsim 3$ keV~\cite{Dobroliubov:1989mr,Davidson:2000hf}.

Additional constraints can be placed on MCPs from indirect cosmology and astrophysics results (See \cite{Davidson:2000hf} and references therein). Photons travelling in a plasma acquire an effective mass and can decay into MCPs. Therefore MCPs produced inside stars can contribute to their cooling. White dwarf and red giant stars bound MCPs for $m_{\rm MCP}~\lsim$ keV 
by requiring that the MCP production rate not exceed the rate of nuclear energy production. 

Constraints from cosmology include bounds from BBN on the number of effective relativistic degrees of freedom. CMB data from WMAP is also an indirect testing ground for new invisible states that inject charged particles into the CMB.  In addition, requiring that the MCP relic density not over-close the Universe excludes $m_{\rm MCP}\sim $ TeV for $Q\sim e$.
For sub-eV masses, light-shining-through-wall-type
setups can even go below some cosmological constraints~\cite{Dobrich:2012sw,Dobrich:2012jd}.

New electron and proton beam dump experiments, planned or proposed to search for light DM, could also cover new parameters space of MCPs, particularly the $m_{\rm MCP}\sim \text{GeV}$ region. The primary modes of production are $p N\to p N Q^+Q^-$ or $p p \to Q^+ Q^-$ at proton beam dump experiments, and $e^- N \to e^- N Q^+Q^-$ at electron beam dump experiments. MCPs produced at the beam dump would exit the dump and travel to the detector, where they could scatter elastically and deposit measurable energy, like 
neutral current events. The detection of MCPs relies on an experiment sensitive to low momentum recoil channels, such as electron recoils and/or coherent nuclear scattering, see \S\ref{subsubsec:nlwcp:other:future:electron}.

\subsection{Proposed and Future Searches}
\label{subsec:nlwcp:other:future}

\subsubsection{Proton Beam Dump Experiments}
\label{subsubsec:nlwcp:other:future:proton}

Proton beam dump experiments have significant potential to search for light DM and other long-lived dark sector states. An intense source of dark sector states can be produced in the primary proton-target collisions and detected through their scattering~\cite{Batell:2009di,deNiverville:2011it,deNiverville:2012ij} or visible decays~\cite{Batell:2009di,Essig:2010gu} in a near detector. Of particular importance to this experimental program are the existing and future Fermilab neutrino experiments such as MiniBooNE, MINOS, NO$\nu$A, MicroBooNE, and LBNE, which have an unprecedented opportunity to search for light DM. 
The studies in~\cite{Batell:2009di,deNiverville:2011it,deNiverville:2012ij} demonstrate the existence of a large DM signal in existing neutrino experiments for motivated regions of DM parameter space. However, numerous experimental challenges remain to maximize the sensitivity to the DM signal, foremost among them contending with the large neutrino neutral current background.

A proposal for a dedicated search for light DM at MiniBooNE is described in~\cite{Dharmapalan:2012xp}. DM particles $\chi$, interacting with the SM through a kinetically mixed dark photon, can be produced through the decays of secondary pseudoscalar mesons, 
$\pi^0, \eta \rightarrow \gamma A'$, $A'\rightarrow \chi \chi^*$.  Such DM particles can travel to the detector and scatter via $A'$ exchange, leaving the signature of a recoiling electron or nucleon. The MiniBooNE sensitivities to DM masses of $1$, $10$, $100$ MeV are represented by the green contours in Fig.~\ref{nlwcp:fig:invisible-A'}. 
MiniBooNE can probe motivated regions of DM parameter space in which the relic density is saturated and the muon 
anomalous magnetic moment discrepancy is explained. 
The signal significance for several operational modes can be found in \cite{Dharmapalan:2012xp}.

In order to mitigate the neutrino background,~\cite{Dharmapalan:2012xp} proposed to run in a beam-off target configuration, in which the protons are steered past the target and onto either 1) the permanent iron absorber located at the end of the 50 m decay volume, or 2) a deployed absorber positioned 25 m from the target. 
A one week test run in the 50 m absorber configuration measured a reduction of the neutrino flux by a factor of 42. Additional improvements in distinguishing $\chi$ signal from the neutrino background are possible by exploiting the ns-level timing resolution between the detector and proton spill, since heavy $O(100\, {\rm  MeV})$  DM particles will scatter out of time. 

The experimental approach to search for light DM employed by MiniBooNE is applicable to other neutrino experiments and intense proton sources, such as  MINOS, NO$\nu$A, MicroBooNE, LBNE, and Project X. For instance, the MicroBooNE LAr detector can also perform a search with comparable sensitivity to that outlined for MiniBooNE with a long enough beam-off-target run.  
More generally, the DM mass range that can be covered is governed by the proton beam energy and the production mechanism, as well as the ability to overcome the neutrino neutral current background. 
$A'$s with masses well above those kinematically accessible in $\pi^0$ decay, or high-mass virtual $A'$s, 
can be produced in hadronic showers and decay to DM at masses up to a few GeV.  
For instance with the FNAL Booster (8.9 GeV) and Main Injector (120 GeV) as well as a future Project X, the accessible DM mass range is a few MeV to a few GeV. Both LBNE~\cite{Adams:2013qkq} and Project X~\cite{Kronfeld:2013uoa} have considered light DM searches to expand the physics reach and help motivate the projects.

The search for light DM provides an additional physics motivation for intense proton beam facilities. Given the significant investment in existing and future infrastructure for neutrino experiments, it is critical to take advantage of the unique opportunity afforded by these experiments to probe the non-gravitational interactions of light DM and more generally explore the possibility of of a dark sector with new,  light, weakly-coupled states.

\subsubsection{B-factories}
\label{subsubsec:nlwcp:other:future:SuperB}

B-factories like \babar\ and Belle and future super-B factories like Belle 2 are powerful probes of light DM with a light mediator.  
An existing mono-photon search by \babar~\cite{Aubert:2008as} already places important constraints on this class of 
models~\cite{Essig:2013vha,Izaguirre:2013uxa} (see also~\cite{Borodatchenkova:2005ct, Fayet:2007ua, Essig:2009nc}), and 
a similar search at a future B-factory can probe significantly more parameter space~\cite{Essig:2013vha}. 
Such searches are more powerful than searches at other collider or fixed-target facilities for 
mediator and hidden-sector particle masses between a few hundred MeV to 10 GeV.  
Mediators produced on-shell and decaying invisibly to hidden-sector particles such as DM can be probed particularly well. 
Sensitivity to light DM produced through an off-shell mediator is more limited, but may be improved with a better theoretical control of backgrounds, allowing background subtraction and a search for kinematic edges.
The implementation of a mono-photon trigger at Belle II would be a necessary step towards providing this crucial window into such light dark sectors.

\subsubsection{Electron beam dump experiments}
\label{subsubsec:nlwcp:other:future:electron}

Electron beam dump experiments enable powerful low-background searches for new light weakly-coupled particles and 
can operate parasitically at several existing facilities \cite{Izaguirre:2013uxa}. 
Electron-nucleus collisions feature a light dark-sector particle production rate comparable to that of neutrino factories, but the production
mechanism is analogous to QED bremsstrahlung. Importantly, beam related neutrino and neutron backgrounds are 
 negligible. Electron beam production also features especially forward-peaked particle kinematics, so for multi-GeV beam energies,
  experimental baselines on a 10~m scale, and 1~m-scale detectors, the signal acceptance is of order one for sub-GeV  
DM masses. This approach is sensitive to any new physics that couples to leptonic currents and is limited only by cosmogenic backgrounds,
 which are both beatable and systematically reducible; even a design with no cosmogenic neutron reduction 
 offers sensitivity to well motivated regions of parameter space.  
 Previous generations of electron beam experiments, such as the MilliQ or E137 experiment at SLAC have already demonstrated 
 sensitivity to light DM~\cite{Diamond:2013oda,E137-invisible}.

The minimal setup requires a $\mathcal{O}$(m$^3$) fiducial volume (or smaller) detector sensitive to neutral current scattering 
 placed 10s of meters downstream of an existing electron beam dump. At low momentum transfers,
DM scattering predominantly yields elastic electron and coherent nuclear recoils in the detector with high 
cross sections but low energy depositions, so potentially high backgrounds.  At higher
  momentum transfers,  inelastic hadro-production and quasi-elastic nucleon ejection dominate the signal yield, have larger energy 
  depositions, so comparatively lower backgrounds.
  The approach can offer comparable sensitivity using either continuous wave (CW) or pulsed electron beams, but CW sensitivity is limited
  by cosmogenic background so background reduction strategies are required to achieve optimal sensitivity; 
  for pulsed beams, integrated luminosities may be substantially lower, but timing cuts can reduce cosmogenic backgrounds 
  significantly.  This approach can be realized parasitically at several existing electron fixed target facilities 
  including SLAC, JLab, and Mainz. It may also be possible to utilize pulsed beams at the SuperKEK linac and at a 
  future ILC. 

Fig.~\ref{nlwcp:fig:invisible-A'} shows the sensitivity projections for a 1~m$^3$ detector placed 20~m downstream of an 
Al~beam dump.  Two lines show the 10-event sensitivity per $5\times 10^{22}$ electrons-on-target, for a 
hypothetical experiment at JLab using a 12 GeV beam (red) and at ILC using a 125 GeV beam (purple).   Excellent sensitivity is 
obtained for light $A'$ and light dark-matter masses.

\subsubsection{Rare Kaon decays}
\label{subsubsec:nlwcp:other:future:kaons}

As discussed in \S\ref{subsubsec:nlwcp:other:pheno:LDM}, rare Kaon decays can already set some limits on invisible 
$A'$ decays.  Future searches for rare Kaon decays, including e.g.~$K^+\to\pi^+ A'$, will have sensitivity to new regions 
in parameter space.  Examples include ORKA~\cite{Essig:2013vha,E.T.WorcesterfortheORKA:2013cya} 
(see Fig.~\ref{nlwcp:fig:invisible-A'}), NA62~\cite{NA62}, and TREK/E36~\cite{TREK/E36,Strauch}.

\section{Chameleons}
\label{sec:nlwcp:chameleons}

\subsection{Theory \& Motivation}
\label{subsec:nlwcp:chameleons:theory}

Cosmological observations are able to pinpoint with great precision details of the Universe on the largest scales, while particle physics experiments probe the nature of matter on the very smallest scales with equally astounding precision. However, these observations have left us with some of the greatest unsolved problems of our time, most notably the remarkable realisation that the most dominant contribution to the energy density of our Universe is also the least well understood. Dark energy, credited with the observed accelerated expansion of the Universe, makes up around 70\% of the total matter budget in the Universe; however, there is no single convincing explanation for this observation nor is there a clear pathway to disting between different models through cosmological observations. If this acceleration is not caused by a cosmological constant then the most convincing explanations come in the form of scalar field models that are phenomenological but with the hope of being effective field theories of ultra-violet physics. If a scalar field is indeed responsible for this observed acceleration, it would need to be very light, $m\sim H_0$, and evolving still today. These light fields should couple to all forms of matter with a coupling constant set by $G_N$. 
A coupling of this kind would cause an as yet unobserved fifth force and should be observable in a plethora of settings from the early Universe through BBN, structure formation, and in all tests of gravity done today. Thus, we are left with a puzzle as to how a scalar field can both be observable as dark energy and yet not be observed to date in all other contexts. 

A solution to this puzzle was presented in  \cite{Khoury_Weltman_2004a,Khoury_Weltman_2004b,Brax_etal_2004} with so-called chameleon fields. Chameleon fields are a compelling dark energy candidate, as they couple to all SM particles without violating any known laws or experiments of physics. Importantly, these fields are testable in ways entirely complementary to the standard observational cosmology techniques, and thus provide a new window into dark energy through an array of possible laboratory and astrophysical tests and space tests of gravity. Such a coupling, if detected, could reveal the nature of dark energy and may help lead the way to the development of a quantum theory of gravity.

A canonical scalar field is the simplest dynamical extension of the SM that could explain dark energy.  In the absence of a self interaction, this field's couplings to matter --- which we would expect to exist unless it is forbidden by some symmetry  --- would lead to a new, fifth fundamental force whose effects have yet to be observed. However, scalar field dark energy models typically require a self interaction, resulting in a nonlinear equation of motion~\cite{Peebles_Ratra_1988,Ratra_Peebles_1988}.  Such a self interaction, in conjunction with a matter coupling, gives the scalar field a large effective mass in regions of high matter density~\cite{Khoury_Weltman_2004a,Khoury_Weltman_2004b}. A scalar field that is massive locally mediates a short-range fifth force that is difficult to detect, earning it the name ``chameleon field.''  Furthermore, the massive chameleon field is sourced only by the thin shell of matter on the outer surface of a dense extended object.  These nonlinear effects serve to screen fifth forces, making them more difficult to detect in certain environments.

Current best theories treat chameleon dark energy as an effective field theory~\cite{Brax_etal_2004,Hui_Nicolis_2010} describing new particles and forces that might be seen in upcoming experiments, and whose detection would point the way to a more fundamental theory. The ultraviolet (UV) behavior of such theories and their connection to fundamental physics are not yet understood, although progress is being made~\cite{Hinterbichler:2010wu, NastaseKhoury, Nastase:2013ik, Nastase:2013los}.  

A chameleon field couples to DM and all matter types, in principle with independent strengths. At the classical level, a chameleon field is not required to couple to photons, though such a coupling is not forbidden.
However, when quantum corrections are included, a photon coupling about three orders of magnitude smaller than the matter coupling is typically generated~\cite{Brax_etal_2010}.  The lowest order chameleon-photon interaction couples the chameleon field to the square of the photon field strength tensor, implying that in a background electromagnetic field, photons and chameleon particles can interconvert through oscillations.  
The mass of chameleon fields produced will depend on the environmental energy density as well as the electromagnetic field strength. This opens the vista to an array of different tests for these fields on Earth, in space, and through astrophysical observations.  Several astrophysical puzzles could also be explained by chameleons, e.g.~\cite{Hui:2009kc}. Their coupling to photons, combined with their light masses in certain environments, allows chameleons to be produced with intense beams of photons, electrons, or protons and detected 
with sensitive equipment.  This makes them, by definition, targets for the intensity frontier. In fact chameleon particles are a natural bridge between the cosmic frontier and the intensity frontier; not only do they hold the possibility of being a dark energy candidate but they are testable through astrophysical and laboratory means.  

The chameleon dark energy parameter space is considerably more complicated than that of axions, but constraints can be provided under some assumptions. With the caveat that all matter couplings are the same but not equal to the photon coupling,  and the assumption of a specific chameleon potential,  $V(\phi) = M_\Lambda^4(1 + M_\Lambda^n/\phi^n)$ in which we set the scale $M_\Lambda = 2.4\times 10^{-3}$~eV to the observed dark energy density and, for concreteness, $n=1$, our constraints and forecasts are provided by Fig.~\ref{f:chameleon_constraints}. Current constraints (solid regions) and forecasts (curves) are discussed below.
\subsection{Current laboratory constraints}
\label{subsec:nlwcp:chameleons:constraints}

\begin{figure}[tb]
\begin{center}
\includegraphics[width=0.7\textwidth]{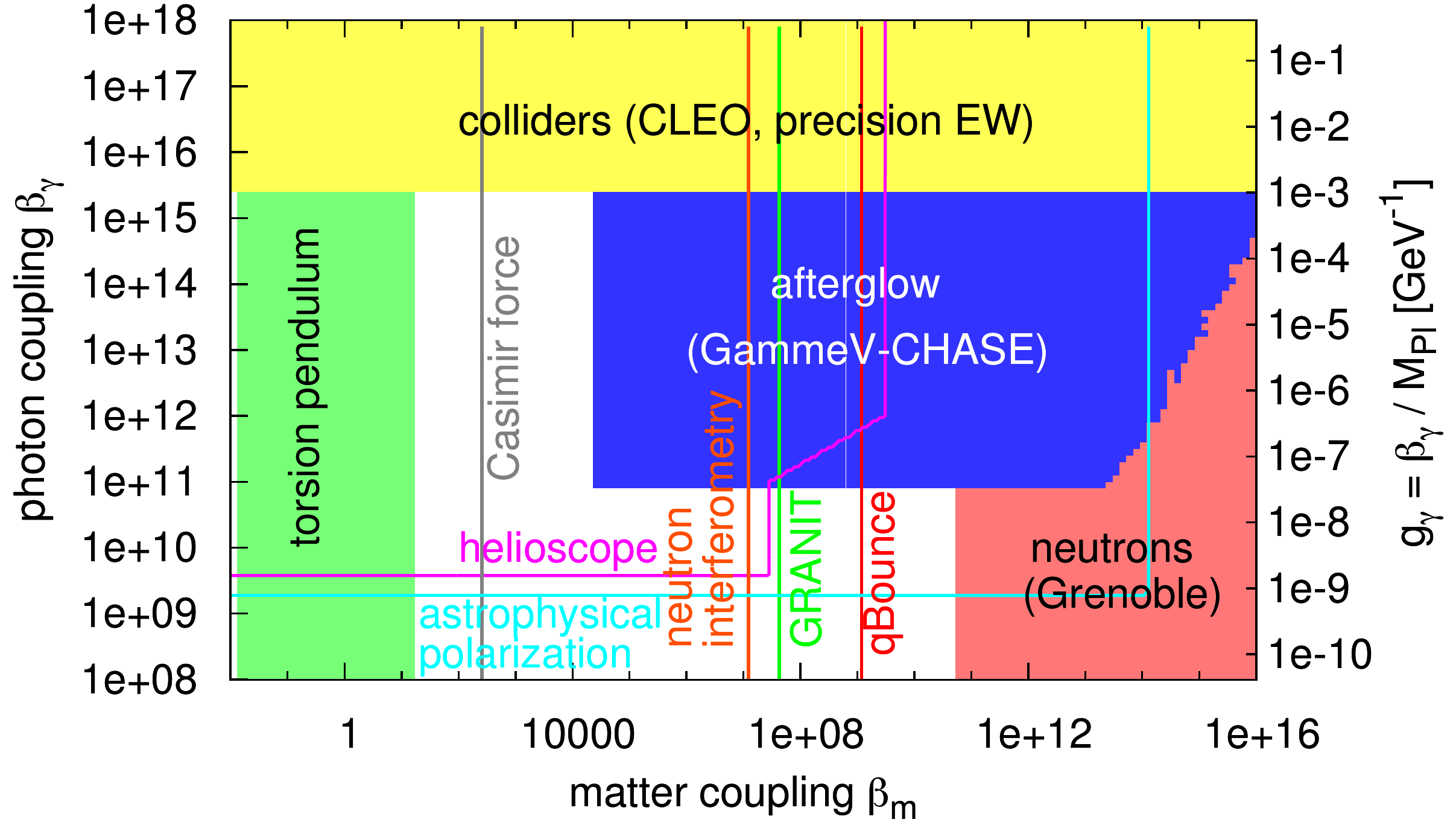}
\caption{Constraints on the matter and photons couplings for a chameleon dark energy model with $V(\phi) = M_\Lambda^4(1+M_\Lambda/\phi)$.  Current constraints are shown as shaded regions, while forecasts are shown as solid lines.  
\label{f:chameleon_constraints}}
\end{center}
\end{figure}

Laboratory constraints on chameleon dark energy come from two different types of experiments: fifth force searches, and photon coupling experiments, both of which are shown as shaded regions in Figure~\ref{f:chameleon_constraints} and listed in Table~\ref{t:expts}. 
 Gravitation-strength fifth forces can be measured directly between two macroscopic objects, such as the source and test masses in a torsion pendulum.  Currently the shortest-range torsion pendulum constraints on gravitation-strength forces come from the \eotwash~experiment~\cite{Kapner_etal_2007}.  The source and test masses in \eotwash~are parallel metal disks a few centimeters in diameter with matched sets of surface features.  As the lower disk is rotated, gravity and any fifth forces induce torques in the upper disk so as to align the surface features.  The separation between the disks can be varied, and the torsional oscillations in the upper disk can be compared with predictions.  Another type of fifth force experiment uses an ultracold gas of neutrons whose bouncing states in the gravitational field of the Earth are quantized, with energy splittings $\sim 1$~peV~\cite{Brax_Pignol_2011}.  If the neutrons feel a fifth force from the experimental apparatus comparable to the gravitational force of the Earth, then the energy splittings will be altered.  The Grenoble experiment measures these energy splittings at the $\sim 10\%$ level, excluding very strong matter couplings $\beta_m \gtrsim 10^{11}$.  

Dark energy may couple to the electroweak sector in addition to matter.  Such a coupling would allow photons propagating through a magnetic field to oscillate into particles of dark energy, which can then be trapped inside a chamber if the dark energy effective mass becomes large in the chamber walls.  An ``afterglow experiment'' produces dark energy particles through oscillation and then switches off the photon source, allowing the population of trapped dark energy particles to regenerate photons which may emerge from the chamber as an afterglow.  Current afterglow constraints from the CHASE experiment \cite{Steffen_etal_2010} exclude photon couplings $10^{11} \lesssim \beta_\gamma \lesssim 10^{16}$ for $\beta_m \gtrsim 10^4$, as shown in Fig.~\ref{f:chameleon_constraints} for an inverse-power-law chameleon potential~\cite{Steffen_etal_2010,Upadhye_Steffen_Weltman_2010,Upadhye_Steffen_Chou_2012}.  At yet higher photon couplings the trapped dark energy particles regenerate photons too quickly for CHASE to detect them.  However, collider experiments can exclude such models, by constraining chameleon loop corrections to precision electroweak observables~\cite{Brax_etal_2009}.  

\begin{table*}[tb]
\begin{center}
\begin{tabular}{|l||c|c|}
\hline
Experiment &   Type           & Couplings excluded                \\
\hline
\hline
\eotwash   & torsion pendulum & $0.01 \lesssim \beta \lesssim 10$ \\
\hline
Lamoreaux  & Casimir          & $\beta \gsim 10^5$ ($\phi^4$)   \\
\hline
Grenoble   & bouncing neutron & $\beta \gsim 10^{11}$           \\
\hline
GRANIT     & bouncing neutron & forecast: $\beta \gsim 10^8$    \\
\hline
NIST       & neutron interferometry & forecast: $\beta \gsim 10^7$ \\
\hline
CHASE      & afterglow        &                              $10^{11} \lesssim \beta_\gamma \lesssim 10^{16}$ subject to \mbox{$10^4 \lesssim \beta_m \lesssim 10^{13}$,}   \\
\hline
ADMX       & microwave cavity & \mbox{$\meff = 1.952~\mu$eV,}                                                   $10^9 \lesssim \beta_\gamma \lesssim 10^{15}$\\
\hline
CAST       & helioscope       & \mbox{forecast: $\beta_m \lesssim 10^{9}$,}                                       $\beta_\gamma > 10^{10}$\\
\hline
\end{tabular}
\end{center}
\caption{Laboratory tests of dark energy.  Approximate constraints on chameleon models with potential $V(\phi) = M_\Lambda^4(1 + M_\Lambda/\phi)$ and $M_\Lambda = 2.4\times 10^{-3}$~eV (unless otherwise noted).  \label{t:expts}}
\end{table*}

\subsection{Forecasts for Terrestrial experiments}
\label{subsec:nlwcp:chameleons:experiments}

Proposed experiments promise to improve constraints on chameleon dark energy by orders of magnitude over the next several years.  Fig.~\ref{f:chameleon_constraints} summarizes forecasts and preliminary constraints, shown as solid lines.
The next-generation \eotwash\space experiment, currently under way, will have an increased force sensitivity and probe smaller distances.  This will allow it to detect or exclude a large class of chameleon models with well-controlled quantum corrections~\cite{Upadhye_Hu_Khoury_2012,Upadhye_2012}.
Improvements to fifth force measurements using neutrons should improve constraints on the chameleon-matter coupling considerably.  Also proposed is a neutron interferometry experiment at NIST, which should be competitive with the bouncing neutron experiments.  A neutron interferometer splits a neutron beam and sends the two through two different chambers, one containing a dense gas which suppresses chameleon field perturbations, and the other a vacuum chamber in which scalar field gradients are large.  These gradients will retard the neutron beam passing through the vacuum chamber, resulting in a phase shift which varies nonlinearly with the gas pressure.  Potentially more powerful are the next-generation Casimir force experiments~\cite{Brax_etal_2007c}.  However, these currently suffer from systematic uncertainties including the proper calculation of thermal corrections to the Casimir effect.  The forecasts shown require that the total uncertainty in the Casimir force be reduced below $1\%$ at distances of $5-10~\mu$m.

Other planned experiments search for photon-coupled chameleons.  Afterglow experiments have been proposed at JLab and the Tore Supra tokamak, while a microwave cavity-based afterglow experiment is under way at Yale.  Since forecasts for these experiments are not available for the chameleon potential assumed in Fig.~\ref{f:chameleon_constraints}, we are unable to include them in the figure.  However, the JLab and Tore Supra experiments are expected to fill in some of the gap between CHASE and torsion pendulum experiments, while the microwave cavity search is a precision experiment capable of targeting a model with a specific mass in response to hints from an afterglow experiment.  Yet another type of experiment is the helioscope, which uses a high magnetic field to regenerate photons from scalar particles produced in the Sun \cite{Brax:2011wp}.  Since such particles do not need to be trapped prior to detection, helioscope forecasts extend down to arbitrarily low matter couplings.  One proposed helioscope adds an X-ray mirror to the CAST axion helioscope at CERN in order to increase its chameleon collecting area; forecasts for this experiment are shown.

\subsection{Tests of the Chameleon Mechanism by Astrophysical Observation}
\label{subsubsec:nlwcp:chameleons:observations}

Complimentary to detector based experiments, chameleons offer a rich phenomenology of unique astrophysical signatures. Combining data from astrophysical observations with laboratory experimental data will allow us to constrain chameleon models. Below we review some of the more intriguing astrophysical signatures predicted in chameleon models. One benefit of observational tests of chameleons is that these observations may be performed complimentarily with observations taken for reasons not related to chameleon gravity. Ordinary matter interacting via a low mass particle ($m\sim H_0$) leading to a new fifth force typically requires a very small coupling. Bounds on any additional fifth force have been set by measuring the frequency shift of photons passing near the Sun from the Cassini satellite on their way to Earth \cite{Bertotti:2003rm}. 

The screening mechanism from chameleons has significant consequences for the formation of structure. These modifications to structure formation include an earlier collapse of density perturbations compared to the prediction from $\Lambda$CDM and clumpier DM halos \cite{Brax:2005ew}. Another effect on structure formation in chameleon gravity is that the critical density required for collapse depends on the comoving size of the inhomogeneity itself \cite{Brax:2010tj}. Also, galactic satellite orbits become modified based on the size of the satellite itself due to a backreaction from the satellite causing a velocity difference of up to 10\% near the thin shell \cite{Pourhasan:2011sm}.

Due to the existence of the two-photon vertex ($\mathcal{L}_{\phi\leftrightarrow\gamma}=F^{\mu\nu}F_{\mu\nu}\phi/4M$), chameleons mix with photons in the presence of a background magnetic field. This mixing is the result of the propagation eigenstates being different from the photon polarization-chameleon eigenstates. The result of this mixing is a non-conservation of photons. In the case of type Ia supernovae, \cite{Burrage:2007ew} demonstrated that photons convert to chameleons in the interior of the supernova, pass through the surface of the supernova, and than convert back to photons in the intergalactic magnetic field. The net result is an observed brightening of supernovae. This scenario provides an explanation for the discrepancy between distance measurements of standard candles and standard rulers beyond $z\sim 0.5$ \cite{Bassett:2003vu}. 

Another prediction of chameleon gravity is that in unscreened environments, (such as voids) stellar structure is modified, 
most notably in the red giant branch of stars. The authors of \cite{Chang:2010xh} found that chameleons affect the size 
and temperature of red giant stars where they tend to be smaller ($\sim 10\%$), and hotter ($\sim100$s of Kelvins). 
Also, observations of circularly polarized starlight in the wavelength range $1-10^3$\AA~could be a strong 
indication of chameleon-photon mixing \cite{Burrage:2008ii}. 
 
Astrophysical tests of chameleons in $f(R)$ theories may be parameterized by how efficiently bodies self-screen ($\chi$) and the strength of the fifth force ($\alpha$) \cite{Jain:2012tn}. For the case of chameleon $f(R)$ gravity, $\chi\equiv df/dR$ is measured at present time. The additional force is parametrized by rescaling Newton's constant $G\rightarrow G(1+\alpha)$ for unscreened objects and $G(r)\rightarrow G[1+\alpha(1-M(r_s)/M(r))]$ for partially screened objects. Fifth forces are screened at radii $r<r_s$, unscreened for radii $r_s<r$, and $M(r)$ is the mass contained within a shell at radius $r$. For an object to be unscreened, $\Phi_N\ll\chi$ where $\Phi_N$ is the Newtonian potential. The Sun and Milky Way (coincidentally) possess a similar gravitational potential: $\Phi_\odot\sim 2\times10^{-6}$ and $\Phi_{MW}\sim10^{-6}$. Stars in the tip of the red giant branch of the HR diagram and Cepheid variables have gravitational potentials $\Phi_N\sim 10^{-7}$. These stars will have their outer layers unscreened provided they reside in smaller galaxies in a shallow gravitational potential. For fifth forces of a strength described by $\alpha=1/3$, values of $\chi$ greater than $5\times10^{-7}$ may be ruled out at $95\%$ confidence. This upper bound is moderately lower for fifth force strength defined by $\alpha=1$, where values of $\chi$ greater than $1\times10^{-7}$ may be ruled out at a $95\%$ confidence level \cite{Jain:2012tn} (also see Fig.(5) of \cite{Jain:2012tn}). These constraints on $\chi$ and $\alpha$ from local Universe observations are stronger than current cosmological constraints on chameleon fifth forces \cite{Schmidt:2009am}-\cite{Lombriser:2012nn} which typically give an upper limit not less than $\chi\sim10^{-6}$.

\subsection{Space tests of Gravity}
\label{subsec:nlwcp:chameleons:space}

Remarkably, the original predictions of signatures in space for chameleon models would still be the most striking \cite{Khoury_Weltman_2004a,Khoury_Weltman_2004b}. The proposed experiments discussed there have not yet taken place. However, the MicroSCOPE \cite{micro} mission and STE-QUEST \cite{quest} are future satellite experiments that hold the promise of testing these theories in a way complementary to the terrestrial and astrophysical methods discussed here. The expected signatures are large and for example an $\mathcal{O}(1) $ observed difference in Newton's constant for unscreened objects would be a smoking gun for these models. 

There is great potential for testing chameleon theories in the laboratory, the sky and through astrophysics ; both at the cosmic and the intensity frontiers. The possibilities for astrophysics are discussed further under the Novel Probes of Dark Energy and Gravity in the Cosmic Frontier.

\section{Conclusions}
\label{sec:nlwcp:conclusions}

Establishing the existence of a dark sector, and the new light weakly-coupled particles it could contain, would revolutionize 
particle physics at the Copernican level: once again our simple conception of Nature would be fundamentally altered, and here 
we would realize that there is much more to the world than just the SM sector.  Searches for dark-sector particles
are strongly motivated by our attempts to understand the nature of the dark matter, the strong CP problem, and puzzling astrophysical 
and particle physics
observations. New physics need not reside exclusively at the TeV scale and beyond; it could well be found at the low-energy frontier 
and be accessible with intensity frontier tools.
Axions, invented to solve the strong CP problem, are a perfect dark matter candidate. 
Dark photons, and any dark-sector particles that they couple to, can be equally compelling dark matter candidates, could 
resolve outstanding puzzles in particle and astro-particle physics, and may also explain dark matter interactions with the 
SM. Other dark-sector particles could account for the Dark Energy.  Discovery of any of these particles would 
redefine our worldview.

Existing facilities and technologies, modest experiments, and experimental cleverness enable the exploration of dark sectors.
Searches for new light weakly-coupled particles depend on the tools and techniques of the intensity frontier, i.e.~intense 
beams of photons and charged particles, on technological means of dealing with high intensities, and on extremely sensitive, 
needle-in-the-haystack detection techniques.  A rich, diverse, and low-cost experimental program is already underway that 
has the potential for one or more game-changing discoveries.  Current ideas for extending  the searches to smaller couplings 
and higher masses increase this potential markedly.  The US high-energy physics program needs to include these experimental 
searches, especially when the investment is so modest, the motives so clear, and the payoff so spectacular. 
At present, nearly all the experimental efforts world-wide  have strong US contributions or significant US leadership, a position 
that should be maintained. 

Axions, ALPS, dark photons, milli-charged particles, and light dark matter are all naturally connected by their dark-sector 
origins, and by the fact that all these particles couple to the photon, either directly, or through couplings induced 
by kinetic mixing.  Microwave cavities and light-shining-through-walls experiments designed to search for axions and ALPs have 
been adapted to search for dark photons. So have helioscopes looking for solar axions.  A series of beam dump experiments, 
originally motivated  as axion searches, have been reinterpreted to set important limits on dark photon couplings and masses.  
More recently, a new series of electron and proton beam dump experiments, the latter capitalizing on existing neutrino detectors 
and eventually Project X beam intensities, will hunt for signs of light dark matter produced in the dump by dark 
photon decays. 

Searches  for new, light, weakly-coupled particles are, compared to typical  contemporary particle physics experiments, small, 
accessible, hands-on, and personal in a way that is impossible in much larger efforts. This environment offers ideal 
educational opportunities for undergraduates, graduate students, and postdocs, and revitalizes more experienced physicists too. 
All must deal with the full breadth of experimental activities: theory, design, proposal writing and defense, 
hardware construction and commissioning, software implementation, data taking, and analysis.  
Theorists and experimentalists have been brought into very close collaboration, to the benefit of both camps and the field as a whole.

A great deal can be done with existing tools and techniques, in searching for QCD axions that could account for the dark 
matter, in extending searches for dark photons throughout the favored parameter space, and in searching for new dark-sector  
particles like light dark matter.  Even more is possible with the addition of relatively modest investments in superconducting 
magnets, more sensitive microwave detection, resonant optical cavities, high rate, highly pixelated silicon detectors, and new
higher energy electron accelerators, high intensity proton facilities, and upgraded  $e^+e^-$ and $pp$ colliding beam facilities.  
Modest investments will pay great dividends. 

In conclusion, the search for dark sectors and new, light, weakly-coupled particles should be 
vigorously pursued in the US and elsewhere.

\acknowledgments
RE is supported in part by the DoE Early Career research program DESC0008061 and by a Sloan Foundation Research Fellowship.  
Support for JJ from DOE HEP at SLAC National Accelerator Laboratory is gratefully acknowledged. 
Part of this work was performed under the auspices of the U.S.~Department of Energy by Lawrence 
Livermore National Laboratory under Contract DE-AC52-07NA27344.


\end{document}

%% file: workshopsymbols.tex
\newcommand{\nc}{\newcommand}  

\def\etal{{\it et al.}}

\def\beq{\begin{equation}}
\def\eeq#1{\label{#1}\end{equation}}
\def\eeqn{\end{equation}}

\newenvironment{Eqnarray}%
   {\arraycolsep 0.14em\begin{eqnarray}}{\end{eqnarray}}
\def\beqa{\begin{Eqnarray}}
\def\eeqa#1{\label{#1}\end{Eqnarray}}
\def\eeqan{\end{Eqnarray}}

\nc{\ra}{\rightarrow}  
\nc{\slsh}{\slash\hspace*{-0.22cm}}
\def\Re{{\cal R \mskip-4mu \lower.1ex \hbox{\it e}\,}}
\def\Im{{\cal I \mskip-5mu \lower.1ex \hbox{\it m}\,}}

\nc{\vev}[1]{ \left\langle {#1} \right\rangle }
\nc{\bra}[1]{ \langle {#1} | }
\nc{\ket}[1]{ | {#1} \rangle }
\nc{\fb}{\,{\rm fb}^{-1}}
\nc{\ev}{{\rm eV}}
\nc{\kev}{{\rm keV}}
\nc{\Mev}{{\rm MeV}}
\nc{\gev}{{\rm GeV}}
\nc{\tev}{{\rm TeV}}
\nc{\mev}{{\rm MeV}}

\def\del{\partial}
\def\Dslash{\not{\hbox{\kern-4pt $D$}}}
\def\dslash{\not{\hbox{\kern-2pt $\del$}}}
\def\pslash{\not{\hbox{\kern-2pt $p$}}}
\def\ETmiss{ \not{\hbox{\kern-4pt $E$}}_T }

\def\msb{{\bar{\ssstyle M \kern -1pt S}}}

\def\babar{\mbox{\sl B\hspace{-0.4em} {\small\sl A}\hspace{-0.37em} \sl B\hspace{-0.4em} {\small\sl A\hspace{-0.02em}R}}}